\documentclass[a4paper,12pt,oneside]{book}
\usepackage{ifthen}
\usepackage{multicol}
\usepackage{alltt}
\usepackage{wrapfig}
\usepackage{epsfig}
\usepackage{subfigure}
\usepackage{cite}
\usepackage{doublespace}

\newlength{\craise}
\newcommand{\McElwaine}{%
  \settoheight{\craise}{M}%
  \addtolength{\craise}{-1ex}%
  M\raise\craise\hbox{c}\kern.1ex Elwaine}

\newtheorem{theorem}{Theorem}[chapter]

\newcommand{\capa}{}\newcommand{\capb}{}
\newcommand{\pairoffigures}[4][tree]{%
  \renewcommand{\capa}{}\renewcommand{\capb}{}
  \ifthenelse{\equal{#1}{tree}}{\renewcommand{\capa}{probability tree}}{}%
  \ifthenelse{\equal{#1}{ds}}{\renewcommand{\capa}{consistency statistics}}{}%
  \ifthenelse{\equal{#1}{pjt}}{\renewcommand{\capa}{projection times}}{}%
  \ifthenelse{\equal{#2}{tree}}{\renewcommand{\capb}{probability tree}}{}%
  \ifthenelse{\equal{#2}{ds}}{\renewcommand{\capb}{consistency  statistics}}{}%
  \ifthenelse{\equal{#2}{pjt}}{\renewcommand{\capb}{projection times}}{}%
  \begin{figure}[ht]
    \mbox{} \hfill
    \subfigure[\capa]{\epsfig{file=figures/#3.#1.ps}}
    \hfill 
    \subfigure[\capb]{\epsfig{file=figures/#3.#2.ps}}
    \hfill \mbox{}
    \caption{#4}
    \label{fig:#3}
  \end{figure}}

\newcommand{\onefigure}[3][tree]{%
  \renewcommand{\capa}{}
  \ifthenelse{\equal{#1}{tree}}{\renewcommand{\capa}{probability tree}}{}%
  \ifthenelse{\equal{#1}{ds}}{\renewcommand{\capa}{consistency statistics}}{}%
  \ifthenelse{\equal{#1}{pjt}}{\renewcommand{\capa}{projection times}}{}%
  \begin{figure}[ht]
    \centering
    \epsfig{file=figures/#2.#1.ps}
    {\small \capa}
    \caption{#3}
    \label{fig:#2}
  \end{figure}}

\newcommand{\includeprogram}[2][m]{
  \tiny%
  \subsection{\texttt{#2.#1}}%
  \label{#2.#1}%
  \begin{multicols}{2}%
    \begin{alltt}%
      \input{programs/#2.#1}%
    \end{alltt}%
  \end{multicols}%
  }

\begin{document}
\frontmatter
\begin{titlepage}
\let\footnotesize\small
\let\footnoterule\relax
\null
\begin{center}
  \vfil
  \mbox{}
  \Huge 
  \textbf{Approximate Consistency and Prediction 
    Algorithms in Quantum Mechanics} 
  \vfil
  \large
  \textbf{J. N. \McElwaine}\\ 
  \emph{Clare College, Cambridge} \par
  \vfil
  \emph{A dissertation submitted for the degree of Doctor of Philosophy \\
  University of Cambridge} \par
  \vspace{1em}
  October 1996
  \mbox{}
\end{center}

\end{titlepage}

\section*{Abstract}

This dissertation investigates questions arising in the consistent
histories formulation of the quantum mechanics of closed systems.
Various criteria for approximate consistency are analysed. The
connection between the Dowker-Halliwell criterion and sphere packing
problems is shown and used to prove several new bounds on the
violation of probability sum rules. The quantum Zeno effect is also
analysed within the consistent histories formalism and used to
demonstrate some of the difficulties involved in discussing
approximate consistency. The complications associated with null
histories and infinite sets are briefly discussed.

The possibility of using the properties of the Schmidt decomposition
to define an algorithm which selects a single, physically natural,
consistent set for pure initial density matrices is investigated. The
problems that arise are explained, and different possible algorithms
discussed. Their properties are analysed with the aid of simple
models.  A set of computer programs is described which apply the
algorithms to more complicated examples.

Another algorithm is proposed that selects the consistent set (formed
using Schmidt projections) with the highest Shannon information. This
is applied to a simple model and shown to produce physically sensible
histories. The theory is capable of unconditional probabilistic
prediction for closed quantum systems, and is strong enough to be
falsifiable. Ideas on applying the theory to more complicated examples
are discussed.
\newpage

\section*{Declaration}  

The research reported in this dissertation was carried out in the
Department of Applied Mathematics and Theoretical Physics at the
University of Cambridge between October 1993 and September 1996. It is
original except where explicit reference is made to the work of
others. Chapters~\ref{chap:prediction} and~\ref{chap:spin} describe
work done in collaboration with Adrian Kent~\cite{McElwaine:Kent}.
Chapter~\ref{chap:acp} is largely the same as~\cite{McElwaine:1}, and
chapter~\ref{chap:max:inf} and part of chapter~\ref{chap:spin} are
similar to~\cite{McElwaine:3}. No part of the dissertation has been or
is being submitted for any degree or other qualification at any other
university.

\section*{Acknowledgments}

I would like to thank my supervisor, Adrian Kent, for his generous
support and constructive criticisms:  I have learned much from his
scientific method and ideas. I would also like to thank Fay Dowker and
Jonathan Halliwell for helpful comments. This work was supported by a
studentship from the United Kingdom Engineering and Physical Sciences
Research Council.

\newpage
\mbox{}
\vfill

\begin{center}
\parbox{3in}{%
  \emph{``If a man does not feel dizzy when he first learns about the
  quantum of action, he has not understood a word.''} \hfil Bohr \hfil}
\end{center}

\vfill
\mbox{}


\tableofcontents
\listoffigures
\mainmatter
\chapter{General introduction}\label{chap:generalintro}
\section{Quantum mechanics}

Quantum mechanics is a tremendously successful theory that makes
predictions with unprecedented accuracy. The standard Copenhagen
interpretation has existed since the
1920's~\cite{WZ:QTM,vonneumann,LL:QM} and has passed every
experimental test. However, it occupies a very unusual place among
physical theories: it contains classical mechanics as a limiting case,
yet at the same time it requires this limiting case for its own
interpretation~\cite{Espagnat,Bohm:quantum,Bell:moral}.

In the beginning quantum mechanics was only tested on microscopic
systems, though some part of it works well for macroscopic systems;
superconductivity, superfluidity, neutron stars and lasers are all
examples where macroscopic behaviour depends on intrinsically quantum
effects. Today, however, experiments involving SQUIDS (Superconducting
Quantum Interference Devices) and other
systems~\cite{SM:cat,MR:reservoir,LG:fiber,KMR:superposition,%
  TAC:kerr,KS:cat,ZGS:linewidth,RK:superposition} aim to create
superpositions of macroscopically distinct states and detect
interference between them, and in quantum cosmology quantum mechanics
is being applied to the entire universe. Both these areas lie outside
the realm of the Copenhagen interpretation, but the difficulties are
particularly acute in quantum cosmology, since there are no external
systems and it is highly unlikely that any systems obeying classical
mechanics existed in the early universe. The Copenhagen interpretation
also suffers from being defined in vague terms: what is an observer,
an observable or a measurement? After seventy years there is still no
consensus, and this suggests to us that quantum mechanics is
incomplete.

We feel that a quantum theory should describe an objective reality
independent of observers and that it should be mathematically
precise\footnote{Even those who believe that an interpretation relying
  on intuitive ideas or verbal prescriptions is acceptable would, we
  hope, concede that it is interesting to ask whether those ideas and
  prescriptions \emph{can} be set out mathematically.}.  That is,
given a closed quantum system the theory should make exact predictions
for the set of possible outcomes and their probabilities. It seems to
us important that the theory should be applicable to individual
systems and require no reference to objects outside the system --- how
else can a theory of quantum cosmology be understood? We discuss below
some of the current programs that seek to extend the Copenhagen
approach and remove its ambiguities, and see how they measure up to
these desiderata.

Current approaches to quantum mechanic can be divided up into those
that are \emph{incompatible} with Copenhagen quantum mechanics --- in
the sense that they fundamentally alter the dynamics of quantum
mechanics and hence make different predictions in some situations ---
and those that are \emph{compatible} with Copenhagen quantum mechanics
--- in the sense that they have the same fundamental dynamics as
quantum mechanics so that they always agree with the probabilistic
predictions of Copenhagen quantum mechanics, though there may be
additional variables and dynamical equations and they may make
additional predictions.

Among the attempts at compatible theories is the very popular
idea of a many-worlds interpretation, which was introduced by
Everett~\cite{Everett:original} and has been extensively
studied~\cite{mwibook,QCSPT:Everett,QCSPT:Deutsch,QCSPT:Shimony}. We
feel the most important criticism of these ideas is not a prejudice
against multiple branching universes, but that no precise formulation
of it has yet been made that is capable of making
predictions~\cite{bellone,belltwo,Adrian:MWI,Stein}. This problem ---
the problem of specifying what constitutes an event and which basis to
use --- recurs in many approaches to quantum mechanics.
 
The de Broglie-Bohm pilot wave theory~\cite{bohm} is compatible.
It is a \emph{realistic} theory that introduces the
positions of particles as \emph{hidden variables} and it continues to
attract
interest~\cite{BDGPZ:survey,BMQT,DDGZ:flux,Callender:Weingard:bohmian,%
  BDGPZ:existence,Ghose:Home:boson,DDGZ:nonlocality,%
  Valentini:1,Valentini:2,Valentini:PhD}. As a result of the
privileging of the position variables quasiclassical dynamics becomes
a corollary of the theory and this approach undoubtedly solves some of
the problems of the Copenhagen interpretation. However, despite much
effort, attempts to produce a relativistic theory have failed, so
though it satisfies our desiderata when applied to
non-relativistic quantum mechanics no relativistic theory exists. In
fact, it seems to us that the theory is intrinsically incapable of a
relativistic extension since it involves action at a distance and
ascribes unbounded velocities to particles.

The de Broglie-Bohm pilot wave theory, as Bell
explains~\cite{Bell:hidden}, highlights the flaws in the assumptions
of the no-hidden-variables proofs of Von Neumann and
others~\cite{vonneumann,Jauch:Piron:NHV,Gudder:NHV}. Realistic,
complete, hidden-variables theories (such as de Broglie-Bohm) do
exist, but his later work~\cite{Bell:EPR}, on the EPR
(Einstein-Podolsky-Rosen) paradox, shows that such theories must be
nonlocal --- in the sense that measurements on one system can effect a
distant system that it interacted with in the past. His proof shows
that certain inequalities (referred to as Bell's inequalities) are
satisfied by quantum mechanics but not by any local hidden-variables
theory, if, roughly\footnote{This can be made mathematically precise.}
speaking, the experimenter can independently specify the experiment.
Bell's inequalities have been experimentally
validated~\cite{Aspect:1,Aspect:2} and Bell's conclusions have
recently been extended by Greenberger
\textit{et.~al.}~\cite{GHSZ,Mermin:on:GHSZ}. They consider three spin
half particles and prove the stronger result that there is a direct
(rather than statistical) contradiction between the EPR axioms
(completeness, reality, locality) and quantum mechanics.

A realistically interpretable, relativistically invariant theory has
been described by Samols~\cite{Samols:model} on a 2d light-cone
lattice. His theory is operationally indistinguishable from standard
quantum theory, though whether this model can usefully be applied to
physical examples is unclear. There is also the question of whether
some suitable continuum limit exists.

In their seminal work~\cite{GRW} Ghirardi, Rimini and Weber rekindled
interest in \emph{stochastic collapse models}.  Their original theory
supplements the Schr\"odinger equation with a stochastic process ---
their theory is incompatible with quantum mechanics. Their ideas have
been developed into a family of
theories~\cite{Kent:Indistinguishability,Buffa:etal,%
Diosi:permanent:reduction,Pearle:Squires:gravity,Pearle:Squires:bound,%
Brun:chaos,NGN:langevin,Jadczyk,BPS:computation,BPS:diffusion,BPS:moving,%
Percival:spacetime,Percival:primary,Gisin:Percival:picture} in which
defects with the original proposal have been removed and extensions have
been made to so called \emph{continuous localisation} models where the
individual jumps become infinitesimal and the states undergo
\emph{quantum state diffusion}.

Ghirardi, Rimini and Weber's original proposal was that for each
particle there was a probability of $1/\tau$ ($\tau \approx
10^{15}\mbox{s}$) per unit term of it undergoing ``wave-function''
collapse. These collapses take place about a randomly chosen point
(probability density function given by $|\psi|^2$) and after the
collapse the new wave function is Gaussian (in the collapsed
coordinate) with characteristic length $a \approx 10^{-7}\mbox{m}$. In
their approach the state vector can be interpreted realistically ---
the state vector is precisely the current physical state --- but its
Hamiltonian evolution is altered by coupling it to a stochastic field.
The individual collapse centres can also be regarded as the ontology
of the theory~\cite{Bell:jumps}.

The stochastic collapse approach has many desirable features.  There
is no need for ill-defined concepts such as measurement, observers,
system or apparatus, and the theory can be successfully applied to
closed systems.  A preferred basis must be chosen (usually position
eigenstates) but it is precisely given.  The preferred basis defines
an operator that couples the wavefunction to the stochastic field such
that the wavefunction evolves into an eigenvector of the preferred
operator. Quantum state diffusion has been suggested as a fundamental
theory but attempts to make these ideas relativistically invariant
have failed~\cite{Gisin:stochastic} --- though perhaps Pearle's latest
theory~\cite{Pearle:relativistic} solves some of the problems. These
theories also generically violate the conservation of energy and other
conserved quantities, though apart from these deficiencies they satisfy
all of our desiderata.

Stochastic theories can also be motivated as potentially arising from
approximations to a quantum theory of 
gravity~\cite{GRW:grav,Diosi:grav,Karolyhazy:1,Karolyhazy:etal,Karolyhazy:2,%
  Pearle:1,Pearle:2,Pearle:3,Pearle:4,Pearle:5}. For example,
Penrose~\cite{Penrose:reduction,Penrose:emperor} believes that as soon
as there is a ``significant'' amount of space-time curvature the rules
of quantum linear superposition must fail --- that is an event occurs
and the wave function ``collapses''. By ``significant'' he means,
roughly speaking, when the difference in gravitational fields is
\emph{one graviton} or more. Though these proposals are interesting,
without a quantum theory of gravity, or an approximation to one, they
cannot be expressed precisely enough to develop a useful theory.

The importance of decoherence is now widely recognised. One of the
earliest papers on decoherence was by Mott~\cite{Mott:alpha} in 1929.
He considered the $\alpha$-decay of nucleus in a cloud chamber and
asked why the $\alpha$-particle produced straight line tracks when it
was a spherical wave. The answer of course is that decoherence has
occured --- the different straight line tracks have become entangled
with environment states that are approximately orthogonal. Decoherence
by interaction with the environment appears to play a fundamental role
in the emergence of classical phenomena.  These ideas have been
studied in a wide range of physical systems in recent
years~\cite{Zeh:Giulini:Kiefer,Zurek:superselection,%
Zurek:Paz,Zurek:preferred:observables,%
PHZ:reduction,Paz:Zurek:classicality,Diosi:Diagonalized,%
Anglin:laflamme:zurek,Zurek:transition,ZHP:coherent,UZ:brownian}.
These papers show that, for a wide range of systems, in a
\emph{suitable} basis, the off-diagonal elements of the reduced
density matrix decay exponentially in time. 

While the significance of decoherence is uncontroversial some authors
claim that the decoherence program contains its own interpretation
(see for example~\cite{Zurek:transition} and references therein).
However, unless some form of non-unitary evolution is proposed this
approach seems unsatisfactory as a fundamental theory to us, since a
closed system can always be regarded as being in a pure state. Without
specifying a preferred basis to use for the reduced density matrix (as
well as a split between system and environment) this approach cannot
describe the events that take place within a closed
system~\cite{Zurek:letters,Bohm:quantum,Penrose:emperor}.

The consistent histories approach to quantum theory was originally
developed by Griffiths~\cite{Griffiths:1}, Omn\`es~\cite{Omnes:1}, and
Gell-Mann \& Hartle~\cite{CEPI:GMH}. Griffiths and Omn\`es see it as
an attempt to remove the ambiguities and difficulties inherent in the
Copenhagen interpretation, but Gell-Mann \& Hartle are motivated by
an attempt to make precise the ideas of a many worlds interpretation
in the context of quantum cosmology. 

The predictions of the consistent histories formalism are identical to
the predictions of standard quantum mechanics where laboratory
experiments are concerned, but they take place within a more general
theory.  The basic objects are sequences of events or
\emph{histories}. A set of histories must include all possibilities
and must be \emph{consistent}. The individual histories can then be
considered physical possibilities with definite probabilities, and
they obey the ordinary rules of probability and logical inference. The
set of projections used and the projection times can be the same for
each history in which case the set of histories is called
\emph{branch-independent}. However, there is no reason to expect
projections which lead to consistent histories in one branch to be
consistent in another: why should projections for the position of the
earth be consistent in another branch of the universe where the solar
system never formed? When the choice of projections and projection
times depend on the earlier projections in a history the set of
histories is called \emph{branch-dependent}. 

For the remainder of this dissertation we are going to focus on the
consistent histories approach, though some of the results highlight
problems that also affect other approaches.

\section{The consistent histories approach}

\subsection{Consistent histories formalism}\label{ssec:CHformalism} 

Let $\rho$ be the initial density matrix of a quantum system.  A
\emph{branch-dependent set of histories} is a set of products of
projection operators indexed by the variables $\alpha = \{ \alpha_n ,
\alpha_{n-1} , \ldots , \alpha_1 \}$ and corresponding time
coordinates $\{ t_n , \ldots , t_1 \}$, where the ranges of the
$\alpha_k$ and the projections they define depend on the values of
$\alpha_{k-1} , \ldots , \alpha_1 $, and the histories take the form:
\begin{equation} \label{histories}
  C_{\alpha} = 
  P_{\alpha_n}^n (t_n ; \alpha_{n-1} , \ldots , \alpha_1) 
  P_{\alpha_{n-1}}^{n-1} (t_{n-1} ; \alpha_{n-2} , \ldots , \alpha_1) 
  \ldots P_{\alpha_1}^1 ( t_1 )\,.
\end{equation}
Here, for fixed values of $\alpha_{k-1} , \ldots , \alpha_1$, the
$P^k_{\alpha_k} (t_k ; \alpha_{k-1} , \ldots , \alpha_1 )$ define a
projective decomposition of the identity\footnote{For brevity, we
refer to projective decompositions of the identity as projective
decompositions hereafter.}  indexed by $\alpha_k$, so that
$\sum_{\alpha_k} P^k_{\alpha_k} (t_k ; \alpha_{k-1} , \ldots ,
\alpha_1 ) = 1 $ and
\begin{equation} \label{decomp}
P^k_{\alpha_k} (t_k ; \alpha_{k-1} , \ldots , \alpha_1 )
P^k_{\alpha'_k} (t_k ; \alpha_{k-1} , \ldots , \alpha_1 ) =
\delta_{\alpha_k \alpha'_k } P^k_{\alpha_k} (t_k ; \alpha_{k-1} ,
\ldots , \alpha_1 )\,.
\end{equation}

These projection operators (hereafter referred to as projections) are
the most basic objects in the consistent histories formulation and
they represent particular states of affairs existing at particular
times~\cite{Griffiths:1}. They are combined into time-ordered strings,
the \emph{class operators} $\{C_\alpha\}$, which are the elementary
events, or \emph{histories}, in the probability sample space.  Here
and later, though we use the compact notation $\alpha$ to refer to a
history, we intend the individual projection operators and their
associated times to define the history.

More general sets of class-operators can be created by
\emph{coarse-graining}. $\mathcal{S}^* = \{C^*_\beta\}$ is a
coarse-graining of $\mathcal{S}$ if $ C^*_\beta = \sum_{\alpha \in
  \overline{\alpha}_\beta} C_\alpha$, where
$\{\overline{\alpha}_\beta\}$ is a partition of $\mathcal{S}$.
Omn\`es defines sets of histories without any coarse-graining as Type
I, and those which have been coarse-grained but where the
class-operators are still strings of projections as Type
II~\cite{Omnes:1}. We shall follow Isham~\cite{Isham:homog} and call
these class operators \emph{homogenous}. We follow Gell-Mann and
Hartle and use completely general
class-operators~\cite{GM:Hartle:classical}, though on occasion we
shall state stronger results which hold for homogenous
class-operators.

Probabilities are defined by the formula
\begin{equation}\label{probdef}
  p(\alpha) = D_{\alpha\alpha},
\end{equation}
where $D_{\alpha\beta}$ is the \emph{decoherence matrix}
\begin{equation}
  D_{\alpha\beta} = \mbox{Tr}\, (C_\alpha \rho C_\beta^\dagger)\,.
  \label{dmdef}
\end{equation}
If no further conditions were imposed these probabilities could
contradict ordinary quantum mechanics: they would be inconsistent.  In
Griffiths' and Omn\`es' original papers~\cite{Griffiths:1,Omnes:1}
they impose the consistency criterion that \emph{probability sum
  rules} must be satisfied for all coarse grainings that result in
homogenous histories. They show that the necessary and sufficient
condition is that  
\begin{equation}
  p(\alpha+\beta) = p(\alpha) + p(\beta), \quad  \forall \alpha \neq \beta 
  \mbox{~s.t.\ $(\alpha + \beta)$ is a homogenous history,}
\end{equation}
which, from eq.~(\ref{probdef}), is equivalent to
\begin{equation}
  \mbox{Re}\,(D_{\alpha\beta}) = 0, \quad  \forall \alpha \neq \beta 
  \mbox{~s.t.\ $(\alpha + \beta)$ is a homogenous history.}
\end{equation}
A slightly stronger and more convenient criterion is that probability
sum rules should be satisfied for \emph{all} coarse grainings. The
necessary and sufficient criterion in this case is~\cite{CEPI:GMH}
\begin{equation}
  \mbox{Re}\,(D_{\alpha\beta}) = 0, \quad \forall \alpha \neq \beta,
  \label{weakcon}
\end{equation}
which Gell-Mann and Hartle call \emph{weak consistency}. A stronger
condition,
\begin{equation}
  D_{\alpha\beta} = 0, \quad \forall \alpha \neq \beta,
  \label{mediumcon}
\end{equation}
is often used in the literature for simplicity, which Gell-Mann and
Hartle call \emph{medium consistency}.  Gell-Mann and Hartle define
two stronger criteria in~\cite{GM:Hartle:classical} and more recently
yet another~\cite{gmhstrong}. In physical examples they usually are
equivalent so we restrict ourselves to the weak (\ref{weakcon}) or
medium (\ref{mediumcon}) criterion according to which is more
convenient.

There are at least two other criteria that have also been proposed.
Kent defines~\cite{Kent:implications} an \emph{ordered consistent} set
to be a consistent set of histories $\mathcal{S}$ such that 
\begin{equation}
  p(\alpha) \leq p(\beta) \mbox{~for all $\alpha \leq \beta$,
  $\alpha$ in some consistent set and $\beta \in \mathcal{S}$,}
\end{equation}
where $\alpha \leq \beta$ means that every projection in history
$\alpha$ projects onto a subspace of the corresponding projection in
history $\beta$. Kent shows that ordered consistency defines a more
strongly predictive version of the consistent histories formulation
that avoids the problems of contrary inferences.

A weaker generalisation is due to Goldstein and
Page~\cite{Goldstein:Page}. They define the probability of a history
$\alpha$ by
\begin{equation}
  \label{linearpos}
  p(\alpha) = \mbox{Re Tr}\,(C_\alpha \rho)\quad 
  \mbox{whenever $ \mbox{Re Tr}\,(C_\alpha \rho) \geq 0$.}
\end{equation}
This agrees with eq.~(\ref{probdef}) for sets of histories that are
weak consistent, but ascribes probabilities to a wider class of sets.

According to the standard view of the consistent histories formalism,
which we adopt here, it is only consistent sets which are of physical
relevance. The dynamics are defined purely by the Hamiltonian, with
no collapse postulate, but each projection in the history can be
thought of as corresponding to a historical event, taking place at the
relevant time.  If a given history is realised, its events correspond
to extra physical information, neither deducible from the initial
density matrix nor influencing it.

Gell-Mann \& Hartle~\cite{GM:Hartle:time:asymmetry} have also
investigated a time neutral version of quantum theory where initial
and final conditions are specified. The decoherence matrix is given by
\begin{equation} \label{dmdeffinal}
  D_{\alpha\beta} = 
  \frac{\mbox{Tr}\, (\rho_f C_\alpha\rho_i {C_\beta}^\dagger)}{%
 \mbox{Tr}\, (\rho_f \rho_i) }\,,
\end{equation}
where $\rho_i$ is the initial density matrix and $\rho_f$ is the final
density matrix.  This time-symmetric formalism may be useful in
quantum cosmology, but it is not developed further here.

Work by Isham and Linden~\cite{Isham:Linden:temporal,Isham:homog}
focuses on general \emph{decoherence operators} which can be applied
to relativistic quantum mechanics or non-standard models.
Hartle~\cite{Hartle:QM:Cosmology} also addresses such issues and
defines a version of the consistent histories formalism in which the
basic objects are particle trajectories. The decoherence\footnote{We
  suspect that this approach is probably contained within the
  standard~(\ref{dmdef}) definition of the decoherence matrix if
  completely general class operators are used.}  matrix is defined
\begin{equation}\label{pathint}
  D_{\alpha\beta} = \int_{\mathbf{q} \in \alpha} \int_{\mathbf{q}' \in \beta}
  \mathcal{D}\mathbf{q}\,\mathcal{D}\mathbf{q}'\, \delta(\mathbf{q}_f
  - \mathbf{q}'_f) \rho(\mathbf{q}_i,\mathbf{q}_i') \exp(i S
  [\mathbf{q}] - i S[\mathbf{q}'])\,,
\end{equation}
where $\rho$ is the initial density matrix and $S[\mathbf{q}]$ denotes
the action for the path $\mathbf{q}$ (units are chosen throughout this
dissertation such that $\hbar = 1$.)  The path integral is taken over
all trajectories that start at $\mathbf{q}_i$ and $\mathbf{q}_i'$,
pass through the regions specified by $\alpha$ and $\beta$ and finish
at a common point $\mathbf{q}_f$. 

Another generalisation is contained in the work of
Rudolph~\cite{Rudolph:1,Rudolph:2}. He adopts an approach where the
basic objects are \emph{effects} or POVMs (positive operator valued
measures.) This approach forms the class operators from strings of
decompositions of the identity into positive Hermitian operators.
Roughly speaking, the basic statements of the theory are then not that
the system is in some particular subspace but that the system is in
some sort of ``density-matrix'' like state.

For the rest of this dissertation we shall concentrate on the simplest
version of the consistent histories formalism where the decoherence
matrix is given by eq.~(\ref{dmdef}). However, many of the problems we
address also arise in the other versions of the formalism.

\subsection{Interpretation}

The consistent histories formalism has given rise to its own,
divergent, interpretations. We start off by explaining some of the
different views and use them to illustrate what we believe the current
difficulties with formalism to be.

Griffiths first set out the theory in~\cite{Griffiths:1} and has
further developed it
in~\cite{Griffiths:EPR,grifflogic,Griffiths:trajectories}.  In his
most recent paper~\cite{griffithschqr} Griffiths calls a set of
consistent histories a \emph{framework} and shows that within a
framework the ordinary rules of (boolean) logic can be applied to
propositions (histories). In Griffiths' theory two frameworks are
\emph{compatible} if all the propositions they both include can be
included as propositions in a larger (more finely grained) framework,
otherwise they are \emph{incompatible}. Propositions in incompatible
frameworks cannot be compared, for instance, if a proposition ``$P$''
is inferred with probability one in one framework and a proposition
``$Q$'' is inferred in a second incompatible framework then ``$P$''
and ``$Q$'' are individually predicted but the proposition \mbox{``$P$
  \emph{and} $Q$''} is declared meaningless. Griffiths' theory is free
from contradictions, but it is incapable of making unconditional
probabilistic predictions. For example, suppose that quasiclassical
variables decohere, then within two incompatible frameworks the
statements ``the universe will continue to be quasiclassical'' and the
``the universe will \emph{not} continue to be quasiclassical'' are
both predicted with probability one. In Griffiths' theory the two
statements cannot be combined so there is no contradiction, but the
theory seems to us to weak to be useful --- the existence and
continued existence of the quasiclassical world are outside the scope
of the theory; it cannot describe why we perceive a quasiclassical
world.

A related interpretation of this can be described as \emph{many
  frameworks}. In this approach, originally mentioned by Griffiths and
described by Dowker and Kent~\cite{Dowker:Kent:approach}, every
framework exists, and from each framework exactly one history occurs.
This is an extravagant theory which offers nothing beyond the
interpretation in the previous paragraph. If a measure could be
defined for the frameworks then one could form a theory where one
framework was realised and from that framework one history with
probabilities proportional to the measure. It seems likely that using
the natural measure (from the Grassmanian manifold) would almost surely
result in non-classical sets~\cite{Kent:quasi}, and no other measure
has yet been proposed.

Omn\`es has published widely on the consistent histories 
approach~\cite{Omnes:truth,Omnes:interpretation,Omnes:1,%
  Omnes:2,Omnes:3,Omnes:4,Omnes:5,Omnes:EPR}.  His papers explain the
branch of mathematics known as \emph{micro-local analysis}, and apply
this to consistent histories, arguing that sets of histories
consisting of quasiclassical projections are approximately consistent.
That is, they suggest that the consistent histories approach is in
agreement with our perceptions of the world.
Other authors~\cite{Dowker:Halliwell,Halliwell:fluctuations,%
DGHP:on:DH:QSD,Anastopoulos:Halliwell,Anderson:Halliwell:thermal,%
Halliwell:uncertainty,CEPI:JJH} have also shown explicitly in simple
examples that approximately consistent sets do appear to accurately
describe the world we perceive.

Omn\`es defines the term \emph{true proposition} to mean a proposition
that holds with probability one in any fine-grained framework and he
bases his interpretation around this idea.
Unfortunately, as pointed out by Dowker and
Kent~\cite{Dowker:Kent:approach}, not only are there are no true
propositions but in fact there are contradictory
propositions~\cite{Kent:contra} --- contradictory propositions can be
retrodicted with probability one in different sets. A point that has
perhaps been insufficiently realised so far is that in a sense the
consistent histories approach is purely algebraic and involves no
dynamics --- consistency is an entirely algebraic statement. The class
of consistent sets depends only on the dimension of the Hilbert space
and the eigenvalues of the initial density matrix, and, in principle,
this class can be explicitly calculated.  From this point of view it
seems obvious that there can be no true facts since there are no
dynamics intrinsic to a consistent set. It is only when the consistent
sets come to be identified with a physical system and times are
associated with the projections that the projections have physical
meaning. This demonstrates that something more is needed if the
consistent histories approach is to make unconditional probabilistic
predictions.  

Gell-Mann and Hartle's discussion of the consistent
histories approach has developed over a long series of
papers~\cite{Hartle:QM:Cosmology,GM:Hartle:time:asymmetry,Hartle:reduction,%
  GM:Hartle:classical,Hartle:spacetime:information,Hartle:quasiclassical,%
  GMH:equivalent,Hartle:Laflamme:Marolf,Hartle:spacetime,Hartle:unitarity,%
  GMH:alternative,hartlethree,gmhstrong,CEPI:GMH}. Two broad themes
run through their work, the notion of IGUS (information gathering and
utilising systems) and the notion of \emph{quasiclassical
  realms}\footnote{Originally called quasiclassical domains.}. Roughly
speaking a quasiclassical realm is a consistent set that is
\emph{complete} --- so that it cannot be non-trivially consistently
extended by more projective decompositions --- and is defined by
projection operators which involve similar variables at different
times and which satisfy classical equations of motion, to a very good
approximation, most of the time.  The notion of a quasiclassical
realm seems natural, though no precise definition of
quasiclassicality has yet been found, nor is any systematic way known
of identifying quasiclassical sets within any given model or theory.
Its heuristic definition is motivated by the familiar example of the
hydrodynamic variables --- densities of chemical species in small
volumes of space, and similar quantities --- which characterise our
own quasiclassical realm.  Here the branch-dependence of the
formalism plays an important role, since the precise choice of
variables (most obviously, the sizes of the small volumes) we use
depends on earlier historical events.  The formation of our galaxy and
solar system influences all subsequent local physics; even present-day
quantum experiments have the potential to do so significantly, if we
arrange for large macroscopic events to depend on their results.  

An IGUS is an object that it complicated enough to perceive the world
around it, process the information and then act upon the data.
Unfortunately Gell-Mann and Hartle appear to us to offer two
contradictory views towards their
formalism~\cite{Dowker:Kent:approach}. On the one hand they believe in
the equality of all consistent sets, and on the other they maintain
that our quasiclassical realm can be deduced --- somehow we as IGUSs
have developed to exploit the quasiclassical realm in which we find
ourselves. Analysing these arguments is beyond the scope of this
dissertation (see for example Dowker \&
Kent~\cite{Dowker:Kent:approach}.)

The problem of deducing the existence of our quasiclassical realm
remains essentially unaltered if the predictions are conditioned on a
large collection of data \cite{Dowker:Kent:approach}, and even if
predictions are made conditional on approximately classical physics
being observed \cite{Kent:quasi}. The consistent histories approach
thus violates both standard scientific criteria and ordinary intuition
in a number of surprising ways
\cite{Dowker:Kent:approach,Kent:quasi,Dowker:Kent:properties,%
  Kent:bohm,Kent:contra,Kent:implications}.  We believe that this
should be taken as a criticism of the formalism, but of course it is
possible to take this as a criticism of standard ideas as do
Griffiths, Omn\'es, Gell-Mann and Hartle.  We believe that the
consistent histories approach gives a new way of looking at quantum
theory which raises intriguing questions and should, if possible, be
developed further.  However, in our view, the present version of the
consistent histories formalism is too weakly predictive in almost all
plausible physical situations to be considered a fundamental
scientific theory.

Whether Gell-Mann and Hartle's program of characterising
quasiclassical sets is taken as a fundamental problem or a
phenomenological one, any solution must clearly involve some sort of
set selection mechanism.  Even without these difficulties we believe
that it is impossible to make such ideas as ``IGUS'' and
``quasiclassical realms'' precise, and that they should play no
fundamental role in a scientific theory. We now turn to a discussion
of possible extensions of the consistent histories approach.

\subsection{Set selection algorithms}

The status of the consistent histories approach remains controversial:
much more optimistic assessments of the present state of the
formalism, can be found, for example, in
refs. \cite{omnesbook,griffithschqr,CEPI:GMH}.  It is, though,
generally agreed that set selection criteria should be investigated.
For if quantum theory correctly describes macroscopic physics, then,
it is believed, real world experiments and observations can be
described by what Gell-Mann and Hartle term \emph{quasiclassical}
consistent sets of histories.

Most projection operators involve rather obscure physical quantities,
so that it is hard to interpret a general history in familiar
language.  However, given a sensible model, with Hamiltonian and
canonical variables specified, one can construct sets of histories
which describe familiar physics and check that they are indeed
consistent to a very good approximation.  For example, a useful set of
histories for describing the solar system could be defined by
projection operators whose non-zero eigenspaces contain states in
which a given planet's centre of mass is located in a suitably chosen
small volumes of space at the relevant times, and one would expect a
sensible model to show that this is a consistent set and that the
histories of significant probability are those agreeing with the
trajectories predicted by general relativity.

It should be stressed that, according to all the developers of the
consistent histories approach, quasiclassicality and related
properties are interesting notions to study within, not defining
features of, the formalism.  On this view, all consistent sets of
histories have the same physical status, though in any given example
some will give more interesting descriptions of the physics than
others.

Identifying interesting consistent sets of histories is presently more
of an art than a science.  One of the original aims of the consistent
histories formalism, stressed in particular by Griffiths and Omn\`es,
was to provide a theoretical justification for the intuitive language
often used, both by theorists and experimenters, in analysing
laboratory setups.  Even here, though there are many interesting
examples in the literature of consistent sets which give a natural
description of particular experiments, no general principles have been
found by which such sets can be identified.  Identifying interesting
consistent sets in quantum cosmological models or in real world
cosmology seems to be still harder, although there are some
interesting criteria stronger than
consistency\cite{Kent:implications,gmhstrong}.

One of the virtues of the consistent histories approach, in our view,
is that it allows the problems of the quantum theory of closed systems
to be formulated precisely enough to allow us to explore possible
solutions.  A natural probability distribution is defined on each
consistent set of histories, allowing probabilistic predictions to be
made from the initial data.  There are infinitely many consistent
sets, which are incompatible in the sense that pairs of sets generally
admit no physically sensible joint probability distribution whose
marginal distributions agree with those on the individual sets.
Indeed the standard no-local-hidden-variables-theorems show that there
is no joint probability distribution defined on the collection of
histories belonging to all consistent
sets~\cite{Goldstein:Page,Dowker:Kent:approach}.  Hence the set
selection problem: probabilistic predictions can only be made
conditional on a choice of consistent set, yet the consistent
histories formalism gives no way of singling out any particular set or
sets as physically interesting. One possible solution to the set
selection problem would be an axiom which identifies a unique
physically interesting set, or perhaps a class of such sets, from the
initial state and the dynamics.

\section{Overview}

Each chapter has its own introduction and conclusions but we give a
brief overview here.

Chapter~\ref{chap:acp} analyses various criteria for approximate
consistency using path-projected states. The connection between the
Dowker-Halliwell criterion and sphere packing problems is shown and
used to prove several bounds on the violation of probability sum
rules. The quantum Zeno effect is also analysed within the consistent
histories formalism and used to demonstrate some of the difficulties
involved in discussing approximate consistency. The complications
associated with trivial histories and infinite sets are briefly
discussed.

These results are used in chapter~\ref{chap:prediction} where
\emph{prediction algorithms} are introduced. The idea here is to
define an algorithm that dynamically generates a set selection rule.
We investigate the possibility of using the properties of the Schmidt
decomposition to define an algorithm which selects a single,
physically natural, consistent set.  We explain the problems which
arise, set out some possible algorithms, and explain their properties.
Though the discussion is framed in the language of the consistent
histories approach, it is intended to highlight the difficulty in
making any interpretation of quantum theory based on decoherence into
a mathematically precise theory.

Chapter~\ref{chap:spin} defines a simple spin model and explicitly
classifies all the exactly consistent sets formed from Schmidt
projections.  The effects of the different selection algorithms from
chapter~\ref{chap:prediction} are explained and deficiencies in the
algorithms are discussed.

In chapter~\ref{chap:random} a random Hamiltonian model is
introduced where sets are only expected to be approximately
consistent. This chapter describes the results of computer simulations
of set selection algorithms on this model. The approximate consistency
parameter that was suggested in chapter~\ref{chap:acp} is discovered
to be inappropriate in this example. Though the algorithms produce
interesting sets of histories we show that they are unstable to
perturbations in the model or the parameters and thus conclude that
the algorithms of chapter~\ref{chap:prediction} cannot be applied to
this model. 

The final chapter 
introduces a new algorithm based on choosing the set from a class of
consistent sets with the largest Shannon information. We show that
this algorithm has many desirable properties and that it produces
natural sets when applied to the spin model of
chapter~\ref{chap:spin}.  Though it has not yet been tested in a wide
range of realistic physical examples, this algorithm appears to
provide a theory of quantum mechanics capable of making unconditional
probabilistic predictions: further investigations would clearly be
worthwhile.


\chapter{Approximate consistency}\label{chap:acp}

\section{Introduction}

Much work has been done on trying to understand the emergence of
classical phenomena within the consistent histories
approach~\cite{GM:Hartle:classical,Caldeira:Leggett,Joos:Zeh,%
Dowker:Halliwell,Pohle,Paz:brownian:motion,%
Zurek:preferred:observables,Anastopoulos:Halliwell,%
Tegmark:Shapiro,Paz:Zurek:classicality,Hartle:spacetime,%
Zurek:transition}. These studies consider closed quantum systems in
which the degrees of freedom are split between an unobserved
environment and distinguished degrees of freedom such as the position
of the centre of mass.

In these and other realistic models it is often hard to find
physically interesting, \emph{exactly} consistent sets, so most
examples studied are only approximately consistent. These models do
show, however, that histories consisting of projections onto the
distinguished degrees of freedom at discrete times are
\emph{approximately} consistent.  This work is necessary for
explaining the emergence of classical phenomena but is incomplete. The
implications of different definitions of \emph{approximate}
consistency have received little research: the subject is more
complicated than has sometimes been realised.  A quantitative analysis
of the quantum Zeno paradox demonstrates some of the problems. A
deeper problem is explaining why quasi-classical sets of histories
occur as opposed to any of the infinite number of consistent,
non-classical sets. Until these problems are understood the program is
incomplete.

In this chapter we examine two different approaches to approximate
consistency and analyse two frequently used criteria. We show a simple
relation with sphere-packing problems and use this to provide a new
bound on probability changes under coarse-grainings.

\subsection{Path-projected states}

A simple way of regarding a set of histories is as a set of
path-projected states or \emph{history states}\footnote{This approach
loses its advantages if a final density matrix is present.}. For a
pure initial density matrix $\rho = |\psi\rangle\langle\psi|$ these
states are defined by
\begin{equation}  \label{PPS}
  {\bf u_\alpha} = C_\alpha|\psi\rangle\, \in \, \mathcal{H}_1,
\end{equation}
where $\mbox{dim}(\mathcal{H}_1) = d$. For a mixed density matrix,
\begin{equation}
  \rho = \sum_{i=1}^n p_i|\psi_i\rangle_1\langle\psi_i|_1\,, \qquad
  |\psi_i\rangle_1 \, \in \, \mathcal{H}_1,
\end{equation}
history states can be defined by regarding $\rho$ as a reduced density
matrix of a pure state in a larger Hilbert space $\mathcal{H}_1
\otimes \mathcal{H}_2$, where $\mathcal{H}_2$ is of dimension
$\mbox{rank}(\rho) = n$ (possibly infinite), with orthonormal basis
$|i\rangle_2$. All operators $A_1$ on $\mathcal{H}_1$, can be extended
to operators on $\mathcal{H}_1\otimes\mathcal{H}_2$ by defining $A =
A_1 \otimes I_2$.  The state in the larger space is
\begin{equation}
  |\psi\rangle = \sum_{i=1}^n \sqrt{p_i} |\psi_i\rangle_1 \otimes
  |i\rangle_2\,, \qquad |\psi\rangle \, \in \, \mathcal{H}_1\otimes \mathcal{H}_2,
\end{equation}
and the history states are again given by equation (\ref{PPS}); but
now they are vectors in an $N = nd$ dimensional Hilbert space.

The decoherence matrix (\ref{dmdef}) is
\begin{equation}
  D_{\alpha\beta} = \mbox{Tr}({\bf u_\alpha}{\bf u_\beta^\dagger}) =
  {\bf u_\beta^\dagger}{\bf u_\alpha},
\end{equation}
so the probability of the history $\alpha$ occurring is $\|{\bf
  u_\alpha}\|^2$. The consistency equations (\ref{weakcon}) are
\begin{equation} \label{weakvectorcon}
  \mbox{Re}\,({\bf u}_\alpha^\dagger{\bf u}_\beta) = 0 \,, \quad
  \forall\,\alpha \neq \beta.
\end{equation}

A complex Hilbert space of dimension $N$ is isomorphic to the real
Euclidean space ${\bf R}^{2N}$. The consistency condition
(\ref{weakvectorcon}) takes on an even simpler form when the history
states are regarded as vectors in the real Hilbert space. We define
the \emph{real history states}
\begin{equation} \label{realsplit}
  {\bf v}_{\alpha} = \mbox{Re}\,({\bf u}_{\alpha}) \oplus \mbox{Im}({\bf
    u}_{\alpha}) \quad \in {\bf R}^{2N},
\end{equation} 
and then the consistency condition (\ref{weakvectorcon}) is that the
set of real history states, $\{{\bf v}_{\alpha}\}$, is orthogonal,
\begin{equation} \label{realweak}
  {\bf v}_\alpha^T{\bf v_\beta} = 0 \,, \quad \forall \, \alpha \neq
\beta.
\end{equation}
The probabilities of history $\alpha$ is $\|{\bf v}_\alpha\|^2$.

For the rest of this chapter we shall only consider pure initial
states since the results can easily be extended to the mixed case by
using the above methods.

\section{Approximate consistency}
In realistic examples it is often difficult to find physically
interesting, \emph{exactly} consistent sets. This rarity impacts upon
the use of consistent histories in studies of dust particles or
oscillators coupled to
environments~\cite{GM:Hartle:classical,Caldeira:Leggett,Joos:Zeh,%
Dowker:Halliwell,Pohle,Paz:brownian:motion,%
Zurek:preferred:observables,Anastopoulos:Halliwell,%
Tegmark:Shapiro,Paz:Zurek:classicality,Hartle:spacetime,%
Zurek:transition}.  Frequently in these studies, the off-diagonal
terms in the decoherence matrix decay exponentially with the time
between projections, but their real parts are never exactly zero, so
the histories are only approximately consistent. Therefore if the
histories are coarse-grained, the probabilities for macroscopic events
will vary very slightly depending on the exact choice of histories in
the set.  Because the probabilities can be measured experimentally,
they should be unambiguously predicted --- at least to within
experimental precision.

In his seminal work Griffiths states that ``violations of [the
consistency criterion (\ref{weakcon})] should be so small that
physical interpretations based on the weights [probabilities] remain
essentially unchanged if the latter are shifted by amounts comparable
with the former''~\cite[sec. 6.2]{Griffiths:1}.
Omn\`es~\cite{Omnes:1,Omnes:interpretation,Omnes:2,Omnes:3,Omnes:4,%
Omnes:5,Omnes:truth}, and Gell-Mann and
Hartle~\cite{GM:Hartle:classical,CEPI:GMH,Hartle:QM:Cosmology} make
the same point. The amount by which the probabilities change under
coarse-graining is the extent to which they are ambiguous. We shall
define the the largest such change in a set to be the \emph{maximum
probability violation} or \mbox{MPV}.

Dowker and Kent~\cite{Dowker:Kent:approach,Dowker:Kent:properties}
argue that more is needed.  Why should \emph{approximately} consistent
sets be used?  They suggest that ``near'' a generic approximately
consistent set there will be an exactly consistent one.  ``Near''
means that the two sets describe the same physical events to order
$\epsilon$; the relative probabilities and the projectors must be the
same to order $\epsilon$. In this chapter we investigate which
criteria will guarantee this, and show that some of the commonly used
ones are not sufficient.

\subsection{Probability violation}

The \mbox{MPV} can be defined equivalently in terms of the decoherence
matrix:
\begin{eqnarray}\label{MPV}
  \mbox{MPV}(D) & = &\max_{\overline \alpha} \left| p(\overline
  \alpha) - \sum_{\alpha \in \overline \alpha} p(\alpha) \right|, \\ &
  = & \max_{\overline \alpha} \left| \sum_{\alpha,\beta \in \overline
  \alpha} D_{\alpha\beta} - \sum_{\alpha \in \overline \alpha}
  D_{\alpha\alpha} \right|, \\ & = & \max_{\overline \alpha} \left |
  \sum_{\alpha \neq \beta \in \overline \alpha}D_{\alpha\beta}
  \right|.
\end{eqnarray} 
The maximum is taken over all possible coarse-grainings $\overline
\alpha$.  For large sets of histories this is difficult to calculate
as the number of possible coarse-grainings is $O(2^n)$.  A simple
criterion that if satisfied to some order $\epsilon(\delta)$ would
ensure that the \mbox{MPV} were less than $\delta$ would be preferable
here.

This is not a trivial problem. The frequently used
criterion~\cite{GM:Hartle:classical,Omnes:5}
\begin{equation}
\label{badcriterion}
  |D_{\alpha\beta}| \leq \epsilon(\delta), \quad \forall \alpha \neq
\beta
\end{equation}
is not sufficient for any $\epsilon(\delta) > 0 $.
Theorem~(\ref{proveuseless}) shows that for any $\epsilon(\delta) >
0 $ there are finite sets of histories satisfying (\ref{badcriterion})
with an arbitrarily large \mbox{MPV}. The example used in the proof
also shows some of the complications that arise when discussing
infinite sets of histories. All sets of histories in the rest of this
chapter will be assumed to be finite unless otherwise stated.

A simple bound\footnote{\raggedright When the class-operators are
  homogenous the bound can be improved to \mbox{$M(D_{\alpha\beta})
  \leq 1/2 \sum_{\alpha \neq \beta }|\mbox{Re}\,(D_{\alpha\beta})|$},
  since $\sum_{\alpha \neq \beta}D_{\alpha\beta} = 0$.} for the
  \mbox{MPV} is
\begin{equation} \label{modupbound}
  \mbox{MPV}(D) \leq \sum_{\alpha \neq \beta }|\mbox{Re}\,
  (D_{\alpha\beta})|.
\end{equation} 
This leads to the criterion for the individual elements
\begin{equation} \label{naivecriterion}
  |\mbox{Re}\,(D_{\alpha\beta})| \leq \frac{\delta}{n(n-1)}, \quad
  \forall \alpha \neq \beta,
\end{equation}
where $n$ is the number of histories.  Equation (\ref{naivecriterion})
ensures that the \mbox{MPV} is less than $\delta$, although the
condition will generally be much stronger than necessary. It would be
preferable however, to have a criterion that only depended on the
Hilbert space and not on the particular set of histories.

\section{The Dowker-Halliwell criterion}
Dowker and Halliwell discussed approximate consistency in their
paper~\cite{Dowker:Halliwell}, in which they introduced a new
criterion\footnote{we have replaced Dowker-Halliwell's $<$ with $\leq$
to avoid problems with histories of zero probability.}
\begin{eqnarray}\label{wic:DHC}
  \left|\mbox{Re}\,(D_{\alpha\beta})\right| & \leq & \epsilon \,
    (D_{\alpha\alpha}D_{\beta\beta})^{1/2}, \quad \forall\,
    \alpha\neq\beta,
\end{eqnarray}
which we shall call the \emph{Dowker-Halliwell criterion} or DHC\@.
Using the central limit theorem and assuming that the off-diagonal
elements are independently distributed, Dowker and Halliwell
demonstrate that (\ref{wic:DHC}) implies
\begin{eqnarray}\label{DHsumlaw}
  \left|\sum_{\alpha \neq \beta \in \overline\alpha}
    D_{\alpha\beta}\right| & \leq & \epsilon \sum_{\alpha \in
    \overline\alpha}D_{\alpha\alpha},
\end{eqnarray}
for most coarse-grainings $\overline\alpha$. This is a natural
generalisation of (\ref{weakcon}) to saying that the probability sum
rules are satisfied to \emph{relative} order $\epsilon$. For
homogenous histories this is a similar but stronger condition than
requiring that the \mbox{MPV} (\ref{MPV}) is less than $\epsilon$,
since
\begin{eqnarray}
  \sum_{\alpha \in \overline\alpha}D_{\alpha\alpha} &\leq & \sum
  D_{\alpha\alpha} = 1.
\end{eqnarray}
But for general class-operators $\sum D_{\alpha\alpha}$ is unbounded
and (\ref{DHsumlaw}) must either be modified or supplemented by a
condition such as
\begin{eqnarray}
  \left|\sum D_{\alpha\alpha}-1\right| & \leq & \epsilon.
\end{eqnarray}
This is only a very small change and for approximately consistent sets
is not significant.

For the sake of completeness, we shall occasionally mention a similar
criterion which we shall call the \emph{medium} \mbox{DHC}
\begin{eqnarray}
  \label{MDHC}
  \left|D_{\alpha\beta}\right| & \leq & \epsilon \,
    (D_{\alpha\alpha}D_{\beta\beta})^{1/2}, \quad \forall\,
    \alpha\neq\beta.
\end{eqnarray}

As Dowker and Halliwell~\cite{Halliwellconversation} point out, the
off-diagonal terms are often not well modelled as independent random
variables. Indeed even when this assumption is valid, the \mbox{MPV}
will usually be much higher. By appropriately choosing $\epsilon$ as a
function of $\delta$, however, it is possible to eliminate these
problems, and to utilise the many other useful properties of the
DHC.

\subsection{Geometrical properties}
The Dowker-Halliwell criterion has a simple geometrical
interpretation. In terms of the real history states (\ref{realsplit})
the DHC can be written (ignoring trivial\footnote{A trivial history is
one with probability equal to $0$.} histories)
\begin{eqnarray} \label{dhconh}
  \frac{|{\bf v}_{\alpha}^{T}{\bf v}_\beta|}{\|{\bf
      v}_{\alpha}\|\,\|{\bf v}_\beta\|} = |\cos(\theta_{\alpha\beta})|
      & \leq & \epsilon\,, \quad \forall\, \alpha\neq\beta,
\end{eqnarray}
where $\theta_{\alpha\beta}$ is the angle between the real history
vectors ${\bf v}_{\alpha}$ and ${\bf v}_\beta$. The DHC requires that
the angle between every pair of histories must be at least
$\cos^{-1}\epsilon$ degrees.

In a $d$ dimensional Hilbert space there can only be $2d$ exactly
consistent, non-trivial histories. Thus, if a set contains more than
$2d$ non-trivial histories, it cannot be continuously related to an
exactly consistent set unless some of the histories become trivial.
Establishing the maximum number of histories satisfying (\ref{dhconh})
in finite dimensional spaces is a particular case from a family of
problems, which has received considerable study.

\subsection{Generalised kissing problem}

The Generalised Kissing Problem is the problem of determining how many
$(k-1)$-spheres of radius $r$ can be placed on the surface of a sphere
with radius $R$ in ${\bf R}^k$.  This problem is equivalent to
calculating the maximum number of points that can be found on the
sphere all at least $\cos^{-1}\epsilon$ degrees apart, where $\epsilon
= 1 - 2r^2(R+r)^{-2}$.

To express these ideas mathematically, we define $M({\bf L}, ({\bf u},
{\bf v}) \leq s)$ to be the size of the largest subset of ${\bf L}$,
such that $({\bf u}, {\bf v}) \leq s$ for all different elements in
the subset, where ${\bf L}$ is a metric space. The Generalised Kissing
Problem is calculating
\begin{equation}
  M({\bf S}^{k-1},\, {\bf u}^T{\bf v} \leq \epsilon),
\end{equation}
where ${\bf S}^{k-1}$ is the set of points on the unit sphere in ${\bf
R}^{k}$.  The greatest number of history vectors satisfying the DHC is
\begin{equation}
  M({\bf CS}^{d-1},\, |\mbox{Re}\,({\bf u}^\dagger{\bf v})| \leq
  \epsilon) \,=\, M({\bf S}^{2d-1},\, |{\bf u}^T{\bf v}| \leq
  \epsilon)
\end{equation}
and for the medium DHC is
\begin{equation}
  M({\bf CS}^{d-1},\, |{\bf u}^\dagger{\bf v}| \leq \epsilon ),
\end{equation}
where ${\bf CS}^{d-1}$ is the set of points on the unit sphere in
${\bf C}^{d}$.

There is a large literature devoted to sphere-packing. Although few
exact results emerge from this work, numerous methods exist for
generating bounds. The tightest upper bounds derive from an
optimisation problem. In appendix (\ref{upperboundappendix}) we prove
that the well known bound
\begin{equation}\label{upperbound}
  M({\bf S}^{2d-1},\, |{\bf u}^T{\bf v}| \leq \epsilon) \leq
  \left\lfloor\frac{2d(1-\epsilon^2)}{1-2d\epsilon^2}\right\rfloor
\end{equation}
is the solution to the optimisation problem when $\epsilon^2 \leq
1/(2d+2)$.

The most important feature of this bound is that for $ 0 \leq \epsilon
\leq 1/(2d)$ it is exact, since for $\epsilon < 1/(2d) $ it gives $2d$
as an upper bound and for $\epsilon =1/(2d)$ it gives $2d +1$, and
there are packings that achieve these bounds\footnote{A packing with
$\epsilon =1/(2d)$ is generated by the rays passing through the $2d+1$
vertices of the regular $(2d)$-simplex.}. This is also the range of
most interest in Consistent Histories since an exactly consistent set
cannot contain more than $2d$ non-trivial histories. This result shows
that if $\epsilon < 1/(2d)$ then there cannot be more than $2d$
histories in a set satisfying the DHC\@.  Deciding when a set of
vectors could be a set of histories is a difficult problem, so this
result does not prove that this bound is optimal, although it is
suggestive.

This bound (\ref{upperbound}) can now be used to prove several upper
bounds on probability sum rules.

\begin{eqnarray}
  \left|\sum_{\alpha \neq \beta \in \overline \alpha}
    D_{\alpha\beta}\right| & \leq & \sum_{\alpha \neq \beta \in
    \overline\alpha} |D_{\alpha\beta}|, \\ & \leq & \epsilon
    \sum_{\alpha \neq \beta \in \overline \alpha } \left(
    D_{\alpha\alpha} D_{\beta\beta} \right)^{1/2},\\ & \leq & \epsilon
    (n-1) \sum_{\alpha \in \overline \alpha }
    D_{\alpha\alpha}.\label{abc}
\end{eqnarray}
But the number of history vectors $n$ is bounded by
$2d(1-\epsilon^2)/(1-2d\epsilon^2)$, so
\begin{eqnarray}
  \left|\sum_{\alpha \neq \beta \in \overline \alpha}
    D_{\alpha\beta}\right|&\leq& \epsilon \,
    \frac{2d-1}{1-2d\epsilon^2}\sum_{\alpha \in \overline \alpha }
    D_{\alpha\alpha}.\label{DHsumlawbound}
\end{eqnarray}
Let
\begin{equation}\label{an1}
  \epsilon(\delta) =
  \frac{-(2d-1)+\sqrt{(2d-1)^2+8d\delta^2}}{4d\delta}
\end{equation}
and then (\ref{DHsumlawbound}) implies
\begin{eqnarray}
  \left|\sum_{\alpha \neq \beta \in \overline \alpha} D_{\alpha\beta}
    \right| & \leq & \delta \sum_{\alpha \in \overline \alpha}
    D_{\alpha\alpha}.
\end{eqnarray}
This is the exact version of Dowker and Halliwell's result
(\ref{DHsumlaw}). For homogenous histories $\sum_{\alpha}
D_{\alpha\alpha} =1$ and then (\ref{DHsumlawbound}) and (\ref{an1})
imply
\begin{equation}\label{mainmpvbound}
  \mbox{MPV} < \delta.
\end{equation}
These results can easily be extended to general class-operators since
the same methods lead to a bound on $\sum_{\alpha} D_{\alpha\alpha}$
in terms of $\epsilon$.
\begin{eqnarray}
  \sum_{\alpha,\beta} D_{\alpha\beta} &=& 1\\ \Rightarrow
  \sum_{\alpha} D_{\alpha\alpha} &=& 1 -\sum_{\alpha \neq \beta}
  D_{\alpha\beta} \\ \Rightarrow \sum_{\alpha} D_{\alpha\alpha} & \leq
  & 1 + \sum_{\alpha \neq \beta}|D_{\alpha\beta}|\\ &\leq&1 + \epsilon
  (n-1) \sum_{\alpha} D_{\alpha\alpha} \\ \Rightarrow \sum_{\alpha}
  D_{\alpha\alpha} &\leq&\frac{1}{1- (n-1)\epsilon}.
\end{eqnarray}
There are sets of histories for which this bound is obtained. In
particular if $\epsilon = 1/(n-1)$ there are finite sets for which
$\sum_{\alpha} D_{\alpha\alpha}$ is arbitrarily large.  Inserting this
result into (\ref{abc}) results in
\begin{eqnarray}
  \left|\sum_{\alpha \neq \beta \in \overline \alpha}
    D_{\alpha\beta}\right| & \leq & \frac{\epsilon (n-1)}{1-
    (n-1)\epsilon}\\ & \leq & \epsilon
    \frac{2d-1}{1+\epsilon-2d\epsilon(1+\epsilon)}.\label{gb}
\end{eqnarray}
Let
\begin{equation}\label{an2}
  \epsilon(\delta) = \frac{-(2d-1)(1+\delta) +\sqrt{(2d-1)^2
      (1+\delta)^2+ 8d\delta^2}} {4d\delta},
\end{equation}
and then (\ref{gb}) becomes
\begin{equation} 
  \left|\sum_{\alpha \neq \beta \in \overline \alpha}
    D_{\alpha\beta}\right| \leq \delta,
\end{equation}
so
\begin{equation}
  \mbox{MPV} \leq \delta.
\end{equation}
For physical situation the probability violation must be small so
$\delta \ll 1$ and these results can be simplified. From
(\ref{upperbound}) if $\epsilon < 1/(2d)$ $n\leq2d$ so (\ref{an1}) and
(\ref{an2}) can be simplified to
\begin{equation}\label{epschoice}
  \epsilon(\delta) = \frac{\delta}{2d}, \delta<1 \Rightarrow
  \mbox{MPV} \leq \delta + O(\delta^2),
\end{equation}
for all types of histories.  This is the main result of this
chapter. If the medium DHC holds \emph{or} the class-operators are
homogenous then (\ref{epschoice}) can be weakened to $\epsilon(\delta)
= \delta/d$ and still imply (\ref{mainmpvbound}). If the medium DHC
holds \emph{and} the class-operators are homogenous then
(\ref{epschoice}) can be further weakened to $\epsilon(\delta) =
2\delta/d$ and still imply (\ref{mainmpvbound}).

In appendix (\ref{dheg}) we give a simple example of a class of sets
of histories, of any size, satisfying the medium DHC with $\mbox{MPV}
= d\epsilon/4$. If $\epsilon$ is chosen according to (\ref{epschoice})
then the $\mbox{MPV} = \delta/8$.  This example illustrates that
equation (\ref{epschoice}) is close to the optimal bound.
Since the example satisfies the \emph{medium} DHC and the
class-operators are homogenous $\epsilon$ can be chosen to be
$2\delta/d$ and the \mbox{MPV} is then $\delta/2$, so for this example
the bound is achieved within a factor of two.

The choice $\epsilon = \delta/(2d)$ in relation to the DHC is
particularly convenient in computer models. Often one constructs a set
of histories by individually making projections, and one desires a
simple criterion which will bound the \mbox{MPV}. The DHC solves
this problem.

The only known lower bounds for the generalised kissing problem derive
{}from an argument of Shannon's~\cite{Shannon} developed by
Wyner~\cite{Wyner}. Shannon proved that
 \begin{equation} \label{slbound}
  M({\bf S}^{2d-1},\, |{\bf u}^T{\bf v}| \leq \epsilon)  \geq 
  (1-\epsilon^2)^{1/2-d}.
\end{equation}
We explain the proof and extend it for the medium DHC in appendix
(\ref{shannonap}).

This simple bound (\ref{slbound}) has an important consequence: the
number of history vectors satisfying the DHC can increase exponentially with
$d$ if $\epsilon$ is constant. So for constant $\epsilon > 0$ by
choosing a large enough Hilbert space the \mbox{MPV} can be
arbitrarily large, therefore $\epsilon$ must be chosen according to
the dimension of the Hilbert space.

When the Hilbert space is infinite-dimensional and separable, and
$\epsilon > 0$, (\ref{slbound}) suggests that there can be an
uncountable number of history vectors satisfying the DHC\@. If so, the DHC
can only guarantee proximity to an exactly consistent set for finite
Hilbert spaces. Though if the system is set up in a Hilbert space of
dimension $d$ and the limit $d\to\infty$, $\epsilon =
O(\sqrt{\frac{\log d}{d}})$ is taken (assuming it exists) then the
bound remains countable and it may be useful even for infinite spaces.
 
If there are $n$ histories satisfying the DHC with $\epsilon =
\delta/(n-1)$, then, from (\ref{abc}), $\mbox{MPV} \leq \delta + O(
\delta^2)$. This result is trivial, but the DHC also ensures that the
histories will span a subspace of dimension at least $n/2$. Therefore,
there will be exactly consistent sets with the same number of
non-trivial histories that span the same subspace.

If
\begin{displaymath}
  \epsilon \leq \left[1-(2d)^{2/(1-2d)}\right]^{1/2} =
    \left[\frac{2\ln 2d}{2d-1}\right]^{1/2} + O\left\{ \left[\frac{\ln
    d}{d}\right]^{3/2}\right\}
\end{displaymath} 
then the lower bound is less than the trivial lower bound $M \geq d$.
Since the upper bounds (\ref{upperbound}) holds only for $\epsilon
\leq O(1/\sqrt d)$ the two sets of bounds are not mutually useful.
The Shannon bound is too poor for small $\epsilon$ because it ignores
the overlap between spherical caps. A more rigorous bound would add
points one by one on the edge of existing caps, and allow for the
overlap between them. Unfortunately, there are no useful results in
this direction.

\subsection{Other properties of the Dowker-Halliwell criterion}
In standard Quantum Mechanics the probability of observing a system in
state $|\phi\rangle$ when it is in state $|\psi\rangle$ is
$|\langle\phi|\psi\rangle|^2 / (\langle\phi|\phi\rangle \,
\langle\psi|\psi\rangle)$. If we take this as a measure of
distinguishability then the set of history states, $\{{\bf
u}_\alpha\}$, are distinguishable to order $\epsilon^2$ only if
\begin{equation}
  \frac{|{\bf u}_\alpha^\dagger {{\bf u}_\beta}|^2} {\|{\bf
    u}_\alpha\|^2\|{\bf u}_\beta\|^2} \, \leq \, \epsilon^2\,, \quad
    \forall \alpha \neq \beta.
\end{equation}
But this is equivalent to the medium DHC (\ref{wic:DHC}) since
\begin{equation}
  \frac{|{{\bf u}_\alpha}^\dagger {\bf u}_\beta|} {\|{\bf
  u}_\alpha\|\,\|{\bf u}_\beta\|} = \frac{|D_{\alpha\beta}|}
  {(D_{\alpha\alpha} D_{\beta\beta})^{1/2}}.
\end{equation}
Histories which only satisfy the weak consistency criterion
(\ref{weakcon}) need not be distinguishable since a pair of histories
may only differ by a factor of $i$ and would be regarded as equivalent
in conventional quantum mechanics. This is one of the few differences
between the medium (\ref{MDHC}) and the standard (\ref{wic:DHC})
DHC.

Outside of quantum cosmology one usually discusses conditional
probabilities: one regards the past history of the universe as
definite and estimates probabilities for the future from it. One does
this in consistent histories by forming the \emph{current density
matrix} $\rho_c$. Let $\{C_\alpha\}$ be a complete set of
class-operators, each of which can be divided into the the past and
the future, $C_\alpha = C^f_{\alpha_f}C^p_{\alpha_p}$. Then the
probability of history $\alpha_f$ occuring given $\alpha_p$ has
occurred is
\begin{eqnarray} 
  p(\alpha_f | \alpha_p) & = & \frac{p(\alpha_f \, \& \,
    \alpha_p)}{p(\alpha_p)}, \\ & = & \frac{\mbox{Tr} (C^f_{\alpha_f}
    C^p_{\alpha_p} \rho C_{\alpha_p}^{p\dagger}
    C_{\alpha_f}^{f\dagger})}{\mbox{Tr} (C^p_{\alpha_p} \rho
    C_{\alpha_p}^{p\dagger})}, \\ & = & \mbox{Tr} (C^f_{\alpha_f}
    \rho_c C_{\alpha_f}^{f\dagger})\label{cdm},
\end{eqnarray}
where $\rho_c = C^p_{\alpha_p} \rho C_{\alpha_p}^{p\dagger} /
\mbox{Tr} (C^p_{\alpha_p} \rho C_{\alpha_p}^{p\dagger})$. Equation
(\ref{cdm}) shows that all future probabilities can be expressed in
terms of $\rho_c$. The DHC in terms of $\rho_c$ is
\begin{equation}\label{dhcb}
  \frac{\mbox{Tr} (C^f_{\gamma_f} \rho_c C_{\beta_f}^{f\dagger})}
  {[\mbox{Tr} (C^f_{\gamma_f} \rho_c C_{\gamma_f}^{f\dagger})
  \mbox{Tr} (C^f_{\beta_f} \rho_c C_{\beta_f}^{f\dagger})]^{1/2}} \leq
  \epsilon, \quad \forall \gamma_f \neq \beta_f.
\end{equation}
This is the same as the DHC applied to the complete histories, given
the past,
\begin{equation}\label{dhca}
  \frac{\mbox{Tr} (C^f_{\gamma_f} C^p_{\alpha_p} \rho
    C_{\alpha_p}^{p\dagger} C_{\beta_f}^{f\dagger})} {[\mbox{Tr}
    (C^f_{\gamma_f} C^p_{\alpha_p} \rho
    C_{\alpha_p}^{p\dagger}C_{\gamma_f}^{f\dagger}) \mbox{Tr}
    (C^f_{\beta_f} C^p_{\alpha_p} \rho
    C_{\alpha_p}^{p\dagger}C_{\beta_f}^{f\dagger})]^{1/2}} \leq
    \epsilon,
\end{equation}
for all $\gamma_f \neq \beta_f$.  This is a property not possessed by
the usual criterion (\ref{badcriterion}) or any other based on
absolute probabilities, such as one that only bounds the
\mbox{MPV}. This is an important property since any
non-trivial\footnote{That is a branch whose probability is non-zero.}
branch of a consistent set of histories (when regarded as a set of
histories in its own right) must also be consistent and one would like
a criterion for approximate consistency that reflects this.

Experiments in quantum mechanics are usually carried out many times,
and the relative frequencies of the outcomes checked with their
probabilities predicted by quantum mechanics. Consider the situation
where an experiment is carried out at $m$ times $\{t_i\}$ with
probabilities $\{p_i\}$. Let $P^i$ be the projector corresponding to
the experiment being performed at $t_i$ and let
$\{C_\alpha^i\}=\{U(-t_i)C_\alpha U(t_i)\}$ be the set of $n$
class-operators corresponding to the different outcomes of the
experiment when it is started at time $t_i$. For simplicity assume
that the probability of an experiment being performed and its results
are independent of other events. This implies $[P^i,P^j] = 0$,
$[P^i,C^j_\alpha] = 0$ and $[C^i_\alpha,C^j_\alpha] = 0$ so $p_i =
\langle\phi|P^i|\phi\rangle$. There are $(1+n)^m$ class-operators and
they are of the form
\begin{equation}\label{egco}
  \overline{P^{i_k}}\ldots\overline{P^{i_1}} \, P^{j_{m-k}} \ldots
  P^{j_1} \, C_{\alpha_{m-k}}^{j_{m-k}} \ldots C_{\alpha_1}^{j_1},
\end{equation}
corresponding to the experiment being performed at times $t_{j_1}
\ldots t_{j_{m-k}}$ and not at times $t_{i_1} \ldots t_{i_k}$ with
results $\alpha_1 \ldots \alpha_{m-k}$. Because of the commutation
relations the only non-zero off-diagonal-elements of the decoherence
matrix contain factors like
\begin{equation}\label{nzod}
  p_i\,\mbox{Re}\,(\langle\psi|C_\beta^\dagger C_\alpha|\psi\rangle),
\end{equation}
where $|\psi\rangle$ is the initial state in which the experiment is
prepared identically each time. When the environment and time between
experiments are large the $P_i$ will commute and this justifies the
usual arguments where the consistency of the experiment alone is
considered rather than the consistency of the entire run of
experiments.

This is a particular case of the result that an inconsistent set
cannot extend a non-trivial branch of a set of histories without
destroying its consistency. A sensible criterion for approximate
consistency should also have this property. By choosing the $p_i$
small enough the off-diagonal elements (\ref{nzod}) can be made
arbitrarily small, thus any criterion for approximate consistency
which uses absolute probabilities will regard the set as consistent,
however inconsistent the experiment itself may be.

An important feature of the DHC is that it has no such disadvantage,
as the $p_i$'s will cancel and the approximate consistency conditions
will be
\begin{equation}
  \frac{|\mbox{Re}\,(\langle\psi|C_\beta^\dagger
    C_\alpha|\psi\rangle)|}{\|C_\alpha|\psi\rangle\|\,
    \|C_\beta|\psi\rangle\|} \leq \epsilon.
\end{equation}

\section{Conclusions}

A set of histories is approximately consistent to order $\delta$, only
if its \mbox{MPV} is less than $\delta$. The often-used criterion
\begin{equation} \label{ccagain}
  \mbox{Re}\,(D_{\alpha\beta}) \leq \epsilon(\delta), \quad \forall
  \alpha \neq \beta
\end{equation}
is not sufficient for any $\epsilon(\delta) > 0$, since there are sets
of histories satisfying (\ref{ccagain}) with arbitrarily large
\mbox{MPV}.  The criterion (\ref{ccagain}) can only be used if
$\epsilon(\delta) = O(1/n^2)$, where $n$ is the number of
histories. The Dowker-Halliwell criterion has no such disadvantage.

If
\begin{equation}\label{tc}
  \mbox{Re}\,(D_{\alpha\beta}) \leq \frac{\delta}{2d}\,
  (D_{\alpha\alpha}D_{\beta\beta})^{1/2} \quad \mbox{$\forall \alpha
  \neq \beta$, $\delta < 1$,}
\end{equation}
then the \mbox{MPV} is less than $\delta + O(\delta^2) $. This is the
chapter's main result.  If the medium DHC holds,
\begin{equation}\label{mtc}
  |D_{\alpha\beta}| \leq \frac{\delta}{d}\,
  (D_{\alpha\alpha}D_{\beta\beta})^{1/2} \quad \mbox{$\forall \alpha
  \neq \beta$, $\delta < 1$,}
\end{equation}
then the \mbox{MPV} is also bounded by $\delta$. For histories
satisfying either criterion, if only homogenous class-operators are
used then the upper bound on the \mbox{MPV} is strengthened to
$\delta/2$.  The bounds are also optimal in the sense that they are
can be achieved (to within a small factor) in any finite dimensional
Hilbert space.  Any improved bound must use the global structure of
the decoherence matrix.

The DHC is particularly suitable for computer models in which a set of
histories is built up by repeated projections. If each history
satisfies (\ref{tc}) as it is added, then the whole set will be
consistent to order $\delta$ and there will be no more than $2d$
histories.

The DHC also leads to a simple, geometrical picture of consistency:
the path-projected states can be regarded as pairs of points on the
surface of a hyper-sphere, all separated by an angle of at least
$\cos^{-1}\epsilon$. This approach can be used to prove that
$\epsilon$ in the DHC must be chosen according to the dimension of the
Hilbert space. Ideally one would like a criterion for approximate
consistency that implied the existence of an exactly consistent set
corresponding to physical events that only differed to order
$\epsilon$. The DHC seems well adapted to defining proximity to an
exactly consistent set and may be useful in constructing a proof that
such a set exists.

This bound (\ref{slbound}) shows that the number of history vectors
satisfying the DHC can increase exponentially with $d$ if $\epsilon$
is constant. So for constant $\epsilon > 0$ by choosing a large enough
Hilbert space the \mbox{MPV} can be arbitrarily large, therefore
$\epsilon$ must be chosen according to the dimension of the Hilbert
space.

If a set is not exactly consistent then it cannot be a subset of an
exactly consistent set (unless the branch is trivial.) The same is
true for approximate consistency when it is defined by the
DHC\@. This is not true, however, for any criterion which
depends solely on the \mbox{MPV}. It is a particularly useful property
when discussing conditional probabilities.

\chapter{Prediction algorithms}\label{chap:prediction}
\section{Introduction}\label{sec:alg:intro} 

It is hard to find an entirely satisfactory interpretation of the
quantum theory of closed systems, since quantum theory does not
distinguish physically interesting time-ordered sequences of
operators. In this chapter, we consider one particular line of attack
on this problem: the attempt to select consistent sets by using the
Schmidt decomposition together with criteria intrinsic to the
consistent histories formalism.  The chapter is exploratory in spirit:
our aims here are to point out obstacles, raise questions, set out
some possible selection principles, and explain their properties.

Our discussion is framed in the language of the consistent histories
approach to quantum theory, but we believe it is of wider relevance.
Many modern attempts to provide an interpretation of quantum theory
rely, ultimately, on the fact that quantum subsystems decohere.
Subsystems considered include the brains of observers, the pointers of
measuring devices, and abstractly defined subspaces of the total
Hilbert space.  Whichever, the moral is intended to be that
decoherence selects the projection operators, or space-time events, or
algebras of observables which characterise the physics of the
subsystem as it is experienced or observed.  There is no doubt that
understanding the physics of decoherence \emph{does} provide a very
good intuitive grasp of how to identify operators from which our
everyday picture of real-world quasiclassical physics can be
constructed and this lends some support to the hope that a workable
interpretation of quantum theory --- a plausible successor to the
Copenhagen interpretation --- \emph{could} possibly be constructed
along the lines just described.

A key question, it seems to us, is whether such an interpretation can
be made mathematically precise.  That is, given a decohering
subsystem, can we find general rules which precisely specify operators
(or other mathematical objects) which allow us to recover the
subsystem's physics as we experience or observe it?  From this point
of view, we illustrate below how one might go about setting out such
rules, and the sort of problems which arise. We use a version of the
consistent histories formalism in which the initial conditions are
defined by a pure state, the histories are branch-dependent and
consistency is defined by Gell-Mann and Hartle's medium consistency
criterion eq.~(\ref{mediumcon}).

\subsection{The Schmidt decomposition} 

We consider a closed quantum system with pure initial state vector
$|\psi (0)\rangle$ in a Hilbert space $\mathcal{H}$ with Hamiltonian
$H$.  We suppose that $\mathcal{H} = \mathcal{H}_1 \otimes
\mathcal{H}_2$; we write $\dim (\mathcal{H}_j ) = d_j$ and we suppose
that $d_1 \leq d_2 < \infty$.  With respect to this splitting of the
Hilbert space, the \emph{Schmidt decomposition} of $|\psi (t) \rangle$
is an expression of the form
\begin{equation} \label{schmidteqn}
  |\psi (t) \rangle = \sum_{i=1}^{d_1} \, [p_i(t)]^{1/2} \, | w_i
  (t)\rangle_1 \otimes |w_i (t)\rangle_2 \, ,
\end{equation}
where the \emph{Schmidt states} $\{ |w_i\rangle_1 \}$ and $\{
|w_i\rangle_2\}$ form, respectively, an orthonormal basis of
$\mathcal{H}_1$ and part of an orthonormal basis of $\mathcal{H}_2$,
the functions $p_i (t)$ are real and positive, and we take the
positive square root.  For fixed time $t$, any decomposition of the
form eq.~(\ref{schmidteqn}) then has the same list of probability
weights $\{ p_i (t) \}$, and the decomposition~(\ref{schmidteqn}) is
unique if these weights are all different. These probability weights
are the eigenvalues of the reduced density matrix.

This simple result, proved by Schmidt in 1907\cite{schmidt}, means
that at any given time there is generically a natural decomposition of
the state vector relative to any fixed split $\mathcal{H} =
\mathcal{H}_1 \otimes \mathcal{H}_2$, which defines a basis on the
smaller space $\mathcal{H}_1$ and a partial basis on $\mathcal{H}_2$.
The decomposition has an obvious application in standard Copenhagen
quantum theory where, if the two spaces correspond to subsystems
undergoing a measurement-type interaction, it describes the final
outcomes~\cite{vonneumann}.

It has more than once been suggested that the Schmidt decomposition
\emph{per se} might define a fundamental interpretation of quantum
theory. According to one line of thought, it defines the structure
required in order to make precise sense of Everett's
ideas\cite{deutsch}.  Another idea which has attracted some attention
is that the Schmidt decomposition itself defines a fundamental
interpretation\cite{dieks,vanf,healey,kochen}.  Some critical comments
on this last program, motivated by its irreconcilability with the
quantum history probabilities defined by the decoherence matrix, can
be found in ref. \cite{Kent:pla}.

Though a detailed critique is beyond our scope here, it seems to us
that any attempt to interpret quantum theory which relies solely on
the properties of the Schmidt decomposition must fail, even if some
fixed choice of $\mathcal{H}_1$ and $\mathcal{H}_2$ is allowed.  The
Schmidt decomposition seems inadequate as, although it allows a
plausible interpretation of the quantum state at a single fixed time,
its time evolution has no natural interpretation consistent with the
predictions of Copenhagen quantum theory.

Many studies have been made of the behaviour of the Schmidt
decomposition during system-environment interactions.  In developing
the ideas of this paper, we were influenced in particular by
Albrecht's
investigations~\cite{Albrecht:decoherence,Albrecht:collapsing} of the
behaviour of the Schmidt decomposition in random Hamiltonian
interaction models and the description of these models by consistent
histories.

\subsection{Combining consistency and the Schmidt decomposition} 

The idea motivating this chapter is that the combination of the ideas
of the consistent histories formalism and the Schmidt decomposition
might allow us to define a mathematically precise and physically
interesting description of the quantum theory of a closed system.  The
Schmidt decomposition defines four natural classes of projection
operators, which we refer to collectively as \emph{Schmidt
projections}. These take the form
\begin{equation} \label{schmidtprojs1}
  \begin{array}{lll}
  P_i^1 (t) = | w_i (t)  \rangle_1 \langle w_i (t)  |_1 \otimes I_2
  &\mbox{and}& \overline P^1 = I_1 \otimes I_2 - \sum_i P_i^1 (t)\, ,
  \\ P_i^2 (t) = I_1 \otimes | w_i (t)  \rangle_2 \langle w_i (t)  |_2
  &\mbox{and}& \overline P^2 = I_1 \otimes I_2 - \sum_i P_i^2 (t) \, ,
  \\ P_i^{3} (t) 
  = | w_i (t)  \rangle_1 \langle w_i (t)  |_1 \otimes | w_i
  (t)  \rangle_2 \langle w_i (t)  |_2 &\mbox{and}& \overline P^3 
  = I_1 \otimes
  I_2 - \sum_i P_i^{3} (t) \, , \\ P_{ij}^{4} (t) = | w_i (t)  \rangle_1
  \langle w_i (t)  |_1 \otimes | w_j (t)  \rangle_2 \langle w_j (t)  |_2
  &\mbox{and}& \overline P^4 = I_1 \otimes I_2 - \sum_{ij} P_{ij}^{4}
  (t)\,.
 \end{array}
\end{equation}
If $\mbox{dim}\mathcal{H}_1 = \mbox{dim}\mathcal{H}_2$ the
complementary projections $ \overline P^1$, $ \overline P^2$ and
$\overline P^4$ are zero.

Since the fundamental problem with the consistent histories approach
seems to be that it allows far too many consistent sets of
projections, and since the Schmidt projections appear to be natural
dynamically determined projections, it seems sensible to explore the
possibility that a physically sensible rule can be found which selects
a consistent set or sets from amongst those defined by Schmidt
projections.

The first problem in implementing this idea is choosing the split
$\mathcal{H} = \mathcal{H}_1 \otimes \mathcal{H}_2$.  In analysing
laboratory experiments, one obvious possibility is to separate the
system and apparatus degrees of freedom.  Other possibilities of more
general application are to take the split to correspond to more
fundamental divisions of the degrees of freedom --- fermions and
bosons, or massive and massless particles, or, one might speculate,
the matter and gravitational fields in quantum gravity.  Some such
division would necessarily have to be introduced if this proposal were
applied to cosmological models.

Each of these choices seems interesting to us in context, but none, of
course, is conceptually cost-free.  Assuming a division between system
and apparatus in a laboratory experiment seems to us unacceptable in a
fundamental theory, reintroducing as it does the Heisenberg cut which
post-Copenhagen quantum theory aims to eliminate.  It seems
justifiable, though, for the limited purpose of discussing the
consistent sets which describe physically interesting histories in
laboratory situations.  It also allows useful tests: if an algorithm
fails to give sensible answers here, it should probably be discarded;
if it succeeds, applications elsewhere may be worth exploring.

Postulating a fundamental split of Hilbert space, on the other hand,
seems to us acceptable in principle.  If the split chosen were
reasonably natural, and if it were to produce a well-defined and
physically sensible interpretation of quantum theory applied to closed
systems, we would see no reason not to adopt it.  This seems a
possibility especially worth exploring in quantum cosmology, where any
pointers towards calculations that might give new physical insight
would be welcome.

Here, though, we leave aside these motivations and the conceptual
questions they raise, as there are simpler and more concrete problems
which first need to be addressed.  Our aim in this chapter is simply
to explain the problems which arise in trying to define consistent set
selection algorithms using the Schmidt decomposition, to set out some
possibilities, and to explain their properties, using simple models of
quantum systems interacting with an idealised experimental device or
with a series of such devices.

The most basic question here is precisely which of the Schmidt
projections should be used.  Again, our view is pragmatic: we would
happily adopt any choice that gave physically interesting results.
Where we discuss the abstract features of Schmidt projection
algorithms below, the discussion is intended to apply to all four
choices. When we consider simple models of experimental setups, we
take $\mathcal{H}_1$ to describe the system variables and
$\mathcal{H}_2$ the apparatus or environment.  Here we look for
histories which describe the evolution of the system state, tracing
over the environment, and so discuss set selection algorithms which
use only the first class of Schmidt projections: the other
possibilities are also interesting, but run into essentially the same
problems.  Thus, in the remainder of the chapter, we use the term
Schmidt projection to mean the system space Schmidt projections
denoted by $P_i^1$ and $\overline P^1$ defined in
eq.~(\ref{schmidtprojs1}).

In most of the following discussion, we consider algorithms which use
only the properties of the state vector $ |\psi (t) \rangle$ and its
Schmidt decomposition to select a consistent set. However, we will
also consider later the possibility of reconstructing a branching
structure defined by the decomposition
\begin{equation}
 |\psi (t) \rangle = \sum_{i=1}^{N(t)} |\psi_i (t) \rangle \, ,
\end{equation} 
in which the selected set is branch-dependent and the distinct
orthogonal components $ |\psi_i (t) \rangle$ correspond to the
different branches at time $t$.  In this case, we will consider the
Schmidt decompositions of each of the $ |\psi_i (t) \rangle$
separately.  Again, it will be sufficient to consider only the first
class of Schmidt projections.  In fact, for the branch-dependent
algorithms we consider, all of the classes of Schmidt projection
select the same history vectors and hence select physically equivalent
consistent sets.

\section{Approximate consistency and non-triviality}\label{sec:approxcon} 

In realistic examples it is generally difficult to find simple
examples of physically interesting sets that are exactly consistent.
For simple physical projections, the off-diagonal terms of the
decoherence matrix typically decay exponentially.  The sets of
histories defined by these projections separated by times much larger
than the decoherence time, are thus typically very nearly but not
precisely
consistent\cite{GM:Hartle:classical,Caldeira:Leggett,Joos:Zeh,%
Dowker:Halliwell,Pohle,Paz:brownian:motion,%
Zurek:preferred:observables,Anastopoulos:Halliwell,%
Tegmark:Shapiro,Paz:Zurek:classicality,Hartle:spacetime,%
Zurek:transition}.  Histories formed from Schmidt projections are no
exception: they give rise to exactly consistent sets only in special
cases, and even in these cases the exact consistency is unstable under
perturbations of the initial conditions or the Hamiltonian.

The lack of simple exactly consistent sets is not generally thought to
be a fundamental problem \emph{per se}.  According to one
controversial view\cite{CEPI:GMH}, probabilities in any physical
theory need only be defined, and need only satisfy sum rules, to a
very good approximation, so that approximately consistent sets are all
that is ever needed.  Incorporating pragmatic observation into
fundamental theory in this way clearly, at the very least, raises
awkward questions.  Fortunately, it seems unnecessary.  There are good
reasons to expect\cite{Dowker:Kent:approach} to find exactly
consistent sets very close to a generic approximately consistent set,
so that even if only exactly consistent sets are permitted the
standard quasiclassical description can be recovered.  Note, though,
that none of the relevant exactly consistent sets will generally be
defined by Schmidt projections.

It could be argued that physically reasonable set selection criteria
should make predictions which vary continuously with structural
perturbations and perturbations in the initial conditions, and that
the instability of exact consistency under perturbation means that the
most useful consistency criteria are very likely to be approximate.
Certainly, there seems no reason in principle why a precisely defined
selection algorithm which gives physically sensible answers should be
rejected if it fails to exactly respect the consistency criterion.
For, once a single set has been selected, there seems no fundamental
problem in taking the decoherence functional probability weights to
represent precisely the probabilities of its fine-grained histories
and the probability sum rules to \emph{define} the probabilities of
coarse-grained histories.  On the other hand, allowing approximate
consistency raises new difficulties in identifying a single natural
set selection algorithm, since any such algorithm would have --- at
least indirectly --- to specify the degree of approximation tolerated.

These arguments over fundamentals, though, go beyond our scope here.
Our aim below is to investigate selection rules which might give
physically interesting descriptions of quantum systems, whether or not
they produce exactly consistent sets.  As we will see, it seems
surprisingly hard to find good selection rules even when we follow the
standard procedure in the decoherence literature and allow some degree
of approximate decoherence.

Mathematical definitions of approximate consistency were first
investigated by Dowker and Halliwell\cite{Dowker:Halliwell}, who
proposed a simple criterion --- the Dowker-Halliwell criterion, or DHC
--- according to which a set is approximately consistent to order
$\epsilon$ if the decoherence functional
\begin{equation} \label{decohfunct} 
 D_{\alpha \beta } = \langle\psi| C_{\beta}^\dagger C_{\alpha}
 |\psi\rangle
\end{equation} 
satisfies the equation
\begin{eqnarray}\label{DHC}
  |D_{\alpha\beta}| & \leq & \epsilon \,
    (D_{\alpha\alpha}D_{\beta\beta})^{1/2}, \quad \forall\,
    \alpha\neq\beta.
\end{eqnarray}
Approximate consistency criteria were analysed further in chapter
\ref{chap:acp}.  As refs. \cite{Dowker:Halliwell,McElwaine:1} and
chapter \ref{chap:acp} explain, the DHC has natural physical
properties and is well adapted for mathematical analyses of
consistency.  We adopt it here, and refer to the largest term,
\begin{equation} \label{dhp} 
\mbox{max} \{\,
  |D_{\alpha\beta}|(D_{\alpha\alpha}D_{\beta\beta})^{-1/2} \, : \,
  \alpha , \beta \in S \, , \alpha \neq \beta \, , \mbox{~and~}
  D_{\alpha\alpha} , D_{\beta\beta} \neq 0 \,\} \, ,
\end{equation} 
of a (possibly incomplete) set of histories $S$ as the
Dowker-Halliwell parameter, or DHP.

\label{nontrivpage}
A \emph{trivial} history $\alpha$ is one whose probability is zero,
$C_\alpha |\psi\rangle = 0$.  Many of the algorithms we discuss
involve, as well as the DHP, a parameter which characterises the
degree to which histories approach triviality.  The simplest
non-triviality criterion would be to require that all history
probabilities must be greater than some parameter $\delta$, i.e.\ that
\begin{equation}\label{absnontriv} 
D_{\alpha\alpha} > \delta \quad \mbox{for all histories $\alpha$.}
\end{equation} 
As a condition on a particular extension $\{P_i : i = 1,2 , \ldots \}$
of the history $\alpha$ this would imply that $\|P_iC_\alpha
|\psi\rangle\|^2 > \delta$ for all $i$.  This, of course, is an
absolute condition, which depends on the probability of the original
history $\alpha$ rather than on the relative probabilities of the
extensions and which implies that once a history with probability less
than $2\delta$ has been selected any further extension is forbidden.

It seems to us more natural to use criteria, such as the DHC, which
involve only relative probabilities.  It is certainly simpler in
practice: applying absolute criteria strictly would require us to
compute from first cosmological principles the probability to date of
the history in which we find ourselves.  We therefore propose the
following relative non-triviality criterion: an extension $\{P_i : i =
1,2 , \ldots \}$ of the non-trivial history $\alpha$ is non-trivial to
order $\delta$, for any $\delta$ with $0 < \delta < 1$, if
\begin{equation} \label{relnontriv} 
\|P_i \, C_\alpha |\psi\rangle\|^2 \geq \delta \| C_\alpha
|\psi\rangle\|^2 \quad \mbox{for all $i$.}
\end{equation}
We say that a set of histories $S$, which may be branch-dependent, is
non-trivial to order $\delta$ if every set of projections, considered
as an extension of the histories up to the time at which it is
applied, is non-trivial to order $\delta$.  In both cases we refer to
$\delta$ as the non-triviality parameter, or NTP.

An obvious disadvantage of applying an absolute non-triviality
criterion to branch-independent consistent sets is that, if the set
contains one history of probability less than or equal to $2 \delta$,
no further extensions are permitted.

Once again, though, our approach is pragmatic, and in order to cover
all the obvious possibilities we investigate below absolute
consistency and non-triviality criteria as well as relative ones.

\section{Repeated projections and consistency}\label{sec:repeated} 

One of the problems which arises in trying to define physically
interesting set selection algorithms is the need to find a way either
of preventing near-instantaneous repetitions of similar projections or
of ensuring that such repetitions, when permitted, do not prevent the
algorithm from making physically interesting projections at later
times.  It is useful, in analysing the behaviour of repeated
projections, to introduce a version of the DHC which applies to the
coincident time limit of sets of histories defined by smoothly
time-dependent projective decompositions.

To define this criterion, fix a particular time $t_0$, and consider
class operators $C_\alpha$ consisting of projections at times ${\bf t}
= (t_1, \ldots, t_n)$, where $t_n > t_{n-1} > \ldots > t_1 >
t_0$. Define the \emph{normalised histories} by
\begin{equation}\label{normhist}
  |\hat \alpha\rangle = \lim_{{\bf t'} \to {\bf t}} \frac{C_\alpha
      ({\bf t}') |\psi\rangle}{\|C_\alpha ({\bf t}') |\psi\rangle\|},
\end{equation}
where the limits are taken in the order $t'_1 \to t_1$ then $t'_2 \to
t_2$ and so on, whenever these limits exist.  Define the limit DHC
between two normalised histories $|\hat \alpha \rangle$ and $| \hat
\beta \rangle$ as
\begin{equation}\label{limitDHC}
  \langle \hat \alpha | \hat \beta\rangle \leq \epsilon \, .
\end{equation}
This, of course, is equivalent to the limit of the ordinary DHC when
the limiting histories exist and are not null.  It defines a stronger
condition when the limiting histories exist and at least one of them
is null, since in this case the limit of the DHC is automatically
satisfied.

If a set of histories is defined by a smoothly time-dependent
projective decomposition applied at two nearby times, it will contain
many nearly null histories, since $P_m P_n = 0$ for all $n \neq m$.
Clearly, in the limit as the time separation tends to zero, these
histories become null, so that the limit of the ordinary DHC is
automatically satisfied.  When do the normalised histories satisfy the
stronger criterion (\ref{limitDHC})?

Let $P(t)$ be a projection operator with a Taylor series at $t=0$,
\begin{equation}\label{taylorproj}
  P(t) = P + t \dot{P} + \frac{1}{2}t^2\ddot{P} + O(t^3) \, ,
\end{equation}
where $P=P(0)$, $\dot{P} = dP(t)/dt|_{t=0}$ and $\ddot{P} =
d^2P(t)/dt^2|_{t=0}$. Since $P^2(t) = P(t)$ for all $t$
\begin{equation} 
\begin{array}{rcl} 
  P + t \dot{P} + \frac{1}{2}t^2\ddot{P} + O(t^3) &=& [P + t \dot{P} +
  \frac{1}{2}t^2\ddot{P} + O(t^3)] [P + t \dot{P} +
  \frac{1}{2}t^2\ddot{P} + O(t^3)] \\ &=& P + t(P\dot{P} + \dot{P}P) +
  \frac{1}{2}t^2(P\ddot{P} + \ddot{P}P + 2\dot{P}^2) + O(t^3) \, .
\end{array}
\end{equation} 
This implies that
\begin{equation} \label{dPidenta}
  \dot{P} = P\dot{P} + \dot{P}P \, ,
\end{equation}
and
\begin{equation} \label{dPidentb}
\frac{1}{2}\ddot{P} = \frac{1}{2}P\ddot{P} + \frac{1}{2}\ddot{P}P +
\dot{P}^2 \,.
\end{equation}
Now consider a projective decomposition $\{P_k\}$ and the matrix
element
\begin{equation}
  \langle \psi | P_{m} P_{k} (t) P_{n} | \psi \rangle = \langle \psi |
  P_{m} P_{k} P_{n} | \psi \rangle + t \langle \psi | P_{m} \dot P_{k}
  P_{n} | \psi \rangle + \frac{1}{2} t^2 \langle \psi | P_{m} \ddot
  P_{k} P_{n} | \psi \rangle + O(t^3)\, . \label{matrixel1}
\end{equation}
Now $P_{m} P_{k} P_{n} = P_k \delta_{km} \delta_{kn}$, since the
projections are orthogonal, and
\begin{equation} 
\begin{array}{rcl} 
  P_{m} \dot P_{k} P_{n} &=& \delta_{km} (1-\delta_{kn}) \dot P_{k}
  P_{n} + \delta_{kn}(1-\delta_{km}) P_{m} \dot P_{k} \\ &=&
  \delta_{km} \dot P_{k} P_{n} + \delta_{kn} P_{m} \dot P_{k} -
  \delta_{km} \delta_{kn} \dot P_{k} \, ,
\end{array}
\end{equation} 
since $\dot P_{k} P_n = P_k \dot P_{k} P_n $ if $ k \neq n$ and $\dot
P_{k} P_k = (1 - P_k ) \dot P_{k} $.  (No summation convention applies
throughout this paper.)  From eq.~(\ref{dPidentb}) we have that
\begin{eqnarray}
  \frac{1}{2} P_{m} \ddot P_{k} P_{n} &=& \frac{1}{2} (\delta_{mk} +
  \delta_{nk}) P_{m} \ddot P_{k} P_{n} + P_{m} \dot P_{k}^2 P_{n} \, .
\end{eqnarray} 
Eq.~(\ref{matrixel1}) can now be simplified. To leading order in $t$
it is
\begin{eqnarray} 
  && \langle \psi | P_k | \psi \rangle + O(t) \qquad \mbox{if
    $k=m=n$,} \\&& t\langle \psi|\dot P_k P_n |\psi \rangle +O(t^2)
    \qquad \mbox{if $k=m$, $k\neq n$,} \\&& t\langle \psi| P_m \dot
    P_k |\psi \rangle +O(t^2) \qquad \mbox{if $k\neq m$, $k=n$, and}
    \\ \label{repproj} && t^2 \langle \psi | P_{m} \dot P_{k}^2 P_{n}
    | \psi \rangle + O(t^3) \qquad \mbox{if $k\neq m$, $k \neq n$.}
\end{eqnarray}

Now consider a smoothly time-dependent projective decomposition,
$\sigma(t) = \{ P(t) , \overline{P} (t) \}$, defined by a
time-dependent projection operator and its complement.  Write $P=
P(0)$, and consider a state $|\phi\rangle$ such that $P|\phi\rangle
\neq 0 $ and $\overline{P}|\phi\rangle \neq 0 $.  We consider a set of
histories with initial projections $P, \overline{P}$, so that the
normalised history states at $t=0$ are
\begin{equation}
  \left\{ \frac{P|\phi\rangle}{\|P|\phi\rangle\|},
      \frac{\overline{P}|\phi\rangle}{\|\overline{P}|\phi\rangle\|}
      \right\} \, ,
\end{equation}
and consider an extended branch-dependent set defined by applying
$\sigma (t)$ on one of the branches --- say, the first --- at a later
time $t$.

The new normalised history states are
\begin{equation}\label{a1}
  \left\{ \frac{P(t)P|\phi\rangle}{\|P(t)P|\phi\rangle\|},
   \frac{\overline{P}(t)P|\phi\rangle}{\|\overline{P}(t)P|\phi\rangle\|},
   \frac{\overline{P}|\phi\rangle}{\|\overline{P}|\phi\rangle\|}
   \right\}.
\end{equation}
We assume now that $\dot{P}P|\phi\rangle \neq 0$, so that the limit of
these states as $t \to 0$ exists.  We have that
\begin{equation}
  \lim_{t\to0} \frac{(\overline P - t \dot{P}) P |\phi\rangle}{(t^2
    \langle \phi | P \dot P^2 P | \phi \rangle)^{1/2}} =
    \frac{-\dot{P} P |\phi\rangle }{\|\dot{P} P |\phi\rangle \|} \, ,
\end{equation}
so that the limits of the normalised histories are
\begin{equation}
  \left\{ \frac{P|\phi\rangle}{\|P|\phi\rangle\|},
   \frac{-\dot{P}P|\phi\rangle}{\|\dot{P}P|\phi\rangle\|},
   \frac{\overline{P}|\phi\rangle}{\|\overline{P}|\phi\rangle\|}
   \right\}.
\end{equation}
The only possibly non-zero terms in the limit DHC are
\begin{equation} \label{dblprojDHCterm}
  -\frac{\langle\phi| \overline{P} \dot{P} P |\phi\rangle }{\|
    \overline{P} |\phi\rangle \| \, \| \dot{P} P |\phi\rangle \|} =
    -\frac{\langle\phi| \overline{P} \dot{P} |\phi\rangle }{\|
    \overline{P} |\phi\rangle \| \, \| \overline{P} \dot{P}
    |\phi\rangle \|} \, ,
\end{equation}
which generically do not vanish.

Consider instead extending the second branch using $P(t)$ again. This
gives the set
\begin{equation}
  \left\{\frac{P|\phi\rangle}{\|P|\phi\rangle\|},
    \frac{\overline{P}|\phi\rangle}{\|\overline{P}|\phi\rangle\|},
    \frac{-P(t)\dot{P}P|\phi\rangle}{\|P(t)\dot{P}P|\phi\rangle\|},
    \frac{-\overline{P}(t)\dot{P}P|\phi\rangle}{\|
    \overline{P}(t)\dot{P}P |\phi\rangle\|} \right\}.
\end{equation}
Since $P\dot{P}P = 0$ the limit $t \to 0 $ exists and is
\begin{equation}
  \left\{\frac{P|\phi\rangle}{\|P|\phi\rangle\|},
    \frac{\overline{P}|\phi\rangle}{\|\overline{P}|\phi\rangle\|},
    \frac{-\dot{P}^2P|\phi\rangle}{\|\dot{P}^2P|\phi\rangle\|},
    \frac{-\dot{P}P|\phi\rangle}{\|\dot{P}P|\phi\rangle\|} \right\}.
\end{equation}
The DHC term between the first and third histories is
\begin{equation}\label{tripleprojDHCterm}
  -\frac{\langle\phi| P\dot{P}^2P |\phi\rangle}{\|P|\phi\rangle\| \,
    \|\dot{P}^2P |\phi\rangle\|} =
    -\frac{\|\dot{P}P|\phi\rangle\|^2}{\| P|\phi\rangle\| \,
    \|\dot{P}^2P |\phi\rangle\|} \, .
\end{equation}
This is always non-zero since $P\dot{P}|\phi\rangle \neq 0$.

For the same reason, extending the first branch again, or the third
branch, violates the limit DHC\@.  Hence, if projections are 
taken from a continuously parameterised set, and the limit DHC is
used, multiple re-projections will generically be forbidden.

The assumption that $\dot P P |\phi\rangle \neq 0$ can be relaxed. It
is sufficient, for example, that there is some $k$ such that
$\|P^{(j)}\|=0$ for all $j<k$ and that $P^{(k)} P |\phi\rangle \neq
0$, where $P^{(j)} = d^jP(t)/dt^j|_{t=0}$.

Note, finally, that it is easy to construct examples in which a single
re-projection is consistent.  For instance, let
\begin{equation}
  P = \left(
    \begin{array}{cc} I_{d_1} & 0 \\ 0 & 0 \end{array}
  \right) \quad \overline{P} = \left(
    \begin{array}{cc} 0 & 0 \\ 0 & I_{d_2} \end{array} 
  \right) \quad \dot{P} = \left(
    \begin{array}{cc} 0 & A^\dagger \\ A & 0 \end{array} \right)
  \quad |\phi\rangle = \left(
    \begin{array}{c} \sqrt{q} \, {\bf x}\\\sqrt{1-q} \, {\bf y},
    \end{array} \right)
\end{equation}
where ${\bf x}$ is a unit vector in $C^{d_1}$, ${\bf y}$ a unit vector
in $C^{d_2}$ and $A$ a $d_2 \times d_1$ complex matrix.
$\|P|\phi\rangle\| \neq 0,1$ implies that $q \neq 0,1$ and
$\dot{P}P|\phi\rangle \neq 0$ implies that $A{\bf x} \neq 0$. So from
eq. (\ref{dblprojDHCterm}) the DHC term is
\begin{equation}\label{tripleeg}
  -\frac{{\bf y}^\dagger A {\bf x}}{\|A {\bf x}\|} \, .
\end{equation}
If $d_2 \geq 2$ then ${\bf y}$ can be chosen orthogonal to $A {\bf x}$
and then eq.~(\ref{tripleeg}) is zero. The triple projection term
however, eq.~(\ref{tripleprojDHCterm}) is
\begin{equation}
  -\frac{\| A {\bf x} \|^2}{\| A^2{\bf x} \|} \, ,
\end{equation}
which is never equal to $0$ since $A{\bf x} \neq 0$.

\section{Schmidt projection algorithms}\label{sec:schmidt:alg} 

We turn now to the problem of defining a physically sensible set
selection algorithm which uses Schmidt projections, starting in this
section with an abstract discussion of the properties of Schmidt
projection algorithms.

We consider here dynamically generated algorithms in which initial
projections are specified at $t=0$, and the selected consistent set is
then built up by selecting later projective decompositions, whose
projections are sums of the Schmidt projection operators, as soon as
specified criteria are satisfied.  The projections selected up to time
$t$ thus depend only on the evolution of the system up to that time.
We will generally consider selection algorithms for branch-independent
sets and add comments on related branch-dependent selection
algorithms.

We assume that there is a set of Heisenberg picture Schmidt projection
operators $\{P_n (t)\}$ with continuous time dependence, defined even
at points where the Schmidt probability weights are degenerate, write
$P_n$ for $P_n (0)$, and let $I$ be the index set for projections
which do not annihilate the initial state, $I = \{ n : P_n
|\psi\rangle \neq 0 \}$.

We consider first a simple algorithm, in which the initial projections
are fixed to be the $P_n$ for $ n \in I$ together with their
complement $( 1 - \sum_n {P_n})$, and which then selects
decompositions built from Schmidt projections at the earliest possible
time, provided they are consistent.  More precisely, suppose that the
algorithm has selected a consistent set $S_k$ of projective
decompositions at times $t_0 , t_1 , \ldots , t_k$.  It then selects
the earliest time $t_{k+1} > t_k$ such that there is at least one
consistent extension of the set $S_k$ by a projective decomposition
formed from sums of Schmidt projections at time $t_{k+1}$.  In generic
physical situations, we expect this decomposition to be unique.
However, if more than one such decomposition exists, the one with the
largest number of projections is selected; if more than one
decomposition has the maximal number of projections, one is randomly
selected.

Though the limit DHC~(\ref{limitDHC}) can prevent trivial projections,
it does not generically do so here. The limit DHC terms between
histories $m$ and $n$ for an extension involving $P_k$ ($k \not \in
I$) are
\begin{equation} \label{initialDHC} 
  \lim_{t \to 0} \frac{| \langle \psi | P_{m} P_{k} (t) P_{n} | \psi
    \rangle | }{\| P_{m} | \psi \rangle \| \, \| P_{k} (t) P_{n} |
    \psi \rangle\|} = t \frac{| \langle \psi | P_{m} \dot P_{k}^2
    P_{n} | \psi \rangle | }{\| P_{m} | \psi \rangle \| \, \| \dot
    P_{k} P_{n} | \psi \rangle \|} = 0,
\end{equation} 
whenever $\| P_{m} | \psi \rangle \| $ and $\| \dot P_{k} P_{n} | \psi
\rangle\|$ are both non zero.  The first is non-zero by assumption;
the second is generically non-zero.  Thus the extension of all
histories by the projections $P_k \, (k \not \in I)$ and $\sum_{n \in
I} P_n$ satisfies the limit DHC.

Hence, if the initial projections do not involve all the Schmidt
projections, and if the algorithm tolerates any degree of approximate
consistency, whether relative or exact, then the DHC fails to prevent
further projections arbitrarily soon after $t=0$, introducing
histories with probabilities arbitrarily close to zero.
Alternatively, if the algorithm treats such projections by a limiting
process, then generically all the Schmidt projections at $t=0$ are
applied, producing histories of zero probability.  Similar problems
would generally arise with repeated projections at later times, if
later projections occur at all.

There would be no compelling reason to reject an algorithm which
generates unexpected histories of arbitrarily small or zero
probability, so long as physically sensible histories, of total
probability close to one, are also generated.  However, as we note in
subsection~\ref{sub:null} below and will see later in the analysis of
a physical example, this is hard to arrange.  We therefore also
consider below several ways in which small probability histories might
be prevented:
\begin{enumerate}
\item{} The initial state could be chosen so that it does not
precisely lie in the null space of any Schmidt projection. (See
subsection~\ref{sub:state}.)
\item{} An initial set of projections could somehow be chosen,
independent of the Schmidt projections, and with the property that for
every Schmidt projection at time zero there is at least one initial
history not in its null space.  (See subsection~\ref{sub:histories}.)
\item{} The algorithm could forbid zero probability histories by fiat
and require that the selected projective decompositions form an
exactly consistent set.  It could then prevent small probability
histories from occurring by excluding any projective decomposition
$\sigma (t)$ from the selected set if $\sigma (t)$ belongs to a
continuous family of decompositions, defined on some semi-open
interval $ ( t - \epsilon , t ]$, which satisfy the other selection
criteria. (See subsection~\ref{sub:adrian}.)
\item{} A parametrised non-triviality criterion could be used.  (See
subsection~\ref{sub:nontriv}.)
\item{} Some combination of parametrised criteria for approximate
consistency and non-triviality could be used. (See
subsection~\ref{sub:approxcon}.)
\end{enumerate}

We will see though, in this section and the next, that each of these
possibilities leads to difficulties.

\subsection{Choice of initial state}\label{sub:state}

In the usual description of experimental situations, ${\cal{H}}_1$
describes the system degrees of freedom, ${\cal{H}}_2$ those of the
apparatus (and/or an environment), and the initial state is a pure
Schmidt state of the form $ |\psi\rangle = |\psi_1 \rangle_1 \otimes
|\psi_2 \rangle_2$.  According to this description, probabilistic
events occur only after the entanglement of system and apparatus by
the measurement interaction.  It could, however, be argued that, since
states can never be prepared exactly, we can never ensure that the
system and apparatus are precisely uncorrelated, and the initial state
is more accurately represented by $ |\psi\rangle = |\psi_1 \rangle_1
\otimes |\psi_ 2\rangle_2 + \gamma |\phi\rangle$, where $\gamma$ is
small and $|\phi\rangle$ is a vector in the total Hilbert space chosen
randomly subject to the constraint that $ \langle \psi | \psi \rangle
= 1$.  A complete set of Schmidt projections $\{ P_n \}$, with $P_n
|\psi\rangle \neq 0 $ for all $n$, is then generically defined at
$t=0$, and any Schmidt projection algorithm which begins by selecting
all initial Schmidt projections of non-zero probability will include
all of the $ P_n $.

An obvious problem here, if relative criteria for approximate
consistency and non-triviality are used to identify subsequent
projections, is that the small probability initial histories constrain
the later projections just as much as the large probability history
which corresponds, approximately, to the Schmidt state
$|\psi_1\rangle_1 \otimes |\psi_2\rangle_2$ and which is supposed to
reproduce standard physical descriptions of the course of the
subsequent experiment.  If a branch-dependent selection algorithm is
used, a relative non-triviality criterion will not cause the small
probability initial histories to constrain the projections selected
later on the large probability branch, but a relative approximate
consistency criterion still will.

There seems no reason to expect the projections which reproduce
standard descriptions to be approximately consistent extensions of the
set defined by the initial Schmidt projections, and hence no reason to
expect to recover standard physics from a Schmidt projection
algorithm.  When we consider a simple model of a measurement
interaction in the next section we will see that, indeed, the initial
projections fail to extend to a physically natural consistent set.

If absolute criteria are used, on the other hand, we would expect
either that essentially the same problem arises, or that the small
probability histories do not constrain the projections subsequently
allowed and hence in particular do not solve the problems associated
with repeated projections, depending whether the probability of the
unphysical histories is large or small relative to the parameters
$\delta$ and $\epsilon^2$.

\subsection{Including null histories}\label{sub:null} 

If the initial state is Schmidt pure, or more generally does not
define a maximal rank Schmidt decomposition, a full set of Schmidt
projections can nonetheless generically be defined at $t=0$ --- which
we take to be the start of the interaction --- by taking the limit of
the Schmidt projections as $t \rightarrow 0^+$.  The normalised
histories corresponding to the projections of zero probability weight
can then be defined as above, if the relevant limits exist, and used
to constrain the subsequent projections in any algorithm involving
relative criteria.  Again, though, there seems no reason to expect
these constraints to be consistent with standard physical
descriptions.
 
\subsection{Redefining the initial conditions}\label{sub:histories}

The projections selected at $t=0$ could, of course, be selected using
quite different principles from those used in the selection of later
projections.  By choosing initial projections which are not
consistently extended by any of the decompositions defined by Schmidt
projections at times near $t=0$, we can certainly prevent any
immediate reprojection occurring in Schmidt selection algorithms.  We
know of no compelling theoretical argument against incorporating
projections into the initial conditions, but have found no natural
combination of initial projections and a Schmidt projection selection
algorithm that generally selects physically interesting sets.
  
\subsection{Exact consistency and a non-triviality criterion}
\label{sub:adrian}

Since many of the problems above arise from immediate reprojections,
it seems natural to look at rules which prevent zero probability
histories.  The simplest possibility is to impose precisely this
constraint, together with exact consistency and the rules that (i)
only one decomposition can be selected at any given time and (ii) no
projective decomposition can be selected at time $t$ if it belongs to
a continuous family of projections $\sigma (t)$, whose members would,
but for this rule, be selected at times lying in some interval $( t -
\epsilon , t ]$.  This last condition means that the projections
selected at $t=0$ are precisely those initially chosen and that no
further projections occur in the neighbourhood of $t=0$.
Unfortunately, as the model studied later illustrates, it also
generally prevents physically sensible projective decompositions being
selected at later times.  If it is abandoned, however, and if the
initial state $| \psi \rangle$ is a pure Schmidt state, then further
projections will be selected as soon as the interaction begins: in
other words, at times arbitrarily close to $t=0$.  Again, these
projections are generally inconsistent with later physically natural
projections.  On the other hand, if $| \psi \rangle$ is Schmidt-impure,
this is generally true of the initial projections.

All of these problems also arise in the case of branch-dependent set
selection algorithms.

\subsection{Exact consistency and a parametrised non-triviality criterion} 
\label{sub:nontriv}

Another apparently natural possibility is to require exact consistency
together with one of the parametrised non-triviality criteria
(\ref{absnontriv}) or (\ref{relnontriv}), rather than simply
forbidding zero probability histories.  A priori, there seem no
obvious problems with this proposal but, again, we will see that it
gives unphysical answers in the model analysed below, whether
branch-dependent or branch-independent selection algorithms are
considered.

\subsection{Approximate consistency and a parametrised non-triviality criterion}
\label{sub:approxcon}

There are plausible reasons, apart from the difficulties of other
proposals, for studying algorithms which use approximate consistency
and parametrised non-triviality.  The following comments apply to both
branch-dependent and branch-independent algorithms of this type.

Physically interesting sets of projective decompositions --- for
example, those characterising the pointer states of an apparatus after
each of a sequence of measurements --- certainly form a set which is
consistent to a very good approximation.  Equally, in most cases
successive physically interesting decompositions define non-trivial
extensions of the set defined by the previous decompositions: if the
probability of a measurement outcome is essentially zero then, it
might plausibly be argued, it is not essential to include the outcome
in the description of the history of the system.  Moreover, a finite
non-triviality parameter $\delta$ ensures that, after a Schmidt
projective decomposition is selected at time $t$, there is a finite
time interval $[ t , t + \Delta t ]$ before a second decomposition can
be chosen.  One might hope that, if the parameters are well chosen,
the Schmidt projective decompositions at the end of and after that
interval will no longer define an approximately consistent extension
unless and until they correspond to what would usually be considered
as the result of a measurement-type interaction occurring after time
$t$.  While, on this view, the parameters $\epsilon$ and $\delta$ are
artificial, one might also hope that they might be eliminated by
letting them tend to zero in a suitable limit.

However, as we have already mentioned, in realistic physical
situations we should not necessarily expect any sequence of Schmidt
projective decompositions to define an exactly consistent set of
histories.  When the Schmidt projections correspond, say, to pointer
states, the off-diagonal terms of their decoherence matrix typically
decay exponentially, vanishing altogether only in the limit of
infinite time
separation\cite{GM:Hartle:classical,Caldeira:Leggett,Joos:Zeh,%
Dowker:Halliwell,Pohle,Paz:brownian:motion,%
Zurek:preferred:observables,Anastopoulos:Halliwell,%
Tegmark:Shapiro,Paz:Zurek:classicality,Hartle:spacetime,%
Zurek:transition}.  An algorithm which insists on exact consistency,
applied to such situations, will fail to select any projective
decompositions beyond those initially selected at $t=0$ and so will
give no historical description of the physics.  We therefore seem
forced, if we want to specify a Schmidt projection set selection
algorithm mathematically, to introduce a parameter $\epsilon$ and to
accept sets which are approximately consistent to order $\epsilon$ and
then, in the light of the preceding discussion, to introduce a
non-triviality parameter $\delta$ in order to try to prevent
unphysical projective decompositions being selected shortly after
$t=0$.  This suggests, too, that the best that could be expected in
practice from an algorithm which uses a limit in which $\epsilon$ and
$\delta$ tend to zero is that the resulting set of histories describes
a series of events whose time separations tend to infinity.

A parameter-dependent set selection algorithm, of course, leaves the
problem of which values the parameters should take.  One might hope,
at least, that there is a range of values for $\epsilon$ and $\delta$
over which the selected set varies continuously and has essentially
the same physical interpretation.  An immediate problem here is that,
if the first projective decomposition selected after $t=0$ defines a
history which only just satisfies the non-triviality condition, the
decomposition will, once again, have no natural physical
interpretation and will generally be inconsistent with the physically
natural decompositions which occur later.  We will see that, in the
simple model considered below, this problem cannot be avoided with an
absolute consistency criterion.
 
Suppose now that we impose the absolute non-triviality condition that
all history probabilities must be greater than $\delta$ together with
the relative approximate consistency criterion that the modulus of all
DHC terms is less than $\epsilon$.  The parameters $\epsilon$ and
$\delta$ must be chosen so that these projections stop being
approximately consistent before they become non-trivial otherwise
projections will be made as soon as they produce histories of
probability exactly $\delta$, in which case the non-triviality
parameter, far from eliminating unphysical histories, would be
responsible for introducing them.

Let $t_\epsilon$ denote the latest time that the extension with
projection $P_k(t)$ is approximately consistent and $t_\delta$ the
earliest time at which the extension is nontrivial. We
see from~(\ref{initialDHC}) that, to lowest order in $t$,
\begin{eqnarray}
  t_\delta &=& \sqrt{\delta} \|\dot P_{k} P_{n} | \psi \rangle \|^{-1}
  \\ t_\epsilon &=& \epsilon \frac{\| P_{m} | \psi \rangle \| \, \|
  \dot P_{k} P_{n} | \psi \rangle \|}{|\langle \psi | P_{m} \dot
  P_{k}^2 P_{n} | \psi \rangle|}.
\end{eqnarray}
$t_\delta > t_\epsilon$ implies
\begin{equation}\label{edinequality}
  \sqrt{\delta} |\langle \psi | P_{m} \dot P_{k}^2 P_{n} | \psi
  \rangle| > \epsilon \| P_{m} | \psi \rangle \|\, \| \dot P_{k} P_{n}
  | \psi \rangle \|^2.
\end{equation}
Thus we require $\delta > \epsilon^2$, up to model-dependent numerical
factors: this, of course, still holds if we use a relative
non-triviality criterion rather than an absolute one.

This gives, at least, a range of parameters in which to search for
physically sensible consistent sets, and over which there are natural
limits --- for example $\mbox{lim}_{\delta \rightarrow 0}
\mbox{lim}_{\epsilon \rightarrow 0}$.  We have, however, as yet only
looked at some model-independent problems which arise in defining
suitable set selection rules.  In order to gain some insight into the
physical problems, we look next at a simple model of
system-environment interactions.


\chapter{A simple spin model}\label{chap:spin}

\section{Introduction}\label{sec:spin:intro} 

We now consider a simple model in which a single spin half particle,
the system, moves past a line of spin half particles, the environment,
and interacts with each in turn.  This can be understood as modelling
either a series of measurement interactions in the laboratory or a
particle propagating through space and interacting with its
environment.  In the first case the environment spin half particles
represent pointers for a series of measuring devices, and in the
second they could represent, for example, incoming photons interacting
with the particle.

Either way, the model omits features that would generally be
important.  For example, the interactions describe idealised sharp
measurements --- at best a good approximation to real measurement
interactions, which are always imperfect.  The environment is
represented initially by the product of $N$ particle states, which are
initially unentangled either with the system or each other.  The only
interactions subsequently considered are between the system and the
environment particles, and these interactions each take place in
finite time.  We assume too, for most of the following discussion,
that the interactions are distinct: the $k^{\mbox{\scriptsize th}}$ is
complete before the $(k+1)^{\mbox{\scriptsize th}}$ begins.  It is
useful, though, even in this highly idealised example, to see the
difficulties which arise in finding set selection algorithms: we take
the success of a set selection algorithm here to be a necessary, but
not sufficient, condition for it to be considered as a serious
candidate.

\subsection{Definition of the model} 

We use a vector notation for the system states, so that if ${\bf u}$
is a unit vector in $R^3$ the eigenstates of $\sigma. {\bf u }$ are
represented by $| \bf \pm u \rangle$.  With the pointer state analogy
in mind, we use the basis $\{ |\uparrow\rangle_k ,
|\downarrow\rangle_k \}$ to represent the $k^{\mbox{\scriptsize th}}$
environment particle state, together with the linear combinations
$|\pm\rangle_k = (|\uparrow\rangle_k \pm i|\downarrow\rangle_k
)/\sqrt{2}$.  We compactify the notation by writing environment states
as single kets, so that for example $ |\uparrow\rangle_1 \otimes
\cdots \otimes |\uparrow\rangle_n $ is written as $| \uparrow_1 \ldots
\uparrow_n \rangle$, and we take the initial state $|\psi(0)\rangle$
to be $|{\bf v}\rangle \otimes | \uparrow_1 \ldots \uparrow_n
\rangle$.

The interaction between the system and the $k^{\mbox{\scriptsize th}}$
environment particle is chosen so that it corresponds to a measurement
of the system spin along the ${\bf u}_k$ direction, so that the states
evolve as follows:
\begin{eqnarray} \label{measurementinteraction}
  |{\bf u}_k\rangle \otimes |\uparrow\rangle_k & \to & |{\bf
    u}_k\rangle \otimes |\uparrow\rangle_k \, , \\ |{\bf -u}_k \rangle
    \otimes |\uparrow\rangle_k & \to & |{\bf -u}_k \rangle \otimes
    |\downarrow\rangle_k.
\end{eqnarray}
A simple unitary operator that generates this evolution is
\begin{equation}\label{Ukdef} 
  U_k( t ) = P({\bf u}_k) \otimes I_k + P({\bf -u}_k) \otimes
  \mbox{e}^{-i\theta_k(t) F_k} \, ,
\end{equation}
where $P({\bf x}) = |{\bf x}\rangle \langle{\bf x}|$ and $F_k =
i|\downarrow\rangle_k \langle\uparrow|_k - i|\uparrow\rangle_k
\langle\downarrow|_k$.  Here $\theta_k(t)$ is a function defined for
each particle $k$, which varies from $0$ to $\pi/2$ and represents how
far the interaction has progressed.  We define $P_k ({ \pm}) = |{ \pm
}\rangle_k \langle{ \pm}|_k $, so that $F_k = P_k (+)-P_k (-)$.

The Hamiltonian for this interaction is thus
\begin{equation}
  H_{k}(t) = i\dot U_k (t) U_k^\dagger (t) \\ = \dot \theta_k(t)
  P({\bf -u}_k) \otimes F_k \, ,
\end{equation}
in both the Schr\"odinger and Heisenberg pictures.  We write the
extension of $U_k$ to the total Hilbert space as
\begin{equation}\label{Vkdef} 
  V_k = P({\bf u}_k) \otimes I_1 \otimes \cdots \otimes I_n + P({\bf
    -u}_k) \otimes I_1 \otimes \cdots \otimes I_{k-1} \otimes
    \mbox{e}^{-i\theta_k(t) F_k} \otimes I_{k+1} \otimes \cdots
    \otimes I_n \,.
\end{equation}
We take the system particle to interact initially with particle $1$
and then with consecutively numbered ones, and there is no interaction
between environment particles, so that the evolution operator for the
complete system is
\begin{equation}
  U(t) = V_n(t) \ldots V_1(t) \, ,
\end{equation}
with each factor affecting only the Hilbert spaces of the system and
one of the environment spins.

We suppose, finally, that the interactions take place in disjoint time
intervals and that the first interaction begins at $t=0$, so that the
total Hamiltonian is simply
\begin{equation} 
 H (t ) = \sum_{k=1}^n H_k (t) \, ,
\end{equation} 
and we have that $\theta_1 (t) > 0 $ for $t > 0$ and that, if
$\theta_k(t) \in ( 0,\pi/2 ) $, then $\theta_i(t) = \pi/2
{\rm~for~all~} i < k$ and $\theta_i(t) = 0 {\rm~for~all~} i >k$.

\section{Classification of Schmidt projection consistent sets in the model} 
\label{sec:spin:analysis}

For generic choices of the spin measurement directions, in which no
adjacent pair of the vectors $\{{\bf v},{\bf u}_1, \ldots ,{\bf
u}_n\}$ is parallel or orthogonal, the exactly consistent
branch-dependent sets defined by the Schmidt projections onto the
system space can be completely classified in this model. The following
classification theorem is proved in this section:

\vspace{.5\baselineskip}
\noindent\emph{Theorem}\qquad 
In the spin model defined above, suppose that no adjacent pair of the
vectors $\{{\bf v},{\bf u}_1, \ldots ,{\bf u}_n\}$ is parallel or
orthogonal.  Then the histories of the branch-dependent consistent
sets defined by Schmidt projections take one of the following forms:
\begin{description} 
\item[(i)] a series of Schmidt projections made at times between the
interactions --- i.e.\ at times $t$ such that $\theta_k (t) = {0
{\rm~or~} \pi/2} {\rm~for~all~} k$.
\item[(ii)] a series as in (i), made at times $t_1 , \ldots , t_n$,
together with one Schmidt projection made at any time $t$ during the
interaction immediately preceding the last projection time $t_n$.
\item[(iii)] a series as in (i), together with one Schmidt projection
made at any time $t$ during an interaction taking place after $t_n$.
\end{description} 
Conversely, any branch-dependent set, each of whose histories takes
one of the forms (i)-(iii), is consistent.  \vspace{.5\baselineskip}

\noindent We assume below that the set of spin measurement directions satisfies 
the condition of the theorem: since this can be ensured by an
arbitrarily small perturbation, this seems physically reasonable.  The
next sections explain, with the aid of this classification, the
results of various set selection algorithms applied to the model.

\subsection{Calculating the Schmidt states}

Eq.~(\ref{Ukdef}) can be written
\begin{equation}
  U_j(t) = e^{-i\theta_j(t) P(-{\bf u}_j)} \otimes P_j(+) +
    e^{i\theta_j(t) P_j(-{\bf u}_j)} \otimes P_j(-).
\end{equation}
Define $x_{+j}(t) = \exp[-i\theta_j(t) P({\bf -u}_j)]$ and $x_{-j}(t) =
x^\dagger_{+j}(t)$ so $U_j(t) = x_{+j}(t) \otimes P_j(+) + x_{-j}(t)
\otimes P_j(-)$. Let ${\bf \pi}$ be a string of $n$ pluses and
minuses, $|\pi\rangle$ denote the environment state $|\pi_1\rangle_1
\otimes \cdots \otimes |\pi_n\rangle_n$, $P(\pi) =
|\pi\rangle\langle\pi|$ and $x_\pi(t) = x_{\pi_nn}(t) \ldots
x_{\pi_11}(t)$.  Then
\begin{equation}\label{Uteq}
  U(t) = \sum_\pi x_\pi(t) \otimes P(\pi).
\end{equation}
The time evolution of the initial state $|\psi(0)\rangle = |{\bf
  v}\rangle \otimes |\uparrow_1 \ldots \uparrow_n\rangle$, the
  corresponding reduced density matrix and the Schmidt decomposition
  can now be calculated,
\begin{equation}
  |\psi(t)\rangle = \sum_\pi x_\pi(t) \otimes P(\pi) |{\bf v}\rangle
  \otimes |\uparrow_1 \ldots \uparrow_n\rangle = 2^{-n/2} \sum_\pi
  x_\pi(t) |{\bf v}\rangle \otimes |\pi\rangle,
\end{equation}
since $P(\pi)|\uparrow_1 \ldots \uparrow_n\rangle = 2^{-n/2}
|\pi\rangle$.  The reduced density matrix is
\begin{equation}\label{rra}
  \rho_r(t) = \mbox{Tr}_E [|\psi(t)\rangle\langle\psi(t)|] = 2^{-n}
  \sum_\pi x_\pi(t) P({\bf v}) x^\dagger_\pi(t).
\end{equation}
This can be further simplified by using the homomorphism between
$SU(2)$ and $SO(3)$. Define the rotation operators
\begin{equation}
  B_{+k}(t) = P({\bf u}_k) + \cos\theta_k(t) \overline{P}({\bf u}_k) -
  \sin\theta_k(t) {\bf u_k} \wedge,
\end{equation}
$B_{-k}(t) = B_{+k}^T(t)$ and $B_{\pi jk}(t) = B_{\pi_kk}(t) \ldots
B_{\pi_jj}(t)$. $B_{+k}(t)$ corresponds to a rotation of angle
$\theta_k(t)$ about ${\bf u}_k$, and $P({\bf u}_k) = {\bf u}_k {\bf
u}_k^T$, a projection operator on $R^3$. Note that $P({\bf u}_k)$ is
also used to indicate a projection in the system Hilbert space --- its
meaning should be clear from the context. $B_{\pi1n}(t)$ will usually
be simplified to $B_{\pi}(t)$. Then $x_{\pi_11}(t) P({\bf v})
x^\dagger_{\pi_11}(t) = P[B_{\pi_11}(t){\bf v}]$.  Eq.~(\ref{rra}) can
then be written
\begin{equation}\label{rrb}
  \rho_r(t) = 2^{-n} \sum_\pi P[B_\pi(t){\bf v}].
\end{equation}
Define $A_j(t) = 1/2[B_{+j}(t) + B_{-j}(t)] = P({\bf u}_j) +
\cos\theta_j(t) \overline P({\bf u}_j)$ and $A_{jk}(t) = A_k(t) \ldots
A_j(t)$, then $2^{-n}\sum_\pi B_\pi(t) = A_{1n}(t)$. $A_{1n}(t)$ will
usually be written $A(t)$.  Since $P[B_\pi(t){\bf v}]$ is linear in
$B_\pi(t)$ the sum in eq.~(\ref{rrb}) can then be done, so
\begin{equation}\label{rrc}
  \rho_r(t) = \frac{1 + \sigma.A(t){\bf v}}{2}.
\end{equation}
Generically this is not a projection operator since $|A(t){\bf v}|$
may not equal $1$. It is convenient however to define $P({\bf y}) =
1/2(1 + \sigma.{\bf y})$ for all ${\bf y} \in C^3$, and this extended
definition will be used throughout the paper. $P({\bf y})$ is a
projection operator if and only if ${\bf y}$ is a real unit vector.
Eq.~(\ref{rrc}) can now be written as $\rho_r(t) = P[A(t){\bf v}]$.

The eigenvalues of eq.~(\ref{rrc}) are $1/2[1\pm N(t)]$ and the
corresponding eigenstates, for $N(t) \neq 0$, are $|\pm {\bf
w}(t)\rangle$, where $N(t) = |A(t){\bf v}|$ and ${\bf w}(t) = A(t){\bf
v} N^{-1}(t)$.

Lemma $1$. \emph{Sufficient conditions that $N(t)\neq0$ for all $t$
  are that $\theta_i(t) \geq \theta_j(t)$ for all $i< j$ and ${\bf
  u}_i.  {\bf u}_{i+1} \neq 0$ for all $i\geq0$.}

Proof. Suppose $\exists t $ s.t.\ $N(t) = 0$, $\Rightarrow \mbox{det}
A(t) = 0$, $\Rightarrow \exists j $ s.t.\ $ \mbox{det} A_j(t) = 0$,
$\Rightarrow \theta_j(t) = \pi/2$. Let $j$ be the largest $j$ s.t.\
$\theta_j(t) = \pi/2$, then $A_i(t) = P({\bf u}_i) \forall i \leq j$
and $\mbox{det} A_i(t) \neq 0 \forall i>j$, $\Rightarrow N(t) = \|
A_{(j+1)n} (t) {\bf u}_j\| \prod_{j>i\geq0} |{\bf u}_i.{\bf u}_{i+1}|$
and $\mbox{det} A_{(j+1)n}(t) \neq 0$, $\Rightarrow \exists i$ s.t.\
$|{\bf u}_i.{\bf u}_{i+1}| = 0$ \#

For the rest of this paper it will be assumed that $\{\theta_i\}$ and
$\{{\bf u}_i\}$ satisfy the conditions of lemma $1$. The condition on
the $\{\theta_i\}$ holds so long as the environment spin particles are
further apart than the range of their individual interactions. The
condition on $\{{\bf u}_i\}$ holds generically and is physically
reasonable since any realistic experiment will not have exact
alignment.

\subsection{Decoherence matrix elements}
The Heisenberg picture Schmidt projection operators are
\begin{equation}
  P^\pm_H(t) = U^\dagger(t) P[{\bf \pm w}(t)] \otimes I_E U(t)
  \label{HSproja}.
\end{equation}
Eq.~(\ref{HSproja}) can be rewritten using eq.~(\ref{Uteq})
\begin{equation}
  P^\pm_H(t) = \sum_\pi x^\dagger_\pi(t) P[{\bf \pm w}(t)]
  x_\pi(t)\otimes P(\pi) = \sum_\pi P[{\bf \pm w}_\pi(t)] \otimes
  P(\pi), \label{HSprojb}
\end{equation}
where ${\bf w}_\pi(t) = B^T_\pi(t) {\bf w}(t)$.

Consider the probability of a history consisting of projections at
time $t$ and then $s$, where the projectors are Schmidt projectors.
\begin{equation}\label{proba}
  p(\pm\pm) = \| P^\pm_H(s) P^\pm_H(t) |\psi(0)\rangle\|^2.
\end{equation}
Eq.~(\ref{proba}) simplifies using eq.~(\ref{HSprojb}) and $ P(\pi)
|\psi(0)\rangle = 2^{-1/2} |{\bf v}\rangle \otimes |\pi\rangle$ to
become
\begin{eqnarray}
  p(\pm\pm) &=& \sum_\pi \| P[{\bf \pm w}_\pi(s)] P[{\bf \pm
    w}_\pi(t)] |{\bf v}\rangle \otimes P(\pi) |\uparrow_1 \ldots
    \uparrow_n\rangle\|^2 \nonumber \\ &=& 2^{-n-2} \sum_\pi [1 \pm
    {\bf w}_\pi(t).{\bf v}] [1 \pm {\bf w}_\pi(t). {\bf
    w}_\pi(s)]. \label{probb}
\end{eqnarray}
The off-diagonal decoherence matrix elements can be calculated
similarly.
\begin{eqnarray} 
  \lefteqn{\langle\psi(0)| P^\pm_H(t) P^\pm_H(s) P^\mp_H(t)
    |\psi(0)\rangle} && \nonumber \\ &=& 2^{-n} \sum_\pi \mbox{Tr} \{
    P({\bf v}) P[{\bf \pm w}_\pi(t)] P[{\bf \pm w}_\pi(s)] P[{\bf \mp
    w}_\pi(t)] \} \nonumber\\ &=& 2^{-n-2} \sum_\pi [{\bf w}_\pi (t)
    \wedge {\bf v}]. [ \pm {\bf w}_\pi (t) \wedge {\bf w}_\pi (s) \pm
    i {\bf w}_\pi (s)] \,. \label{twooffa}
\end{eqnarray}

For a general set of vectors $\{{\bf u}_k\}$ and time functions
$\{\theta_k\}$ eqs. (\ref{probb}) and (\ref{twooffa}) are very
complicated. However, with a restricted set of time functions a
complete analysis is possible. The functions $\{\theta_k\}$ are said
to describe a \emph{separated interaction} if, for all $t$, there
exists $k$ s.t. $\theta_j(t) = \pi/2$ for all $j<k$, and $\theta_j(t)
= 0$ for all $j>k$. For separated interactions a projection time $t$
is said to be \emph{between} interactions $j$ and $j+1$ when
$\theta_i(t) = \pi/2 $ for all $i \leq j$ and $\theta_i(t) = 0$ for
all $i > j$.  A projection time $t$ is said to be \emph{during}
interaction $j$ when $\theta_i(t) = \pi/2$ for all $i < j$,
$\theta_i(t) = 0$ for all $i > j$ and $0 < \theta_j(t) < \pi/2$.
Separated interactions have a simple physical meaning: the
interactions with the environment spins occur distinctly, and in
sequence.

Under this restriction a complete classification of all the consistent
sets, both branch dependent and branch independent, is possible. This
classification has a particularly simple form for generic ${\bf v}$
and $\{{\bf u}_k\}$ satisfying ${\bf u}_k.{\bf u}_{k+1} \neq 0$, and
${\bf u}_k \wedge {\bf u}_{k+1} \neq 0$ for all $k = 0, \ldots, n-1$.
Recall ${\bf u}_0 = {\bf v}$.  For weak consistency the second
requirement is stronger $({\bf u}_k \wedge {\bf u}_{k+1}).  ({\bf
u}_{k+2} \wedge {\bf u}_{k+1}) = {\bf u}_k \overline P ({\bf u}_{k+1})
{\bf u}_{k+1} \neq 0$. These assumptions will be assumed to hold
unless stated otherwise.

\subsection{Classification theorem}
The proof first considers projections at two times and shows that a
pair of times gives rise to non-trivial consistent histories only when
the earlier time is between interactions or the earlier time is during
an interaction and the later time between this interaction and the
next. The second part of the proof shows that any set of
branch-independent histories consisting of branches that satisfy this
rule for all pairs of projections is consistent. The proof holds for
weak and medium consistency criteria.

\subsubsection{Allowed histories}
Let $t$ be a time during interaction $j$. Define $\omega =
\theta_j(t)$ and $\phi = \theta_j(s)$. Define ${\bf x} = A_{1(j-1)}(s)
{\bf v} = A_{1(j-1)}(t) {\bf v}$ and ${\bf y} = A^T_{(j+1)n}(s)
A_{1n}(s) {\bf v}$.  Note $B_{\pi 1n}(t) = B_{\pi 1j}(t)$ and $B_{\pi
1(j-1)}(t) = B_{\pi 1(j-1)}(s)$ since $t<s$. With this notation and
using simple vector identities the off-diagonal elements of the
decoherence matrix (from eq.~\ref{twooffa}) are
\begin{equation} \label{twooffb}
  2^{-(n+2)} \sum_\pi [ {\bf w}(t) \wedge B_\pi(t){\bf v}].  [\pm {\bf
  w}(t) \wedge B_\pi(t) {\bf w}_\pi(s) \pm i B_\pi(t) {\bf w}_\pi(s)].
\end{equation}
Now
\begin{equation}\label{twooffba}
  B_\pi(t) {\bf w}_\pi (s) = B_{\pi j}(t) B_{\pi 1(j-1)}(t) B^T_{\pi
    1(j-1)}(s) B^T_{\pi jn}(s) {\bf w}(s) = B_{\pi j}(t) B^T_{\pi
    jn}(s) {\bf w}(s),
\end{equation}
which only depends on $\pi_i$ for $i \geq j$. Since $B_{\pi 1j}(t){\bf
  v}$ only depends on $\pi_i$ for $i \leq j$ the sum
  eq.~(\ref{twooffba}) can be done over all $\pi_i$, $i \neq j$.
\begin{eqnarray}
  2^{1-j} \sum_{\pi_i,\, i<j} B_{\pi 1j}(t){\bf v} &=& [A_j (t) -
  \pi_j \sin\omega\, {\bf u}_j \wedge] A_{1(j-1)}(t) {\bf v}\\ &=&
  {\bf w}(t) N(t) - \pi_j \sin\omega\, {\bf u}_j \wedge {\bf x},
\end{eqnarray}
\begin{eqnarray}
  2^{-(n-j)} \sum_{\pi_i,\, i>j} B_{\pi j}(t) B^T_{\pi jn}(s){\bf
  w}(s) &=& N^{-1}(s) B_{\pi j}(t) B_{\pi j}^T(s) A^T_{(j+1)n}(s)
  A_{1n}(s){\bf v} \\ &=& N^{-1}(s) B_{\pi j}(t) B_{\pi j}^T(s) {\bf
  y}.
\end{eqnarray}
Substitute these last two results into eq.~(\ref{twooffb}) which
becomes
\begin{eqnarray}\nonumber
  \lefteqn{2^{-3} N^{-1}(s) \sum_{\pi_j} \{{\bf w}(t) \wedge [{\bf
    w}(t) N(t) - \pi_j \sin\omega {\bf u}_j \wedge {\bf x}]\}}
    \hspace{2in} \\ &&.  \label{twooffc} [\pm {\bf w}(t) \wedge B_{\pi
    j}(t) B_{\pi j}^T(s){\bf y} \pm i B_{\pi j}(t) B_{\pi j}^T(s){\bf
    y}].
\end{eqnarray}
This can easily be simplified since ${\bf w}(t) \wedge {\bf w}(t) =
0$. The only remaining term in the first bracket is then linear in
$\pi_j$, so when the sum over $\pi_j$ is taken only the terms linear
in $\pi_j$ in the second bracket remain. Eq.~(\ref{twooffc}) is
therefore
\begin{equation} \label{twooffd}
  1/4 N^{-1}(s) \sin\omega \sin(\omega-\phi) [{\bf w}(t) \wedge ({\bf
    u}_j \wedge {\bf x})]. [{\bf w}(t) \wedge ({\bf u}_j \wedge {\bf
    y}) \pm i {\bf u}_j \wedge {\bf y}].
\end{equation}
Now ${\bf w}(t) = [P({\bf u}_j) + \cos\omega \overline P({\bf u}_j)]
{\bf x} N^{-1}(t)$ so ${\bf w} (t). ({\bf x} \wedge {\bf u}_j) =
0$. Therefore
\begin{equation}\label{s1}
  [{\bf w}(t) \wedge ({\bf u}_j \wedge {\bf x})]. [{\bf w}(t) \wedge
  ({\bf u}_j \wedge {\bf y})] ={\bf x}^T \overline P ({\bf u}_j) {\bf
  y}.
\end{equation}
Also ${\bf u}_j. {\bf w}(t) = {\bf u}_j. {\bf x}N^{-1}(t)$ so
\begin{equation}\label{s2}
  [{\bf w}(t) \wedge ({\bf u}_j \wedge {\bf x})]. ({\bf u}_j \wedge
  {\bf y}) = - N^{-1}(t) ({\bf u}_j. {\bf x}) {\bf x}. ({\bf u}_j
  \wedge {\bf y}).
\end{equation}
Eq.~(\ref{twooffc}) can be simplified using eq.~(\ref{s1}) and
eq.~(\ref{s2}) to
\begin{equation} \label{twooffe}
  1/4 N^{-1}(s) \sin\omega \sin(\phi-\omega) \{ \pm {\bf x}^T
  \overline P ({\bf u}_j) {\bf y} \pm i N^{-1}(t) ({\bf u}_j. {\bf x})
  {\bf u}_j. ({\bf x} \wedge {\bf y}) \}
\end{equation}
The probabilities can be calculated during the same results. Summing
all the terms $i \neq j$ in eq.~(\ref{probb}) results in
\begin{eqnarray}\nonumber
  &&2^{-3} \sum_{\pi_j} \{1 \pm {\bf w}(t).[{\bf w}(t) N(t) - \pi_j
  \sin\omega {\bf u}_j \wedge {\bf x}]\} \left\{ 1 \pm \frac{{\bf x}^T
  A_j(\omega) B_{\pi j}(t) B^T_{\pi j}(s) {\bf y}}{N(s)N(t)}\right\}
  \\ &=& 2^{-2} [1 \pm N(t)] \left\{1 \pm \frac{{\bf x}^T [P({\bf
  u}_j) + \cos\omega\cos(\phi-\omega) \overline P({\bf u}_j)] {\bf
  y}}{N(s)N(t)}\right\}\label{probc}
\end{eqnarray}
$N^2(s) = |A_{1n}(s){\bf v}| = {\bf x}^T A_j(\phi) {\bf y}$ and
  $\cos(\omega-\phi) \cos\omega - \cos\phi =
  \sin\omega\sin(\phi-\omega)$, so eq.~(\ref{probc}) is
\begin{equation}\label{probd}
  1/4[1 \pm N(t)] \left[1 \pm \frac{N^2(s) + \sin\omega
    \sin(\phi-\omega){\bf x}^T \overline P({\bf u}_j) {\bf y}}
    {N(s)N(t)}\right]
\end{equation}

To write the decoherence matrix without using ${\bf x}$ and ${\bf y}$
it is necessary to consider three cases: when times $s$ and $t$ are
during the same interaction, when they are during adjacent
interactions and when they are during separated interactions. If $t$
is during interaction $j$ and $s$ during interaction $k$ the three
cases are $k=j$, $k=j+1$ and $k>j+1$. For the remainder of this
section let $\phi = \theta_k(s)$,
\begin{equation}
  N_j(\omega) = |A_j(t) {\bf u}_{j-1}| \mbox{~and~} \lambda_{ij} =
  \prod_{j>k\geq i} |{\bf u}_k. {\bf u}_{k+1}|\,.
\end{equation}
Then
\begin{eqnarray}
  {\bf x} &=& \lambda_{0(j-1)} {\bf u}_{j-1} \\ N(t) &=&
  \lambda_{0(j-1)} N_j(\omega) \\ N(s) &=& \lambda_{0(k-1)} N_k(\phi)
  \\ {\bf y} &=& \left\{
  \begin{array}[c]{ll}
    \lambda_{0(j-1)} A_j(s) {\bf u}_{j-1} & \mbox{for $k=j$} \\
    \lambda_{0j} A_{j+1}^2(s) {\bf u}_{j} & \mbox{for $k=j+1$} \\
    \lambda_{(j+1)(k-1)} \lambda_{0(k-1)} N^2_k(\phi) {\bf u}_{j+1} &
    \mbox{for $k>j+1$}
  \end{array} \right.
\end{eqnarray}
The probabilities of the histories (eq.~\ref{probc}) are
\begin{eqnarray}
  p(\pm\pm) = 1/4 [1 \pm \lambda_{0(j-1)} N_j(\omega)] [1 \pm a]
\end{eqnarray}
where
\begin{eqnarray}
  a = \left \{
    \begin{array}[c]{ll}
      \frac{N_j^2(\phi) + \sin\omega \cos\phi \sin(\phi-\omega) |{\bf
          u}_{j-1} \wedge {\bf u}_j|^2}{ N_j(\omega) N_j(\phi)} &
          \mbox{for $k=j$} \\ \frac{ \lambda_{(j-1)j} N_{j+1}^2(\phi)
          + \cos\omega \sin\omega \lambda^2_{j(j+1)} \sin^2\phi {\bf
          u}_{j-1}^T \overline P({\bf u}_j) {\bf u}_{j+1}}{
          N_j(\omega) N_{j+1}(\phi)} &\mbox{for $k=j+1$} \\
          N_{k}(\phi)\frac{ \lambda_{(j-1)(k-1)} +
          \lambda_{(j+1)(k-1)} \cos\omega \sin\omega {\bf u}_{j-1}^T
          \overline P({\bf u}_j) {\bf u}_{j+1}}{ N_j(\omega) }
          &\mbox{for $k>j+1$}
    \end{array} \right..
\end{eqnarray}
The nonzero off-diagonal terms are (eq.~\ref{twooffe})
\begin{equation} \label{twoofff}
\left \{
  \begin{array}[c]{ll}
    \frac{\lambda_{0(j-1)} \sin\omega \sin(\phi-\omega) \cos\phi |{\bf
        u}_{j-1} \wedge {\bf u}_j|^2}{4 N_j(\phi)} & \mbox{for
        $k=j$}\\ \frac{\lambda_{0(j-1)} \lambda_{j(j+1)} \sin\omega
        \cos \omega \sin^2 \phi[ N_j(\omega) {\bf u}_{j-1}^T \overline
        P({\bf u}_j) {\bf u}_{j+1} \pm i \lambda_{(j-1)j} {\bf
        u}_{j-1}.({\bf u}_j \wedge {\bf u}_{j+1})]}{4 N_j(\omega)
        N_{j+1}(\phi)} & \mbox{for $k=j+1$}\\ \frac{\lambda_{0(j-1)}
        \lambda_{(j+1)(k-1)} N_k(\phi) \sin\omega \cos \omega[
        N_j(\omega) {\bf u}_{j-1}^T \overline P({\bf u}_j) {\bf
        u}_{j+1} \pm i \lambda_{(j-1)j} {\bf u}_{j-1}.({\bf u}_j
        \wedge {\bf u}_{j+1})]}{4 N_j(\omega)} & \mbox{for $k>j+1$.}
  \end{array} \right.
\end{equation}

The off-diagonal terms can be zero for two reasons, either there is a
degeneracy in the measurement spin directions, or $s$ and $t$ take
special values. The necessary and sufficient conditions for the
measurement spin directions not to be degenerate is that for all $j$
${\bf u}_j. {\bf u}_{j+1} \neq 0$ and ${\bf u}_j \wedge {\bf u}_{j+1}
\neq 0$. The first condition ensures that $\lambda_{ij} \neq 0$ for
all $i$ and $j$ and that the Schmidt states are well defined.  These
cases do not need to be considered when we are interested in exact
consistency because they have measure zero and \emph{almost surely}
under any perturbation the degeneracy will be lifted. If weak
consistency is used only the real part needs to vanish and the
measurement direction need to satisfy the stronger condition ${\bf
u}_{j-1}^T \overline P({\bf u}_j) {\bf u}_{j+1} \neq 0 $ for all
$j$. This is still of measure zero. If approximate consistency is
being considered the situation is more complicated as the histories
will remain approximately consistent under small enough perturbations.
This will not be considered in this letter.  Unless said otherwise it
will be assumed that the measurement spin direction are not
degenerate.

Therefore from eqs.~(\ref{twoofff}) the only pairs of times giving
rise to consistent projections are repeated projections (that is $s=t$
which implies $j=k$ and $\omega=\phi$), projections in between
interactions and any later time (that is $\omega = 0$ or $\pi/2$), and
a projection during an interaction and a projection at the end of the
same interaction (that is $j=k$ $\omega \in [0,\pi/2]$ and $\phi=
\pi/2$.)

\subsubsection{Probabilities of allowed histories}
The model is invariant under strictly monotonic reparameterisations of
time, $t \to f(t)$. Therefore for separated interactions no generality
is lost by choosing the time functions $\{\theta_j\}$ such that the
$j^{\mbox{\scriptsize th}}$ interaction finishes at $t=j$, that is
$\theta_i(j) = \pi/2$ for all $i \leq j$ and $\theta_i(j) = 0$ for all
$i>j$. It is convenient to define $R_{\pi ij} = [P({\bf u}_i) - \pi_i
{\bf u}_i \wedge] \ldots [P({\bf u}_i) - \pi_i {\bf u}_i \wedge]$.
Then $B_\pi(m) = R_{\pi 1m}$.

Consider the history $\alpha$ that consists of projections at times
$\{m_i: i = 1,2, \ldots l\}$, then at time $t \in (k-1,k)$ and then at
time $k$, where $\{m_i,k\}$ is an ordered set of positive integers.
This history means that the particle spin was in direction $\pm{\bf
u}_{m_i}$ at time $m_i$, $i = 1, \ldots ,l$, direction $\pm{\bf w}(t)$
at time $t$ and direction $\pm{\bf u}_k$ at time $k$.  Define ${\bf
u}_0 = {\bf v}$ and $m_0 = 0$.

Using the same method as for two projections the probability for
history $\alpha$ is
\begin{eqnarray} \nonumber
  p_\alpha &=& 2^{-n} 2^{-(l+2)} \sum_\pi \prod_{i=0}^{l-1} [1 +
  \alpha_{i} \alpha_{i+1} {\bf w}_\pi(m_{i}).  {\bf w}_\pi(m_{i+1})]
  \\ && \mbox{} \times [1 + \alpha_l \alpha_{t} {\bf w}_\pi(m_{l}).
  {\bf w}_\pi(t)] \times [1 + \alpha_{t} \alpha_{k} {\bf w}_\pi(t).
  {\bf w}_\pi(m_k)] \label{probsa}
\end{eqnarray}
Now
\begin{eqnarray}
  {\bf w}_\pi(m_{i}). {\bf w}_\pi(m_{i+1}) = {\bf u}_{m_{i}}^T R_{\pi
  1 m_i} R^T_{\pi 1 m_{i+1}} {\bf u}_{m_{i+1}} = {\bf u}_{m_{i}}^T
  R_{\pi (m_{i}+1) m_{i+1}} {\bf u}_{m_{i+1}},
\end{eqnarray}
which only depends on $\pi_j$ for $m_{i+1} \geq j > m_i$.  Also
\begin{equation}
  {\bf w}_\pi(t).{\bf w}_\pi(k) = N^{-1}_k(t) {\bf u}_{k-1}^T A_k(t)
  B_{k\pi_k}(t) {\bf u}_{k} = N^{-1}_k(t) ({\bf u}_{k-1}. {\bf
  u}_{k}),
\end{equation}
which is independent of $\pi$ and
\begin{equation}
  {\bf w}_\pi(t).{\bf w}_\pi(m_l) = N^{-1}_k(t) {\bf u}_{k-1}^T A_k(t)
  B_{\pi_k k}(t) R_{\pi (m_l+1) (k-1)} {\bf u}_{m_l},
\end{equation}
which only depends on $\pi_j$ for $j > m_l$.  These last three
equations show that each $B_{\pi_i i}$ is linear so the sum over $\pi$
is trivial and each $B_{\pi_i i}$ can be replaced by $A_i$.
\begin{equation}
  2^{m_{i}-m_{i+1}-1}
  \sum_{\makebox[0in][c]{\scriptsize$\pi_j,\,m_{i+1}>j>m_i$}} {\bf
  w}_\pi(m_{i}). {\bf w}_\pi(m_{i+1}) = {\bf u}_{m_{i}}^T P({\bf
  u}_{m_{i}+1}) \cdots P({\bf u}_{m_{i+1}-1}) {\bf u}_{m_{i+1}} =
  \lambda_{m_{i}m_{i+1}},
\end{equation}
\begin{equation}
  2^{m_l-k} \sum_{\makebox[0in][c]{\scriptsize$\pi_i,\,k \geq i >
  m_l$}} {\bf w}_\pi(t). {\bf w}_\pi(m_l) = N^{-1}_k(t) {\bf
  u}_{k-1}^T A^2_k(t) {\bf u}_{k-1} \lambda_{m_l(k-1)} =
  \lambda_{m_l(k-1)} N_k(t)
\end{equation}
Using these results to do the sum over all $\pi$ eq.~(\ref{probsa}) is
\begin{equation}\label{probsb}
  p_\alpha = 2^{-(l+2)} [1 + \alpha_l \alpha_{t} \lambda_{m_l(k-1)}
  N_k(t)] [1 + \alpha_t \alpha_{k} N^{-1}_k(t) ({\bf u}_{k-1}. {\bf
  u}_k)] \prod_{i=0}^{l-1} [1 + \alpha_i \alpha_{i+1}
  \lambda_{m_im_{i+1}}].
\end{equation}
 
\subsubsection{Consistency of allowed histories}
Since a coarse graining of a consistent set is consistent it is
sufficient to only consider the off-diagonal decoherence matrix
elements between the most finely grained allowed histories, which are
those that consist of projections between all interactions and one
projection during the interaction before the final projection. The
off-diagonal elements of the decoherence matrix arise from only three
forms, which depend on where the two branches separate, that is the
earliest projector where they differ.

First consider the case where two histories differ at a projection in
between interactions and all projections up to that point have also
been in between interactions. Let $C_\alpha = Q_\alpha P_H(k) \ldots
P_H(1)$ and $C_\beta = Q_\beta \overline P_H(k) \ldots P_H(1)$. The
decoherence matrix element between them is
\begin{eqnarray}\nonumber
  2^{-n} \sum_\pi \mbox{Tr} \{Q_\pi P({\bf u}_k) x_\pi(k) P[{\bf
    w}_\pi(k-1)] \ldots P[{\bf w}_\pi(1)] P({\bf v}) \\ \times P[{\bf
    w}_\pi(1)] \ldots P[{\bf w}_\pi(k-1)] x^\dagger_\pi(k) P(-{\bf
    u}_k)\} \label{offcase1a}
\end{eqnarray}
where $Q_\pi = \langle\pi| x_\pi(k)Q^\dagger_\alpha Q_\beta
x^\dagger_\pi(k) |\pi\rangle$. Since $Q_\alpha$ and $Q_\beta$ only
contain projections after interaction $k$ has completed $Q_\pi$ is
independent of $\pi_j$ for all $j \leq k$.  Now $P[{\bf w}_\pi(j)]
P[{\bf w}_\pi(j-1)] P[{\bf w}_\pi(j)] = 1/2(1+ {\bf u}_{j-1}. {\bf
u}_j) P[{\bf w}_\pi(j)]$. Let $\mu = 2^{1-m}\prod_{0<j<m} (1+ {\bf
u}_{j-1}.{\bf u}_j)$ and eq.~(\ref{offcase1a}) is
\begin{equation} \label{offcase1b}
  \mu 2^{-n} \sum_\pi \mbox{Tr} \{ Q_\pi P({\bf u}_k) P[B_\pi(k) {\bf
    w}_\pi(k-1)] P({\bf -u}_k)\}
\end{equation}
But $1/2\sum_{\pi_k} P[B_\pi(k) {\bf w}_\pi(k-1)] = P[{\bf u}_k({\bf
  u}_k.{\bf u}_{k-1})]$ and $P({\bf u}_k) P[{\bf u}_k({\bf u}_k.{\bf
  u}_{k-1})] P({\bf -u}_k) = 0$ so eq.~(\ref{offcase1b}) is zero.

Now consider $C_\alpha = P_H(k) P_H(t) P_H(k-1) \ldots P_H(1)$ and
$C_\beta = P_H(k) \overline P_H(t) P_H(k-1) \ldots P_H(1)$. The
decoherence matrix element between them is
\begin{equation} \label{offcase2a}
  \mu 2^{-n} \sum_\pi \mbox{Tr} \{P[{\bf w}_\pi(k)] P[{\bf w}_\pi(t)]
  P[{\bf w}_\pi(k-1)] P[{\bf -w}_\pi(t)] P[{\bf w}_\pi(k)]\},
\end{equation}
which, because $B_{\pi_kk} {\bf u}_k = {\bf u}_k$ equals
\begin{equation}
  \mu 2^{-n} \sum_\pi \mbox{Tr} \{P({\bf u}_k) P[{\bf w}(t)]
  P[B_{\pi_kk}(t) {\bf u}_{k-1}] P[{\bf -w}(t)] P({\bf w}(k)\}.
  \label{offcase2b}
\end{equation}
The sum over $\pi_k$ can be done to give $P[{\bf w}(t)] P[A_k(t) {\bf
  u}_{k-1}] P[{\bf -w}(t)]$, and since ${\bf w}(t)$ is parallel to
  $A_k (t) {\bf u}_{k-1}$, eq.~(\ref{offcase2b}) is zero.

The final case to consider is when then the histories $\alpha$ and
$\beta$ differ in their final projection. They will be trivially
consistent.

\section{Application of selection algorithms to the spin model}\label{app}

We can define a natural consistent set which reproduces the standard
historical account of the physics of the separated interaction spin
model by selecting the Schmidt projections at all times between each
successive spin measurement.  A set of this type ought to be produced
by a good set selection algorithm, either as the selected set itself
or, perhaps, a subset.  The first three subsections below describe the
results actually produced by various set selection algorithms applied
to the spin model.  All of these algorithms are dynamical, in the
sense that the decision whether to select projections at time $t$, and
if so which, depends only on the evolution of the state vector up to
time $t$.  The following two subsections discuss how these results are
affected by altering the initial conditions of the model.  In the next
subsection we consider a selection algorithm which is quasi-dynamical,
in the sense that the decisions at time $t$ depend on the evolution of
the state vector up to and just beyond $t$.  We summarise our
conclusions in the last subsection.

\subsection{Exact limit DHC consistency} 

Since any projective decomposition at time $t$ defines an exactly
consistent set when there is only one history up to that time, a
Schmidt projection selection algorithm without a non-triviality
criterion will immediately make a projection.  The normalised
histories are defined as
\begin{equation} 
  \lim_{t \to 0} P_\pm(t) |\psi\rangle / \|P_\pm(t) |\psi\rangle \| \,
  ,
\end{equation} 
where $P_\pm(t)$ denotes the Schmidt projections at time $t$.  The
Schmidt states to first order in $\omega = \theta_1(t)$ are
\begin{equation}
 |{\bf v}\rangle \otimes |\uparrow_1\ldots\uparrow_n\rangle - i
  \omega/2 (1 - {\bf u}_1. {\bf v}) |{\bf v}\rangle \otimes
  |\downarrow_1\uparrow_2\ldots\uparrow_n\rangle
\end{equation}
and
\begin{equation}
 |{\bf u}_1 \wedge {\bf v}| |{\bf -v}\rangle \otimes
  |\downarrow_1\uparrow_2\ldots\uparrow_n\rangle + i \omega/2 \sqrt{
  \frac{ 1 - {\bf u}_1. {\bf v} }{ 1 + {\bf u}_1. {\bf v} } } |{\bf
  -v}\rangle \otimes |\uparrow_1\ldots\uparrow_n\rangle \, ,
\end{equation}
so the normalised histories are
\begin{equation}
  \{ |{\bf v}\rangle \otimes
  |\uparrow_1\uparrow_2\ldots\uparrow_n\rangle , |{\bf -v}\rangle
  \otimes |\downarrow_1\uparrow_2\ldots\uparrow_n\rangle \}\,.
\end{equation}
The limit DHC term for one projection at time $0$ and another during
interaction $k$ at time $t$ (from eq.~\ref{twoofff}) is
\begin{equation}
\begin{array}{ll}  \displaystyle
  \cos\phi & \mbox{for $k = 1$,} \\ \displaystyle \frac{\sin^2\phi\,
  |{\bf u}_1.  {\bf u}_2| |{\bf v} \wedge ( {\bf u}_1 \wedge {\bf u}_2
  )|}{N_2(\phi) [1 - ({\bf v}. {\bf u}_1)^2 N_2^2(\phi)]^{1/2}} &
  \mbox{for $k = 2$,} \\[2ex] \displaystyle \frac{ \lambda_{2 (k-1)}
  N_k (\phi) | {\bf v} \wedge ( {\bf u}_1 \wedge {\bf u}_2 )|}{ [1 -
  \lambda^2_{0(k-1)} N^2_k (\phi)]^{1/2}} & \mbox{for $k > 2$} \, ,
\end{array} 
\end{equation}
where $\phi = \theta_k(t)$.

Whether the algorithm is taken to be branch-dependent or
branch-independent, the only future Schmidt projections which are
consistent with the initial projections are thus those between the
first and second interactions, and the projections selected will be at
the end of the first interaction. The state at this time is
\begin{equation}
  |\psi(1)\rangle = |{\bf u_1} \rangle \langle{\bf u}_1 | {\bf
  v}\rangle \otimes |\uparrow_1\ldots \uparrow_n\rangle + |{\bf -u}_1
  \rangle \langle{\bf -u}_1 | {\bf v}\rangle \otimes
  |\downarrow_1\uparrow_2\ldots\uparrow_n\rangle,
\end{equation}
The time evolved histories are
\begin{eqnarray}
  |h_1(t)\rangle &=& |{\bf u_1} \rangle \langle{\bf u}_1 | {\bf
  v}\rangle \otimes |\uparrow_1 \ldots \uparrow_n\rangle + |{\bf -u}_1
  \rangle\langle{\bf -u}_1 | {\bf v}\rangle \otimes |\uparrow_1 \ldots
  \uparrow_n\rangle \\ |h_2(t)\rangle &=& |{\bf u_1} \rangle
  \langle{\bf u}_1 | {\bf -v}\rangle \otimes |\downarrow_1\uparrow_2
  \ldots \uparrow_n\rangle - |{\bf -u}_1 \rangle \langle{\bf -u}_1 |
  {\bf -v}\rangle \otimes |\downarrow_1\uparrow_2 \ldots
  \uparrow_n\rangle
\end{eqnarray}
so the new normalised histories are
\begin{eqnarray}
  \{ |{\bf u}_1\rangle \otimes |\uparrow_1\ldots \uparrow_n\rangle,
  |{\bf u}_1\rangle \otimes |\downarrow_1\uparrow_2 \ldots
  \uparrow_n\rangle,\\ |{\bf -u}_1\rangle \otimes
  |\uparrow_1\rangle\ldots\uparrow_n\rangle, |{\bf -u}_1\rangle
  \otimes |\downarrow_1\uparrow_2 \ldots \uparrow_n\rangle\}.
\end{eqnarray}
Since no future Schmidt projections are consistent with those
selected, the algorithm clearly fails to produce the correct set.

\subsection{Exact consistency and non-triviality} 

Suppose that, instead of using the limit DHC, we consider only sets
defined by decompositions at different times and require exact
consistency.  As explained earlier, without a non-triviality criterion
this leads to an ill-defined algorithm: the initial projections at
$t=0$ produce a null history, and the Schmidt projections at all times
greater than zero are consistent with these initial projections, so
that no minimal non-zero time is selected by the algorithm.

Introducing a non-triviality criterion removes this problem.  Suppose,
for example, we impose the absolute criterion $D_{\alpha\alpha} \geq
\delta$ for all histories $\alpha$.  Since any physically reasonable
$\delta$ would have to be extremely small, let us assume $\delta \ll |
{\bf u}_i \wedge {\bf u}_j |$.  The first projections after $t=0$ are
then selected at the first time when $D_{\alpha\alpha} = \delta$,
which occurs during the first interaction.  Whether or not
branch-dependent projections are allowed, the only other Schmidt
projections which can consistently be selected then take place at the
end of the first interaction, and it again follows from the
classification theorem that no further projections can take place.
Again, by making projections too early, this algorithm fails to
produce the correct consistent set.

A suitably large value of $\delta$ could ensure that no extension will
occur until later interactions but, generically, the first extension
made after $t=0$ will take place during an interaction rather than
between interactions, and the classification theorem ensures that no
more than four histories will ever be generated.

The same problems arise if the non-triviality criterion is taken to be
relative rather than absolute.  It is possible to do better by
fine-tuning the parameters: for example, if branch independent
histories are used, a relative non-triviality criterion is imposed and
$ \delta = (1 - |{\bf u}_k. {\bf u}_{k+1}|)/2$ for all $k =
0,\ldots,n-1$, then projections will occur at the end of each
interaction producing the desired set of histories.  This, though, is
clearly not a satisfactory procedure.

\subsection{Approximate consistency and non-triviality} 

One might wonder if these problems can be overcome by relaxing the
standards of consistency, since a projection at a very small time will
be approximately consistent --- according to absolute measures of
approximate consistency, at least --- with projections at the end of
the other interactions.  However, this approach too runs into
difficulties, whether relative or exact criteria are used.

Consider first a branch-dependent set selection algorithm which uses
the absolute non-triviality criterion $D_{\alpha\alpha} \geq \delta$
for all $\alpha$, and the absolute criterion for approximate
consistency $|D_{\alpha\beta}| \leq \epsilon$ for all $\alpha \neq
\beta$.  No history with probability less than $2\delta$ will thus be
extended, since if it were one of the resultant histories would have
probability less than $\delta$.

Any history $\alpha$ with a probability less than or equal to
$\epsilon^2$ will automatically be consistent with any history $\beta$
according to this criterion, since $|D_{\alpha\beta}| \leq
(D_{\alpha\alpha} D_{\beta\beta})^{1/2} \leq (\epsilon^2 \cdot
1)^{1/2} = \epsilon$.  Therefore if $\delta \leq \epsilon^2$ then
histories of probability $\delta$ will be consistent with all other
histories.  The first projection after $t=0$ will be made as soon as
the non-triviality criterion permits, when the largest Schmidt
eigenvalue is $1-\delta$.  Other projections onto the branch defined
by the largest probability history will follow similarly as the
Schmidt projections evolve.  The final set of histories after $n$
projections will thus consist of one history with probability
$1-n\delta$ and $n$ histories with probability $\delta$ --- clearly
far from the standard picture.

Suppose now that $\delta > \epsilon^2$.  The probabilities for
histories with projection in the first interval, at time $t$ with
$\theta_1 (t) = \omega$, are
\begin{equation} 
   1/2 [1 - \sqrt{1- \sin^2\omega |{\bf v} \wedge {\bf u}_1|^2}].
\end{equation}
The first projection will therefore be made when
\begin{equation} \label{tdef} 
\theta_1 (t) = \omega \simeq 2\sqrt{\delta} |{\bf v} \wedge {\bf
  u}_1|^{-1} \, ,
\end{equation} 
producing histories of probabilities $\delta$ and $(1- \delta )$.  The
next projections selected will necessarily extend the history of
probability $(1- \delta )$, since the absolute non-triviality
criterion forbids further extensions of the other history.  We look
first at projections taking place at a later time $t'$, with $\theta_1
(t') = \phi$, during the first interaction, and define $N_1 (\omega )
= ( 1 - \sin^2\omega |{\bf v} \wedge {\bf u}_1|^2 )^{1/2} $.  Of the
probabilities of the extended histories, the smaller is
\begin{eqnarray}\nonumber
  && 1/4 [ 1 + N_1(\omega) ] \bigg\{ 1 - N^{-1}_1(\omega)
  N^{-1}_1(\phi) [({\bf v}. {\bf u}_1)^2 + \cos\phi \cos\omega
  \cos(\phi-\omega) |{\bf v} \wedge {\bf u}_1 |^2 ] \bigg\} \\ &=& 1/4
  |{\bf v} \wedge {\bf u}_1|^2 (\omega-\phi)^2 [ 1 + O(\omega) +
  O(\phi)] \, ,
\end{eqnarray} 
Therefore this extension will be non-trivial when
\begin{equation}
  \phi \simeq \omega + 2\sqrt{\delta} |{\bf v} \wedge {\bf u}_1|^{-1}
= 4\sqrt{\delta} |{\bf v} \wedge {\bf u}_1|^{-1} + O(\delta).
\end{equation}
The largest off-diagonal element in the decoherence matrix for this
extension is
\begin{eqnarray} \label{soreproj}
  1/4 N_1^{-1}(\phi) |{\bf v} \wedge {\bf u}_1|^2 \cos\phi \sin\omega
  \sin(\phi-\omega) = \delta + O(\delta^3).
\end{eqnarray}
Unless $\delta > \epsilon$, then, this extension is selected together,
again, with a series of further extensions generating small
probability histories.

Suppose now that $\delta > \epsilon$.  The term on the left hand side
of eq.~(\ref{soreproj}) increases monotonically until $\phi \simeq
\pi/4$, and then decreases again as $\phi \rightarrow \pi/2$.  For
$\phi \simeq \pi/2$, it equals
\begin{eqnarray}
 1/2 \sqrt{\delta} \cos \phi |{\bf v} \wedge {\bf u}_1| |{\bf v}. {\bf
  u}_1|^{-1} [1 + O(\cos\phi)] \,.
\end{eqnarray}
Hence the approximate consistency criterion is next satisfied when
\begin{equation}
  \phi = \pi/2 - \frac{2\epsilon |{\bf v}. {\bf u}_1|}{\sqrt{\delta}
        |{\bf v} \wedge {\bf u}_1|} + O(\epsilon^2/\delta) \, ,
\end{equation}
and this extension is also non-trivial unless $ \bf v $ and $ {\bf
u}_1 $ are essentially parallel, which we assume not to be the case.
In this case, then, projections are made towards the beginning and
towards the end of the first interaction, and a physically reasonable
description of the first measurement emerges.

This description, however, cannot generally be consistently extended
to describe the later measurements.  If we consider the set of
histories defined by the Schmidt projections at time $t$, given by
eq.~(\ref{tdef}) above, together with the Schmidt projections at time
$t''$ such that $\theta_k (t'') = \phi$ for some $k>1$, we find that
the largest off-diagonal decoherence matrix element is
\begin{equation} \label{tsep} 
1/2 \sqrt{\delta} \lambda_{2(k-1)} N_k(\phi) |{\bf v} \wedge {\bf
  u}_1| | {\bf v} \wedge ( {\bf u}_1 \wedge {\bf u}_2)|[ 1 +
  O(\sqrt\omega)] \,.
\end{equation}

Since we have chosen $\epsilon < \delta$ to prevent multiple
projections, and since the other terms are not small for generic
choices of the vectors, the set generally fails to satisfy the
criterion for approximate consistency.  Note, however, that if all the
measurement directions are apart by an angle greater than equal to
some $\theta >0$, then $\lambda_{2(k-1)}$ decreases exponentially with
$k$.  After a large enough number (of order $O(-\log \epsilon)$) of
interactions have passed the algorithm will select a consistent
extension, and further consistent extensions will be selected at
similar intervals.  The algorithm does thus eventually produce
non-trivial consistent sets, though the sets produced do not vary
smoothly with $\epsilon$ and do not describe the outcome of most of
the spin measurements.

The reason this algorithm, and similar algorithms using approximate
consistency criteria, fail is easy to understand.  The off-diagonal
decoherence matrix component in a set defined by the Schmidt
projections at time $t$ together with Schmidt projections during later
interactions is proportional to $\sin \omega \cos \omega$, together
with terms which depend on the angles between the vectors.  The
decoherence matrix component for a set defined by the projections at
time $t$, together with Schmidt projections at a second time $t'$ soon
afterwards is proportional to $\sin^2(\phi-\omega)$.  The obstacle to
finding non-triviality and approximate consistency criteria that can
prevent reprojections in the first interaction period, yet allow
interactions in later interaction periods, is that when $(\phi -
\omega)$ is small the second term is generally smaller than the first.

Using a relative non-triviality criterion makes no difference, since
the branchings we consider are from a history of probability close to
$1$, and using the DHC instead of an absolute criterion for
approximate consistency only worsens the problem of consistency of
later projections, since the DHC alters eq.~(\ref{tsep}) by a factor
of $1/\sqrt{\delta}$, leaving a term which is generically of order
unity.  Requiring branch-independence, of course, only worsens the
problems.

\subsection{Non-zero initial Schmidt eigenvalues}

We now reconsider the possibility of altering the initial conditions
in the context of the spin model.  Suppose first that the initial
state is not Schmidt degenerate.  For example, as the initial
normalised histories are $\{ |{\bf v}\rangle \otimes
|\uparrow_1\ldots\uparrow_n\rangle, |{\bf -v}\rangle \otimes
|\downarrow_1\uparrow_2\ldots\uparrow_n\rangle\}$ a natural ansatz is
\begin{equation}
  |\psi(0)\rangle = \sqrt{p_1}|{\bf v}\rangle
  \otimes|\uparrow_1\ldots\uparrow_n\rangle + \sqrt{p_2} |{\bf
  -v}\rangle \otimes |\downarrow_1\uparrow_2\ldots\uparrow_n\rangle
  \,.
\end{equation}
Consider now a set of histories defined by Schmidt projections at
times $0$ and a time $t$ during the $k^{\mbox{\scriptsize th}}$
interaction for $k >2$, so that $\theta_1 (t) = \theta_2 (t) = \pi/2$.
The moduluses of the non-zero off-diagonal elements of the decoherence
matrix are
\begin{equation}
  1/2 \sqrt{p_1p_2} | {\bf v} \wedge [{\bf u}_1 \wedge {\bf u}_2 ] |
\lambda_{2k} \,.
\end{equation}
Generically, these off-diagonal elements are not small, so that the
perturbed initial conditions prevent later physically sensible
projections from being selected.

\subsection{Specifying initial projections}

We consider now the consequence of specifying initial projections in
the spin model. Suppose the initial projections are made using $P({\bf
\pm h}) \otimes I_E$. The modulus of the non-zero off-diagonal
elements of the decoherence matrix for a projection at time $t$ during
interaction $k$, for $k>2$, is
\begin{equation} 
 1/4 | {\bf h} \wedge {\bf v}|\, |{\bf h} \wedge {\bf u}_1 |
\lambda_{1(k-1)} N_k ( \theta_k (t) ) \, ,
\end{equation} 
and again we see that physically natural projections generically
violate the approximate consistency criterion.
 
It might be argued that the choice of initial projections given by
${\bf h} = {\bf \pm v}$ is particularly natural.  This produces an
initial projection on to the initial state, with the other history
undefined unless a limiting operation is specified.  If the limit of
the normalised histories for initial projections ${\bf h'} \rightarrow
{\bf h}$ is taken, the normalised histories are simply $|{\bf \pm
h}\rangle$. If an absolute consistency criterion is used the null
history will not affect future projections and the results will be the
same as if no initial projection had been made. If, on the other hand,
the limit DHC is used then the consistency criterion is the same as
for general ${\bf h}$, that is ${\bf h}$ must be parallel to ${\bf
u}_1$.  This requires that the initial conditions imposed at $t=0$
depend on the axis of the first measurement, and still fails to permit
a physically natural description of later measurements.

\subsection{A quasi-dynamical algorithm}\label{quasi} 

For completeness, we include here an algorithm which, though not
strictly dynamical, succeeds in selecting the natural consistent set
to describe the spin model.  In the spin model as defined, it can be
given branch-dependent or branch-independent form and selects the same
set in either case.  In the branch-independent version, the Schmidt
projections are selected at time $t$ provided that they define an
exactly consistent and non-trivial extension of the set defined by
previously selected projections \emph{and} that this extension can
itself be consistently and non-trivially extended by the Schmidt
projections at time $t+ \epsilon$ for every sufficiently small
$\epsilon > 0$.\footnote{Alternatively, a limiting condition can be
used.}  In the branch-dependent version, the second condition must
hold for at least one of the newly created branches of non-zero
probability in the extended set.

It follows immediately from the classification theorem that no Schmidt
projections can be selected during interactions, since no exactly
consistent set of Schmidt projections includes projections at two
different times during interactions.  The theorem also implies that
the Schmidt projections are selected at the end of each interval
between interactions, so that the selected set describes the outcomes
of each of the measurements.

\subsection{Comments}

The simple spin model used here illustrates the difficulty in encoding
our physical intuition algorithmically.  The model describes a number
of separated interactions, each of which can be thought of as a
measurement of the system spin.  There is a natural choice of
consistent set, given by the projections onto the system spin states
along the measured axes at all times between each of the
measurements.\footnote{Strictly speaking, there are many equivalent
consistent sets, all of which include the Schmidt projections at some
point in time between each measurement and at no time during
measurements, and all of which give essentially the same physical
picture.}  This set does indeed describe the physics of the system as
a series of measurement events and assigns the correct probabilities
to those events.  Moreover, the relevant projections are precisely the
Schmidt projections.

We considered first a series of Schmidt projection set selection
algorithms which are dynamical, in the sense that the projections
selected at time $t$ depend only on the physics up to that time.
Despite the simplifying features of the models, it seems very hard to
find a dynamical Schmidt projection set selection algorithm which
selects a physically natural consistent set and which is not
specifically adapted to the model in question.

It might be argued that the very simplicity of the model makes it an
unsuitable testing ground for set selection algorithms.  It is
certainly true that more realistic models would generally be expected
to allow fewer exactly consistent sets built from Schmidt projections:
it is not at all clear that any non-trivial exactly consistent sets of
this type should be expected in general.  However, we see no way in
which all the problems encountered in our discussion of dynamical set
selection algorithms can be evaded in physically realistic models.

We have, on the other hand, seen that a simple quasi-dynamical set
selection algorithm produces a satisfactory description of the spin
model.  However, as we explain in the next section, there is another
quite general objection which applies both to dynamical set selection
algorithms and to this quasi-dynamical algorithm.

\section{The problem of recoherence} \label{sec:recoherence} 

The set selection algorithms above rely on the decoherence of the
states of one subsystem through their interactions with another.  This
raises another question: what happens when decoherence is followed by
recoherence?

For example, consider a version of the spin model in which the system
particle initially interacts with a single environment particle as
before, and then re-encounters the particle, reversing the
interaction, so that the evolution takes the form
\begin{equation} \label{recoh} 
\begin{array}{rcl} 
a_1 | {\bf u}\rangle \otimes |\uparrow_1\rangle + a_2 |{\bf -u}
\rangle \otimes |\uparrow_1\rangle & \to & a_1 | {\bf u}\rangle
\otimes |\uparrow_1\rangle + a_2 |{\bf -u} \rangle \otimes
|\downarrow_1\rangle \\ & \to & a_1 | {\bf u}\rangle \otimes
|\uparrow_1\rangle + a_2 |{\bf -u} \rangle \otimes |\uparrow_1\rangle
\, ,
\end{array} 
\end{equation}
generated by the unitary operator
\begin{equation}
  U( t ) = P({\bf u}) \otimes I + P({\bf -u}) \otimes
  \mbox{e}^{-i\theta(t) F} \, ,
\end{equation}
where
\begin{equation}
\theta (t) = \left\{ \begin{array}{ll} t & \mbox{for $0 \leq t \leq
\pi/2$,} \\ \pi/2 & \mbox{for $\pi/2 \leq t \leq \pi$,} \\ 3 \pi/2 - t
& \mbox{for $\pi \leq t \leq 3\pi/2$.}
\end{array} \right.
\end{equation}

We have taken it for granted thus far that a dynamical algorithm makes
selections at time $t$ based only on the evolution of the system up to
that time.  Thus any dynamical algorithm which behaves sensibly,
according to the criteria which we have used so far, will select a
consistent set which includes the Schmidt projections at some time
between $\pi/2$ and $\pi$, since during that interval the projections
appear to describe the result of a completed measurement.  These
projections cannot be consistently extended by projections onto the
initial state $ a_1 | {\bf u}\rangle + a_2 |{\bf -u} \rangle $ and the
orthogonal state $ \overline{a_2} | {\bf u}\rangle - \overline{a_1}
|{\bf -u} \rangle $ at time $3 \pi/2$, so that the algorithm will not
agree with the standard intuition that at time $\pi$ the state of the
system particle has reverted to its initial state.  In particular, if
the particle subsequently undergoes interactions of the form
(\ref{measurementinteraction}) with other environment particles, the
algorithm cannot reproduce the standard description of these later
measurements.  The same problem afflicts the quasi-dynamical algorithm
considered in subsection~\ref{quasi}.

In principle, then, dynamical set selection algorithms of the type
considered so far imply that, following any experiment in which exact
decoherence is followed by exact recoherence and then by a
probabilistic measurement of the recohered state, the standard
quasiclassical picture of the world cannot generally be recovered.  If
the algorithms use an approximate consistency criterion --- as we have
argued is necessary for a realistic algorithm --- then this holds true
for experiments in which the decoherence and recoherence are
approximate.

We know of no experiments of precisely this type.  Several neutron
interferometry experiments have been performed in which one or both
beams interact with an electromagnetic field before
recombination\cite{mds,wb,abr,wbrs,nr,brs,brt,ackokw} and measurement.
In these experiments, though, the electromagnetic field states are
typically superpositions of many different number states, and are
largely unaffected by the interaction, so that (\ref{recoh}) is a poor
model for the process.\footnote{See, for example,
ref. \cite{summhammer} for a review and analysis.}  Still, it seems
hard to take seriously the idea that if a recoherence experiment were
constructed with sufficient care it would jeopardise the
quasiclassicality we observe, and we take the recoherence problem as a
conclusive argument against the general applicability of the
algorithms considered to date.

\section{Retrodictive algorithms}\label{sec:retrodiction}

We have seen that dynamical set selection algorithms which run
forwards in time generally fail to reproduce standard physics.  Can an
algorithm be developed for reconstructing the history of a series of
experiments or, in principle, of the universe?

\subsection{Retrodictive algorithms in the spin model} 

We look first at the spin model with separated interactions and
initial state
\begin{equation} 
|\psi (0) \rangle = |{\bf v}\rangle \otimes |\uparrow_1 \ldots
\uparrow_n \rangle \, ,
\end{equation}
and take the first interaction to run from $t=0$ to $t=1$, the second
second from $t=1$ to $t=2$, and so on.  The final state, in the 
Schr\"odinger picture, is
\begin{equation}
  |\psi(n)\rangle = \sum_{\bf \alpha} \sqrt{p_{\bf \alpha}} |\alpha_n
  {\bf u}_n\rangle \otimes | \beta_1 \ldots \beta_n \rangle \,.
\end{equation}
Here ${\bf \alpha} = \{ \alpha_1 , \ldots , \alpha_{n} \}$ runs over
all strings of $n$ plusses and minuses, we write $ \beta_i = \,
\uparrow$ if $\alpha_i = 1$ and $ \beta_i = \,\downarrow $ if
$\alpha_i = -1$, and
\begin{equation} 
  p_{\bf \alpha} = 2^{-n}(1 + \alpha_n \alpha_{n-1} {\bf u}_n. {\bf
  u}_{n-1}) \ldots (1 + \alpha_1 {\bf u_1. u_0}) \,.
\end{equation} 

Consider now a set selection algorithm which begins the selection
process at $t=n$ and works backwards in time, selecting an exactly
consistent set defined by system space Schmidt projections.  The
algorithm thus begins by selecting projections onto the Schmidt states
$|\pm{\bf u}_n\rangle$ at $t=n$.  The classification theorem implies
that any Schmidt projection during the time interval $[n-1,n)$ defines
a consistent and non-trivial extension to the set defined by these
projections. If the algorithm involves a parametrised non-triviality
condition with sufficiently small non-triviality parameter $\delta$,
the next projection will thus be made as soon as the non-triviality
condition is satisfied, which will be at some time $t=n-\Delta t$,
where $\Delta t$ is small.

If a non-triviality condition is not used but the limit DHC is used
instead, then a second projection will be made at $t=n$, but the
normalised path projected states will be the same (to lowest order in
$\Delta t$) as for projection at $t=n-\Delta t$.  The classification
theorem then implies that the only possible times at which further
extensions can consistently be made are $t= n-1, \ldots, 1$ and, if
$\delta$ is sufficiently small and the measurement axes are
non-degenerate, the Schmidt projections at all of these times will be
selected.

In fact, this algorithm gives very similar results whether a
non-triviality condition or the limit DHC is used.  We use the limit
DHC here for simplicity of notation.  Since the Schmidt states at the
end of the $k^{\mbox{\scriptsize th}}$ interaction are $|\pm{\bf
u}_k\rangle$, the histories of the selected set are indexed by strings
$\{ \alpha_1 , \ldots , \alpha_{n+1} \}$ consisting of $n+1$ plusses
and minuses.  The corresponding class operators are defined in terms
of the Heisenberg picture Schmidt projections as
\begin{equation}
  P_H^{\alpha_{n+1}}(n) P_H^{\alpha_{n}}(n) P_H^{\alpha_{n-1}} (n-1)
\ldots P_H^{\alpha_{1}}(1) \, .
\end{equation}
Define $C_{\bf \alpha} = P_H^{\alpha_{n}}(n) \ldots
P_H^{\alpha_{1}}(1)$.  Then
\begin{equation}
\begin{array}{rcll} 
P_H^{\alpha_{n+1}}(n) C_{\bf \alpha} &=& C_{\bf \alpha} & \mbox{if
$\alpha_{n+1} = \alpha_n$,}\\ P_H^{\alpha_{n+1}} (n) C_{\bf \alpha}
&=& 0 & \mbox{if $\alpha_{n+1} = -\alpha_n$,}
\end{array}
\end{equation}  
and to calculate the limit DHC eq.~(\ref{limitDHC}) we note that
eq.~(\ref{dPidenta}) implies that
\begin{equation}
\begin{array}{rcl} 
\lim_{\epsilon \to 0} \epsilon^{-1}P_H^{-\alpha_{n}}(n)
  P_H^{\alpha_{n}}(n-\epsilon) \ldots P_H^{\alpha_{1}}(1) &=&
  P_H^{-\alpha_{n}}(n) \dot P_H^{\alpha_{n}}(n) \ldots
  P_H^{\alpha_{1}}(1) \\ &=& \dot P_H^{\alpha_n}(n)
  P_H^{\alpha_{n}}(n) \ldots P_H^{\alpha_{1}}(1)\\ &=& \dot
  P_H^{\alpha_n}(n) C_{\bf \alpha} \,.
\end{array}
\end{equation}
The complete set of class operators (up to multiplicative constants)
is $\{C_{\bf \alpha}, \dot P_H^+(n) C_{\bf \alpha}\}$ and the set of
normalised histories is therefore
\begin{equation}
  \{ |\alpha_n{\bf u}_n\rangle \otimes |{\bf \alpha}\rangle,
    |-\alpha_n{\bf u}_n\rangle \otimes |{\bf \alpha}\rangle\} \, .
\end{equation}

Of these histories, the first $2^n$ have probabilities $p_{\bf \alpha}
= 2^{-n}(1 + \alpha_n \alpha_{n-1} {\bf u}_n. {\bf u}_{n-1}) \ldots (1
+ \alpha_1 {\bf u_1. u_0})$ and have a simple physical interpretation,
namely that the particle was in direction $\alpha_i {\bf u} _i$ at
time $t=i$, for each $i$ from $1$ to $n$, while the second $2^n$ have
zero probability.  Thus the repeated projections that the algorithm
selects at $t=n$, while non-standard, merely introduce probability
zero histories, which need no physical interpretation.  The remaining
projections reproduce the standard description so that, in this
example, at least, retrodictive algorithms work.  While this is
somewhat encouraging, the algorithm's success here relies crucially on
the simple form of the classification of consistent sets in the spin
model, which in turn relies on a number of special features of the
model.  In order to understand the behaviour of retrodictive
algorithms in more generality, we look next at two slightly more
complicated versions of the spin model.

\subsection{Spin model with perturbed initial state} 

Consider now the spin model with a perturbed initial state
$|\psi\rangle + \gamma|\phi\rangle$.  For generic choices of $\phi$
and $\gamma$, there is no non-trivial exactly consistent set of
Schmidt projections, but it is easy to check that the set selected in
the previous section remains approximately consistent to order
$\gamma$, in the sense that the DHC and limit DHC parameters are $O(
\gamma )$.

This example nonetheless highlights a difficulty with the type of
retrodictive algorithm considered so far.  Some form of approximate
consistency criterion is clearly required to obtain physically
sensible sets in this example.  However, there is no obvious reason to
expect that there should be any parameter $\epsilon$ with the property
that a retrodictive algorithm which requires approximate consistency
(via the limit DHC and DHC) to order $\epsilon$ will select a
consistent set whose projections are all similar to those of the set
previously selected.  The problem is that, given any choice of
$\epsilon$ which selects the right projections at time $n$, the next
projections selected will be at time $(n-1) + O ( \gamma )$ rather
than at precisely $t = n-1$.  The level of approximate consistency
then required to select projections at times near $n-2$, $n-3$, and so
forth, depends on the projections already selected, and so depends on
$\gamma$ only indirectly and in a rather complicated way.

We expect that, for small $\gamma$ and generic $\phi$, continuous
functions $\epsilon_k(\gamma, \phi )$ exist with the properties that
$\epsilon_k (\gamma , \phi) \rightarrow 0 $ as $\gamma \rightarrow 0$
and that some approximation to the set previously selected will be
selected by a retrodictive algorithm which requires approximate
consistency to order $\epsilon_k(\gamma, \phi )$ for the
$k^{\mbox{\scriptsize th}}$ projection.  Clearly, though, since the
aim of the set selection program is to replace model-dependent
intuition by a precise algorithmic description, it is rather
unsatisfactory to have to fine-tune the algorithm to fit the model in
this way.

\subsection{Delayed choice spin model}  

We now return to considering the spin model with an unperturbed
initial state and look at another shortcoming.  The interaction of the
system particle with each successive environment particle takes the
form of a spin measurement interaction in which the axis of each
measurement, $\{ {\bf u}_i \}$, is fixed in advance.  This is a
sensible assumption when modelling a natural system-environment
coupling, such as a particle propagating past a series of other
particles.  As a model of a series of laboratory experiments, however,
it is unnecessarily restrictive.  We can model experiments with an
element of delayed choice simply by taking the axis $\{ {\bf u}_i \}$
to depend on the outcome of the earlier measurements.

If we do this, while keeping the times of the interactions fixed and
non-overlapping, the measurement outcomes can still be naturally
described in terms of a consistent set built from Schmidt projections
onto the system space at times $t=1,2,\ldots n$, so long as both the
Schmidt projections and the consistent set are defined to be
appropriately branch-dependent.  Thus, let
\begin{equation} 
|\psi (0) \rangle = |{\bf v}\rangle \otimes |\uparrow_1 \ldots
\uparrow_n \rangle \,
\end{equation} 
be the initial state and let $P_H^{\alpha_{1}}(1)$, for $\alpha_1 =
\pm$, be the Schmidt projections onto the system space at time $t=1$.
We define a branch-dependent consistent set in which these projections
define the first branches and consider independently the evolution of
the two states $ P_H^{+} (1) | \psi(0) \rangle $ and $ P_H^{-} (1) |
\psi(0) \rangle $ between $t=1$ and $t=2$.  These evolutions take the
form of measurements about axes ${\bf u}_{2; \alpha_1 }$ which depend
on the result of the first measurement.  At $t=2$ the second
measurements are complete, each branch splits again, and the
subsequent evolutions of the four branches now depend on the results
of the first two measurements.  Similar splittings take place at each
time from $1$ to $n$, so that the axis of the $m^{\mbox{\scriptsize
th}}$ measurement in a given branch, ${\bf u}_{m; \alpha_{m-1},
\ldots, \alpha_1 } $, depends on the outcomes $\alpha_{m-1} , \ldots ,
\alpha_1$ of the previous $(m-1)$ measurements.  Thus, the evolution
operator describing the $m^{\mbox{\scriptsize th}}$ interaction is
\begin{eqnarray*}
\lefteqn{ V_m( t ) = } \\ & \displaystyle \sum_{\alpha_{m-1}, \ldots ,
\alpha_1} & \{ P({\bf u}_{m; \alpha_{m-1}, \ldots, \alpha_1 } )
\otimes P_1 ( \beta_1 ) \otimes \ldots \otimes P_{m-1} ( \beta_{m-1} )
\otimes I_m \otimes \ldots \otimes I_n \, + \\ & & \hspace{-.7ex} P( -
{\bf u}_{m; \alpha_{m-1}, \ldots, \alpha_1} ) \otimes P_1 ( \beta_1)
\otimes \ldots \otimes P_{m-1} ( \beta_{m-1}) \otimes
\mbox{e}^{-i\theta_m (t) F_m } \otimes I_{m+1} \otimes \ldots \otimes
I_n \} \, .
\end{eqnarray*} 
Again we take $\beta_i = \, \uparrow$ if $\alpha_i = +$ and $\beta_i =
\, \downarrow$ if $\alpha_i = - $.  The full evolution operator is
\begin{equation}
  U(t) = V_n (t) \ldots V_1 (t) \, .
\end{equation} 
During the interval $(m-1,m)$ we consider the Schmidt decompositions
on each of the $2^{m-1}$ branches defined by the states
\begin{eqnarray*} 
  && U(t) P_H^{\alpha_{m-1} ; \alpha_{m-2} , \ldots, \alpha_1} (m-1)
  \ldots P_H^{\alpha_1} (1) | \psi(0) \rangle \\ &=& V_m(t) [ P(
  \alpha_{m-1} {\bf u}_{m-1; \alpha_{m-2} , \ldots, \alpha_1}) \ldots
  P( \alpha_1 {\bf u}_1 ) | {\bf v} \rangle] \otimes |\beta_1 \ldots
  \beta_{m-1} \uparrow_m \ldots \uparrow_n\rangle\,
\end{eqnarray*} 
with $\alpha_1 , \ldots , \alpha_{m-1}$ independently running over the
values $\pm$.  Here
\begin{equation} 
P_H^{\alpha_m ; \alpha_{m-1}, \ldots , \alpha_1} (t) = U^{\dagger} (t)
P( \alpha_{m} {\bf u}_{m; \alpha_{m-1} , \ldots, \alpha_1}) \otimes I
U(t) \, ,
\end{equation} 
that is, the Heisenberg picture projection operator onto the
branch-dependent axis of measurement.  The branches, in other words,
are defined by the branch-dependent Schmidt projections at times from
$1$ to $m-1$.

It is not hard, thus, to find a branch-dependent consistent set, 
built from the branch-dependent Schmidt projections at times $1$ 
through to $n$, which describes the delayed-choice spin model
sensibly.\footnote{This sort of branch-dependent Schmidt decomposition
could, of course, be considered in the original spin model, where all
the axes of measurement are predetermined, but would not affect the
earlier analysis, since the Schmidt projections in all branches are
identical.}  However, since the retrodictive algorithms considered so
far rely on the existence of a branch-independent set defined by the
Schmidt decompositions of the original state vector, they will not
generally reproduce this set (or any other interesting set).
Branch-dependent physical descriptions, which are clearly necessary in
quantum cosmology as well as in describing delayed-choice experiments,
appear to rule out the type of retrodictive algorithm we have
considered so far.

\section{Branch-dependent algorithms} \label{sec:branchdep}

The algorithms we have considered so far do not allow for
branch-dependence, and hence cannot possibly select the right set in
many physically interesting examples.  We have also seen that it is
hard to find good Schmidt projection selection algorithms in which the
projections selected at any time depend only on the physics up to that
time, and that the possibility of recoherence rules out the existence
of generally applicable algorithms of this type.

This suggests that \emph{retrodictive} branch-dependent algorithms
should be considered.  Such algorithms, however, seem generally to
require more information than is contained in the evolution of the
quantum state.  In the delayed-choice spin model, for example, it is
hard to see how the Schmidt projections on the various branches,
describing the delayed-choice measurements at late times, could be
selected by an algorithm if only the entire state $\psi(t)$ --- summed
over all the branches --- is specified.

The best, we suspect, that can be hoped for in the case of the
delayed-choice spin model is an algorithm which takes all the final
branches, encoded in the $2^n$ states $| \pm {\bf v} \rangle \otimes |
\beta_1 \ldots \beta_n \rangle$, where each of the $\beta_i$ is one of
the labels $\uparrow$ or $\downarrow$, and attempts to reconstruct the
rest of the branching structure from the dynamics.

One possibility, for example, is to work backwards from $t=n$, and at
each time $t$ search through all subsets $Q$ of branches defined at
that time, checking whether the sum $| \psi^Q (t) \rangle$ of the
corresponding states at time $t$ has a Schmidt decomposition with the
property that the Schmidt projections, applied to $| \psi^Q(t)
\rangle$, produce (up to normalisation) the individual branch states.
If so, the Schmidt projections are taken to belong to the selected
branch-dependent consistent set, the corresponding branches are
unified into a single branch at times $t$ and earlier, and the state
corresponding to that branch at time $t'$ is taken to be $ U(t')
U(t)^{\dagger} | \psi^Q (t) \rangle$, where $U$ is the evolution
operator for the model.  Clearly, though, by specifying the final
branch states we have already provided significant information ---
arguably most of the significant information --- about the physics of
the model.  Finding algorithmic ways of supplying the branching
structure of a natural consistent set, given all of its final history
states, may seem a relatively minor accomplishment.  It would
obviously be rather more useful, though, if the final history states
themselves were specified by a simple rule.  For example, if the
system and environment Hilbert spaces are both of large dimension, the
final Schmidt states would be natural candidates.  It would be
interesting to explore these possibilities in quantum cosmology.

\section{Conclusions}

John Bell, writing in 1975, said of the continuing dispute about
quantum measurement theory that it ``is not between people who
disagree on the results of simple mathematical manipulations.  Nor is
it between people with different ideas about the actual practicality
of measuring arbitrarily complicated observables.  It is between
people who view with different degrees of concern or complacency the
following fact: so long as the wave packet reduction is an essential
component, and so long as we do not know exactly when and how it takes
over from the Schr\"odinger equation, we do not have an exact and
unambiguous formulation of our most fundamental physical
theory.''\cite{bell:hepp}

New formulations of quantum theory have since been developed, and the
Copenhagen interpretation itself no longer dominates the debate quite
as it once did.  The language of wave packet reduction, in particular,
no longer commands anything approaching universal acceptance ---
thanks in large part to Bell's critiques.  But the fundamental dispute
is still, of course, very much alive, and Bell's description of the
dispute still essentially holds true.  Many approaches to quantum
theory rely, at the moment, on well-developed intuition to explain,
case by case, what to calculate in order to obtain a useful
description of the evolution of any given physical system.  The
dispute is not over whether those calculations are correct, or even as
to whether the intuitions used are helpful: generally, both are.  The
key question is whether we should be content with these successes, or
whether we should continue to seek to underpin them by an exact and
unambiguous formulation of quantum theory.

Consensus on this point seems no closer than it was in 1975.  Many
physicists take the view that we should not ever expect to find a
complete and mathematically precise theory of nature, that nature is
simply more complex than any mathematical representation.  If so, some
would argue, present interpretations of quantum theory may well
represent the limit of precision attainable: it may be impossible, in
principle, to improve on imprecise verbal prescriptions and intuition.
On the other hand, this doubt could be raised in connection with any
attempt to tackle any unsolved problem in physics.  Why, for example,
should we seek a unified field theory, or a theory of turbulence, if
we decide a priori not to look for a mathematically precise
interpretation of quantum theory?  Clearly, too, accepting the
impossibility of finding a complete theory of nature need not imply
accepting that any definite boundary to precision will ever be
encountered.  One could imagine, for example, that every technical and
conceptual problem encountered can eventually be resolved, but that
the supply of problems will turn out to be infinite.  And many
physicists, of course, hope or believe that a complete and compelling
theory of nature will ultimately be found, and so would simply reject
the initial premise.

Complete agreement on the desiderata for formulations of quantum
theory thus seems unlikely.  But it ought to be possible to agree
whether any given approach to quantum theory actually does supply an
exact formulation and, if not, what the obstacles might be.  Our aim
in this paper has been to help bring about such agreement, by
characterising what might constitute a precise formulation of some of
the ideas in the decoherence and consistent histories literature, and
by explaining how hard it turns out to be to supply such a
formulation.

Specifically, we have investigated various algorithms that select one
particular consistent set of histories from among those defined by the
Schmidt decompositions of the state, relative to a fixed
system-environment split.  We give examples of partial successes.
There are several relatively simple algorithms which give physically
sensible answers in particular models, and which we believe might
usefully be applied elsewhere.  We have not, though, found any
algorithm which is guaranteed to select a sensible consistent set when
both recoherence and branch-dependent system-environment interactions
are present.

Our choice of physical models is certainly open to criticism.  The
spin model, for example, is a crudely simplistic model of real world
decoherence processes, which supposes both that perfect correlations
are established between system and environment particles in finite
time and that these interactions do not overlap.  We would not claim,
either, that the delayed-choice spin model necessarily captures any of
the essential features of the branching structure of quasiclassical
domains, though we would be very interested to know whether it might.
We suspect that these simplifications should make it easier rather
than harder to find set selection algorithms in the models, but we
cannot exclude the possibility that more complicated and realistic
models might prove more amenable to set selection.

The type of mathematical formulation we have sought is, similarly,
open to criticism.  We have investigated what seem particularly
interesting classes of Schmidt projection set selection algorithms,
but there are certainly others which may be worth exploring.  There
are also, of course, other mathematical structures relevant to
decoherence apart from the Schmidt decomposition, and other ways of
representing historical series of quantum events than through
consistent sets of histories.

Our conclusion, though, is that it is extraordinarily hard to find a
precise formulation of non-relativistic quantum theory, based on the
notions of quasiclassicality or decoherence, that is able to provide a
probabilistic description of series of events at different points in
time sufficiently rich to allow our experience of real world physics
to be reconstructed.  The problems of recoherence and of
branch-dependent system-environment interaction, in particular, seem
sufficiently serious that we doubt that the ideas presented in the
literature to date are adequate to provide such a formulation.
However, we cannot claim to have exhaustively investigated every
possibility, and we would like to encourage sceptical readers to
improve on our attempts.

It is possible that this model has too many symmetries and the form of
the interaction is too limited for an algorithm to work. In the next
chapter we consider a very general model, but we discover that
although some of the difficulties encountered in this chapter
disappear other problems arise and the conclusions of the previous
paragraph still hold.


\chapter{A random Hamiltonian model}\label{chap:random}
\section{Introduction}
Consider a simple quantum system consisting of a finite Hilbert space
$\mathcal{H} = \mathcal{H}_1 \otimes \mathcal{H}_2$
($\mbox{dim}\mathcal{H}_i = d_i$), a pure initial state
$|\psi(0)\rangle$ and a Hamiltonian drawn from the GUE (Gaussian
Unitary Ensemble), which is defined by
\begin{equation}
  \label{GUEa}
  P(H) = N^{-1} \exp\{-\mbox{Tr}[(\lambda H + \mu)^2]\},
\end{equation}
where $N$ is a normalisation constant.

The GUE is the unique ensemble of Hermitian matrices invariant under
$U(d)$ with independently distributed matrix elements, where $d = \mbox{dim}
{\cal H} =d_1d_2$.  The GUE is also the unique ensemble with maximum entropy,
$-\int dH\, P(H) \log P(H)$, subject to $E[\mbox{Tr}(H)] = \mu$ and
$E[\mbox{Tr}(H^2)] = \lambda$.  The book by Mehta~\cite{Mehta}
contains a short proof of this as well as further analysis of the GUE
and related ensembles. All the results concerning the GUE in this
thesis can be found in this book or in appendix~(\ref{app:gue}).

This model is not meant to represent any particular physical system,
though Hamiltonians of this from are used in models of nuclear
structure and have often been studied in their own right
(see~\cite{Mehta,Simons:Altshuler} and references therein), and a
large class of other ensembles approximate the GUE in the large $d$
limit.

Because $H$ is drawn from a distribution invariant under $U(d)$ there
is no preferred basis, no distinction between system and environment
degrees of freedom and no time asymmetry. In other words the model is
chosen so that there is no obvious consistent set: we do not already
know what the answer should be.  Moreover it does not single out a
pointer basis that one might associate with classical states, so that
the Copenhagen interpretation cannot make any predictions about a
model like this in the $t \to \infty$ limit.  If an algorithm works
for this model, when there are no special symmetries, it should work
for a wide variety of models.  The question whether a pointer basis
can arise dynamically using Schmidt states was addressed by Albrecht
in~\cite{Albrecht:decoherence,Albrecht:collapsing}, but no general
prescription emerged from his study. Albrecht also studied the
relationship between Schmidt states and consistent histories, and his
studies suggested that the relationship was complicated.  

The model
considered here generalises Albrecht's model: the Hamiltonian for the
entire Hilbert space is chosen from a random ensemble. Albrecht also
used a different distribution, but as we explain below the GUE seems
more natural, though this it probably makes little difference.
 
Without loss of generality, we take $\mu=0$ and
$\lambda=1/2$. With this choice and using the Hermiticity property of
$H$ eq.~(\ref{GUEa}) becomes
\begin{equation}
  \label{GUEb}
   P(H) = \pi^{-n^2/2} 2^{-n/2} N^{-1} \prod_{i<j} e^{-|H_{ij}|^2}
   \prod_{i} e^{-|H_{ii}|^2/2}.
\end{equation}
Therefore the diagonal elements are independently distributed, \emph{real}
normal random variables with mean $0$ and variance $1$ and the
off-diagonal elements are independently distributed, \emph{complex}
normal random variables with mean $0$ and variance $1$.

Since the Hamiltonian is invariant under $U(d)$ the only significant
degrees of freedom in the choice of initial state are the initial
Schmidt eigenvalues (the eigenvalues of the initial reduced density
matrix.)  The usual choice in an experimental situation is an initial
state of the form $|\psi\rangle = |u\rangle_1 \otimes |v\rangle_2$
which corresponds to a pure initial density matrix. A more general
choice in the spirit of the model is to draw the initial state from
the $U(d)$ invariant distribution subject to fixed rank $n$. This is
equivalent to choosing the first $n$ eigenvalues to be components of a
random unit vector in $R^n$ and the remaining $d_1-n$ components to be
zero.

\section{Analysis}\label{sec:rand:analysis}
The calculations in this section are an attempt to gain insight into
the expected properties of prediction algorithms applied to the random
model. These calculations rest on a large number of assumptions and
are at best approximations, but the conclusions are borne out by
numerical simulations and the calculations do provide a rough feel for
the results that different algorithms can be expected to produce.  In
particular they suggest that there are only narrow ranges of values
for the approximate consistency parameter which are likely to produce
physically plausible sets of histories.  These calculations may also
be applicable to other models since this model makes so few
assumptions and the interaction is completely general.

In a random model there is no reason to expect exactly consistent sets
of histories formed from Schmidt projections to exist, so only
parameterised approximate consistency 
criteria~(\ref{badcriterion},~\ref{wic:DHC},~\ref{DHC}) are considered and
throughout this chapter $\epsilon$ will always be the consistency
parameter in these equations.  We shall only discuss medium consistency
criteria: the results for weak consistency are qualitatively the same.

If an absolute approximate consistency criterion is being used there
are strong theoretical reasons for imposing a parameterised
non-triviality criterion (see section~\ref{sec:approxcon}). However,
if the approximate DHC is being used one is not needed, though it is
convenient to introduce one for computational reasons. The
non-triviality parameter (which we shall always write as $\delta$ in
this chapter) can be taken very small if the DHC is used and is not
expected to influence the results --- except possibly for the first
projections --- and the numerical simulations show that this is indeed
the case.  We shall refer to histories with probability less than or
equal to $\delta$ (relative or absolute) as \emph{trivial} histories
and a projection that gives rise to a \emph{trivial} history as a
\emph{trivial} projection. There are no absolute reasons for rejecting
set of histories containing trivial histories --- if $\delta$ is
sufficiently small and there are not too many they are physically
irrelevant --- though obviously sets are preferable if all the
histories are non-trivial. However, an algorithm must produce results
that are approximately the same for a range of parameter values if it
is to make useful predictions, and trivial histories will almost
certainly vary as $\delta$ is changed.  If the DHC is used,
generically all the later projections will also change, since trivial
histories can significantly influence the consistency of later
projections.  If an absolute consistency criterion is used trivial
projections are more likely to be consistent than non-trivial
projections so for many values of the parameters only trivial
projections are made.

\subsection{Repeated projections and relative consistency}
Consider a history $\alpha$ extended by the projective
decomposition $\{P, \overline P\}$ and the further extension of
history $P|\alpha\rangle$ by $\{P(t), \overline P(t)\}$. This was
discussed for the DHC in section~(\ref{sec:repeated}) and the DHP for
this case was shown to be (\ref{dblprojDHCterm})
\begin{equation}\label{repeatedDHCa}
  \frac{|\langle\alpha| \overline{P} \dot{P} |\alpha\rangle| }{\|
    \overline{P} |\alpha\rangle \| \, \| \overline{P} \dot{P}
    |\alpha\rangle \|}.
\end{equation}
The reprojection will occur unless $\epsilon$, the approximate
consistency parameter, is smaller than (\ref{repeatedDHCa}).  From
eq.~({\ref{schmidtPHevol}}), the
time evolution of Heisenberg picture Schmidt projections is 
\begin{equation}\label{anl:pevol}
  \dot P = i[H-B \otimes I,P],
\end{equation}
where $H$ is the Hamiltonian,
\begin{equation}
  B = \sum_{k \neq m} \frac{Q_k \dot \rho_r Q_m}{p_m-p_k} \, ,
\end{equation}
$Q_k$ are projection operators (in $H_1$) on to the Schmidt
eigenspaces, $p_k$ their respective (distinct) eigenvalues and
$\dot \rho_r$ the derivative of the reduced density matrix. 

In analysing (\ref{repeatedDHCa}) and similar expressions we
make the following assumptions. First that $|\alpha\rangle$ is
uncorrelated with the Schmidt states --- this generally is a good
approximation when there are a large number of histories.  Second
that $B \otimes I$ is an operator drawn from the GUE with unspecified
variance independent of the other variables --- in some situations
this assumption is exact but it generically is not.

Let $G = H-B \otimes I$, an element of the GUE with variance $\sigma$,
then using eq.~(\ref{anl:pevol}) (\ref{repeatedDHCa}) is
\begin{equation}\label{repeatedDHCb}
  \frac{|\langle\alpha| \overline P G P |\alpha\rangle| }{\|
    \overline P |\alpha\rangle \| \, \| \overline P G P |\alpha\rangle
    \|}.
\end{equation}
Because $G$ is drawn from a distribution invariant under $U(d)$ and is
independent of $P|\alpha\rangle$ and $\overline
P|\alpha\rangle$, (\ref{repeatedDHCb}) can be simplified by
choosing a basis in which $P|\alpha\rangle/\|P|\alpha\rangle\| =
(1,\ldots,0)$ and $\overline P|\alpha\rangle/\|\overline
P|\alpha\rangle\| = (0,1,\ldots,0)$. (\ref{repeatedDHCb}) becomes
\begin{equation}\label{repeatedDHCc}
  \frac{|Z_{1}|}{[\sum_{r \geq k \geq 1} |Z_k|^2]^{1/2}}\,,
\end{equation}
where $r =\mbox{rank}(\overline P)$ and $Z_k =G_{1(k+1)}$. Since
$\{G_{ij},i < j\}$ is a set of independent, complex, normal random
variables, (\ref{repeatedDHCc}) is the square root of a $B(1,r-1)$
random variable\footnote{$B(p,q)$ :- a beta random variable with
  parameters $p$ and $q$. This has a density function $\propto
  t^{p-1}(1-t)^{q-1}$. A B(1,r-1) random variable has the same
  distribution as that of the inner-product squared between two
  independent unit vectors in $C^r$.}.  

Suppose we choose $\epsilon$ so that reprojections will occur with
some small probability $q$ --- note that only choosing $\epsilon = 0$
will definitely prevent all repeated projections.  The probability of
(\ref{repeatedDHCc}) being less than $\epsilon$ is
\begin{equation}
  1 - (1-\epsilon^2)^{r-1}.
\end{equation}
Therefore if
\begin{eqnarray}\nonumber
  \epsilon &\approx& [1-(1-q)^{1/(r-1)}]^{1/2}
  = \left[\frac{-\log(1-q)}{r}\right]^{1/2}
  + O \left[\frac{\log(1-q)}{r}\right]^{3/2} 
  \\ &&\leq d^{-1/2} \sqrt{-\log(1-q)}\,,\label{repeatedDHCd} 
\end{eqnarray}
a reprojection will occur with probability $\approx q$.

However, we show in subsection~(\ref{sub:approxcon}) that the DHC
cannot prevent trivial reprojections on the initial state if the
initial density matrix has less than full rank. If the initial density
matrix has rank one then the first projection will always be made with
probability $\delta$. A non-triviality criterion can then work in
conjunction with the DHC to prevent further trivial extensions.
Suppose either that the initial density matrix has rank greater than
$1$ and $P_n$ and $P_m$ are two projections onto the non-zero
eigenspaces, or assume that the rank is one and $P_n$ is a projection
onto the initial state and $P_m$ is a projection making a history of
probability $\delta$. In either case, let $P_k$ be projection onto the
null space.  To prevent the trivial projection $P_k$ being made the
parameters $\delta$ and $\epsilon$ must be chosen to satisfy
eq.~(\ref{edinequality})
\begin{equation}\label{edinequalityr}
  \sqrt{\delta} |\langle \psi | P_{m} \dot P_{k}^2 P_{n} | \psi
  \rangle| > \epsilon \| P_{m} | \psi \rangle \|\, \| \dot P_{k} P_{n}
  | \psi \rangle \|^2.
\end{equation}
Though the probability distribution for this is complicated, the
approximate relation between $\delta$ and $\epsilon$ can be estimated
by squaring both sides of eq.~(\ref{edinequalityr}) and then taking
the expectation. Note that treating $G$ as an element of the GUE is
exact in this case as the terms involving $B \otimes I$ are
identically zero. Using results from eq.~(\ref{expectgue}),
eq.~(\ref{edinequalityr}) becomes
\begin{equation}\label{preventinitial:rel}
  \delta > \epsilon^2 (r+1) \|P_n|\psi\rangle\|^2\,,
\end{equation}
where $r$ is the rank of $P_k$. By assumption $\|P_n|\psi\rangle\|$ is
order one and $r<d$, so if $\delta > d \epsilon^2$ initial
reprojections will not occur. The results are the same for a relative
non-triviality criterion since instead of
eq.~(\ref{preventinitial:rel}) we have $ \delta > \epsilon^2 (r+1)$.

\subsection{Repeated projections and absolute consistency}
An algorithm using an absolute parameterised consistency criterion
will make nothing but trivial projections unless a parameterised
non-triviality criterion is also used, so only algorithms with a
non-triviality criterion are considered.

Let $t_\epsilon$ denote the latest time that the reprojection is
approximately consistent and $t_\delta$ the earliest time at which the
extension is absolutely nontrivial. We see from
section~(\ref{sec:repeated}) and eq.~(\ref{initialDHC}) that, to
lowest order in $t$,
\begin{eqnarray}
  t_{\delta} &=& \sqrt{\delta} \|\dot P P | \alpha \rangle \|^{-1} \\
  t_{\epsilon} &=& \epsilon |\langle \alpha | \overline P \dot P P |
  \alpha \rangle|^{-1}\,.
\end{eqnarray}
$t_{\delta} > t_{\epsilon}$ implies
\begin{equation}\label{absoluteta}
  \sqrt{\delta} |\langle \alpha | \overline P \dot P P | \alpha
  \rangle| > \epsilon \| \dot P P |\alpha\rangle \|.
\end{equation}
Again we choose $\epsilon$ so that reprojections occur with
probability q and assume that $\| P |\alpha\rangle \|$ and $\|
\overline P |\alpha\rangle \|$ are order one, so that
eq.~(\ref{absoluteta}) can be written
\begin{equation}\label{absolutetb}
  \frac{|Z_{1}|}{[\sum_{r \geq k \geq 1} |Z_k|^2]^{1/2}} >
  \epsilon/\sqrt{\delta}.
\end{equation}
The l.h.s.\ is the same random variable as in eq.~(\ref{repeatedDHCc})
so $\delta$ and $\epsilon$ must be chosen so that 
\begin{equation}\label{prevrepabsa}
  \epsilon/\sqrt \delta = d^{-1/2} \sqrt{-\log(1-q)}\,.
\end{equation}
The assumption that $\| P |\alpha\rangle \|$ and $\| \overline P
|\alpha\rangle \|$ are order one will obviously not always be valid.
As more projections are made the probabilities of the histories will
decrease. When both probabilities  are
$\delta$ eq.~(\ref{absoluteta}) is
\begin{equation}\label{absolutetc}
  \frac{|Z_{1}|}{[\sum_{r \geq k \geq 1} |Z_k|^2]^{1/2}} >
  \epsilon/\delta,
\end{equation}
so $\epsilon/\delta = d^{-1/2} \sqrt{-\log(1-q)}$. If reprojections of
smaller probability histories are to be prevented this choice of
parameters is clearly more appropriate than eq.~(\ref{prevrepabsa}).

This analysis has picked a very conservative upper bound for
$\epsilon$ to prevent repeated projections, since decoherence matrix
terms with the other histories will tend to reduce the likelihood of
repeated projections, and thus allow larger values of $\epsilon$ to be
used. A more detailed analysis suggests that for relative and absolute
consistency $r$ can be treated as much smaller than $d$ so that
choosing $\delta$ a small factor larger than $\epsilon^2$ or
$\epsilon$ respectively, are sufficient conditions.

\subsection{Projections in the long time limit}\label{subsec:longtime}
The previous subsection has shown how $\epsilon$ and $\delta$ affect
the probability of repeated projections: this subsection calculates
how they affect the probability of projections as $t \to \infty$. In
infinite dimensional systems, off-diagonal terms of the decoherence
matrix for quasiclassical projections often tend to zero as $t$
increases~\cite{Zurek:superselection}. In the limit $d \to \infty$ one
would also expect this for Schmidt projections in this model ---
though the limit only exists for initial density matrices of finite
rank. 

Consider the DHP for a Schmidt projection extending history $\alpha$
from the set of normalised exactly consistent histories
$\{|\alpha\rangle,|\beta_i\rangle, i = 1,\ldots, k\}$ as $t\to\infty$.
For $t\to \infty$ and for large $k$ the Schmidt states are
approximately uncorrelated with the histories. The DHP for an
extension $\{P, \overline P\}$ of history $\alpha$ is
\begin{equation}\label{rellongtime}
 \max \left \{
 \frac{|\langle\beta_i|P|\alpha\rangle|}{\|P|\alpha\rangle\|},
 \frac{|\langle\beta_i|\overline P|\alpha\rangle|}{%
 \|\overline P|\alpha\rangle\|}, i = 1,\ldots,k \right \}. 
\end{equation}
Since $\langle\beta_i|\alpha\rangle = 0$ for all $i$,
eq.~(\ref{rellongtime}) is equal to
(within a factor of $\sqrt 2$) 
\begin{equation}\label{rellongtimeb}
  \max_{i = 1,\ldots,k} 
    \frac{|\langle\beta_i|P|\alpha\rangle|}{%
      \|P|\alpha\rangle\|\|\overline P|\alpha\rangle\|}\,. 
\end{equation}
The  cumulative frequency distribution for (\ref{rellongtimeb})
squared is calculated in appendix~(\ref{sec:probdists}) as
\begin{equation}
  P \left( \max_{i = 1,\ldots,k} 
    \frac{|\langle\beta_i|P|\alpha\rangle|^2}{%
      \|P|\alpha\rangle\|^2 \|\overline P|\alpha\rangle\|^2} 
    < \lambda \right ) =  
  \sum_m (-1)^m {k \choose m} (1-m\lambda)^{d-1} 1_{m\lambda<1},
\end{equation}
which approximately equals $[1-e^{-d \lambda}]^k$ when $dk^2\lambda^2
= o(1)$. This is the probability that a pair of projections acting on
one history in a set of $k+1$ consistent histories satisfies the
medium DHC with parameter $\epsilon = \sqrt\lambda$.  There are $n_p =
2^{\min(d_1-1,d_2)}-1$ distinct choices for the projections so the DHP
(to within a factor of $\sqrt2$) for extending $\alpha$ with any
Schmidt projection is
\begin{equation}\label{rellongtimec}
  \min_{j = 1,\ldots,n_p} \max_{i = 1,\ldots,k} 
    \frac{|\langle\beta_i|P_j|\alpha\rangle|}{%
      \|P_j|\alpha\rangle\|\|\overline P_j|\alpha\rangle\|}\,, 
\end{equation}
where $\{P_j, \overline P_j\}$ range over all $n_p$ binary partitions
of the basis Schmidt projections.  The distribution for this random variable
is hard to calculate but if we assume that the DHP's for each $\{P_j,
\overline P_j\}$ are independent the cumulative distribution function
is
\begin{equation}\label{rellongtimed}
  1-\{1-[1-e^{-d \lambda + o(1)}]^k\}^{n_p}\,.
\end{equation}
This assumption is obviously very approximate since the different
projections are all formed using the same basis. However, treating the
$\{P_j, \overline P_j\}$ as independent in (\ref{rellongtimec}) will be
a lower bound for the exact result and (\ref{rellongtimeb}) will be an
upper bound for the exact result.

Suppose now we wish to choose $\epsilon$ so that the probability of
making a projection at a large time is $p$, where $p$ is close to one. Then
from eq.~(\ref{rellongtimed})
\begin{equation}
  \epsilon^2 = -1/d\log\{1- [1-(1-p)^{1/n_p}]^{1/k}\}\,.
\end{equation}
For large $k$
\begin{equation}
 \epsilon^2 \approx -1/d\log\{-1/k\log[1-(1-p)^{1/n_p}]\} \approx
 1/d\log(k)\,.
\end{equation}
This calculation has involved a lot of assumptions and approximations,
but it should accurately reflect the behaviour for $d \gg k^2 \gg 1$.
The logarithmic dependence on $k$ is a generic feature of extreme
order statistics and so is the independence of the answer from the other
factors $p$ and $n_p$. The $1/\sqrt d$ dependence is also expected because
the mean value of the square of the inner product between two random
vectors in a $d$ dimensional space is $1/d$. 

A generic feature of asymptotic extreme order statistics such as the
previous calculation, is the slow rate of convergence and this
calculation is only expected to be accurate for very large $d$ and
$k$. For actual application to particular models the exact
distribution can be calculated using Monte-Carlo methods. 

The two programs (\ref{calclongtime.m}~and~\ref{mcpercentileplot.m})
calculate the distribution and plot $\epsilon$ as a function of $p$
and $k$, for the weak or medium case.  Program~(\ref{calclongtime.m})
generates samples from the exact distribution for the DHP\@. For each
sample it randomly picks a set of exactly consistent histories, and
then calculates the DHP for all $n_p$ combinations of projections on
one particular history. Ten thousand  samples were sufficient to produce
smooth cumulative frequency distributions.  These can be inverted to
calculate $\epsilon(p,k)$ as in fig.~(\ref{fig:percentile}), where
for example $\epsilon(.99,k)$ is the solid curve.

The same arguments apply for absolute consistency as $t \to \infty$,
but since the consistency requirement is not normalised the expected
DHP values will be reduced by a factor of $1/\sqrt k$ --- the
average value for the length of a history when there are $k$
histories.

These calculations suggest that choosing $\epsilon = O(1/\sqrt d)$ is
most likely to produce histories with a complicated branching
structure and many non-trivial projections. In the next section we
discuss the results of simulations for all values of the parameters
and show that they agree with these theoretical calculations.

\section{Computer simulations}
\begin{wrapfigure}{I}{2.25in}%
  \epsfig{file=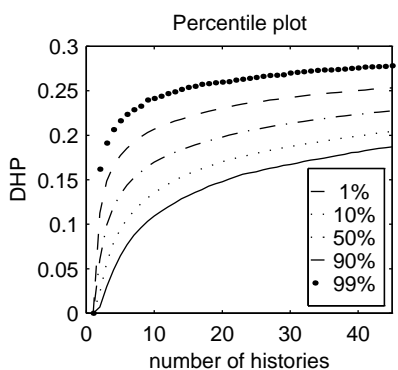}%
  \caption{estimating $\epsilon$}\label{fig:percentile}%
\end{wrapfigure}
The computer programs are explained and listed in
appendix~(\ref{app:programs}). The results described here were carried
out with a system of dimension $3$, with an environment of dimension
$15$ and with either medium absolute consistency or medium relative
consistency (DHC). Fig.~(\ref{fig:percentile}) gives the probability
distribution for the DHP plotted via percentile curves as a function
of the number of histories in the long time limit.  For example, this
graph shows that with $\epsilon \geq 0.3 \approx 2/\sqrt d$
projections will almost certainly be consistent for any number of
histories, whereas for $\epsilon \leq 0.05 \approx.3/\sqrt d$
projections will probably only be consistent when there are two or
three histories. When there are twenty histories it shows that for
$98\%$ of the time the DHP will be between $0.15$ and $0.25$.
Fluctuations will only occasionally ($1\%$ of the time) lie below the
solid line and if $\epsilon$ has this value (for a particular number
of histories) then although projections will probably occur they will
occur as a result of large fluctuations from the mean. Therefore one
would expect the projections to occur at widely separated times and if
$\epsilon$ is changed only slightly the times generically to change
completely, and indeed computer simulations show this.

The simulations described here were run for ten thousand program steps
or until thirty histories had been generated. For a given set of
parameters, many simulations with different Hamiltonians and initial
states were carried out and were found generically to produce
qualitatively the same results, though only individual
simulations are described here.

\pairoffigures[tree]{ds}{example}{example} One way to look at the
results of a simulation is to look at the \emph{probability tree}
associated with the set of histories such as
fig.~(\ref{fig:example}a). The root node on the far left represent the
initial state, the terminal nodes represent the histories and the
other nodes represent intermediate path-projected states. Each node
has a probability and the lines linking the nodes have an associated
projection operator and projection time. The projections associated
with lines emanating to the right from the same node form a projective
decomposition and all occur at the same time. The scaling of the axis
and relative positions between the nodes is arbitrary, only the
topology is relevant. For example, in fig.~(\ref{fig:example}a) the
probabilities for the histories are $0.42$, $0.25$, $0.05$, $0.02$,
$0.02$, $0.01$, $0.08$ and $0.14$ --- the probabilities of the
terminal nodes.

Another useful interpretative aid is a graph of the \emph{consistency
  statistics} fig.~(\ref{fig:example}b). This graph shows the DHP for
the most consistent non-trivial extension. At times where no Schmidt
projections result in non-trivial histories no points are plotted,
though there are no such times in fig.~(\ref{fig:example}b).  The
program makes a projection when this value is $\epsilon$. The flat
line indicates $\epsilon$ and the crosses indicate when projections
have occurred --- in this case at times $0$, $1$, $11$, $12$, $36$ and
$65$ (approximately). A graph of the projection times will also be used
sometimes, for example fig.~(\ref{fig:releg2}b).

When any Schmidt eigenvalues are equal their eigenspaces becomes
degenerate and the corresponding Schmidt projections are not uniquely
defined. The reduced density matrix varies continuously in this model
and it will only be degenerate for a set of times of
measure zero so generically it is possible to define the Schmidt
states so that they are continuous functions of $t$ for all $t$.
This was not found to be necessary in the simulations.

\subsection{Results for relative consistency}
\subsubsection{Rank one initial density matrix} 
\pairoffigures{ds}{releg1}{relative consistency, $\epsilon =
  0.03 \approx 0.2/\sqrt{d}$} Fig.~(\ref{fig:releg1}) shows the
probability tree and minimum consistency statistics for a simulation
with $\epsilon=0.03 \approx 0.2/\sqrt d$ and a rank one initial
density matrix. As expected there is one almost immediate trivial
projection. Three more projections occur and then no more. From
fig.~(\ref{fig:percentile}) the probability of a projection with five
histories and $\epsilon = 0.03$ is less than 1\% so this result is as
expected. The simulation was run for longer than is shown in the
figure (until $t=100$) but no further projections occurred.

\pairoffigures{pjt}{releg2}{relative consistency, $\epsilon = 0.15
  \approx 1/\sqrt{d}$} \pairoffigures{pjt}{releg3}{relative
  consistency, $\epsilon = 0.16 \approx 1/\sqrt{d}$}
Fig.~(\ref{fig:releg2}) shows the probability tree and projection
times for a simulation with $\epsilon=0.15 \approx 1/\sqrt d$. Again
there is the initial trivial projection but no other trivial
projections occur. Projections then occurred at roughly equal equal
time intervals until there were fifteen histories.  The time between
projections then rapidly increased. This is in accord with
fig.~(\ref{fig:percentile}) as the probability for a projection with
fifteen histories and $\epsilon=0.15$ is around $5\%$. Projections after
this time only occur for large deviation away from the mean and
therefore occur extremely erratically. These later projections are
extremely unlikely to vary smoothly for a range of $\epsilon$. The
simulation was run until $t\approx100$ and no further projections
occurred. This simulation has produced an interesting set of histories
with a complicated branching structure.

\begin{wrapfigure}{i}{2.25in}
  \epsfig{file=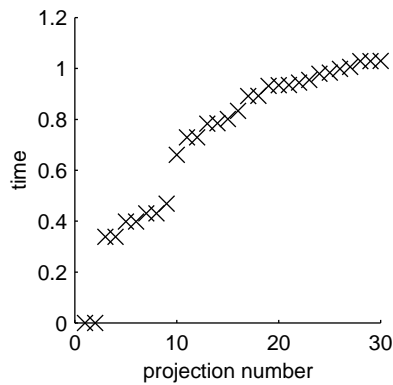}%
  \caption{relative consistency, $\epsilon=0.3\approx 2/\sqrt d$}
  \label{fig:releg4}
\end{wrapfigure}
The next pair of figures fig.~(\ref{fig:releg3}) shows the results of
a simulation with all the parameters unchanged except for $\epsilon$ which is
now $0.16$. The qualitative description is the same and the
first eight or so projections are similar. After that however the two sets of
histories are very different. This is the problem with the algorithm
applied to this model: interesting sets of histories are produced, but they
change dramatically for small changes in $\epsilon$.

From fig.~(\ref{fig:percentile}) choosing $\epsilon=0.3\approx 2\sqrt
d$ looks large enough so that projections will always be made before
the background level is reached. The theoretical analysis also
suggests that for such a large value of $\epsilon$ some repeated
projections will occur. Indeed fig.~(\ref{fig:releg4}) demonstrates
that nine repeated projections occurred each giving a history of
probability $\delta$. 

\onefigure[ds]{releg5}{relative consistency, $\epsilon$ chosen at
  $50\%$} An interesting alternative is to choose $\epsilon$ as a
percentile from fig.~(\ref{fig:percentile}), that is $\epsilon(k)
\approx \epsilon(e) \log k$ where $k$ is the number of histories.
Fig.~(\ref{fig:releg5}) demonstrates the consistency statistics for a
run with $\epsilon(k)$ chosen at the $50\%$ level. All of the
probabilities except for the initial projection were non-trivial.
Rather than the projections being made in regimes where the DHP
fluctuates about its mean value most of the projections have been made
at times when the DHP is monotonically decreasing, so that the
histories are much more likely to vary continuously with $\epsilon$.
Two other advantages of choosing $\epsilon$ this way are that larger
sets of histories are produced, and if an algorithm is designed to
produce a set of histories of a certain size choosing $\epsilon$ in
this way will produce a more consistent set than choosing $\epsilon$
to be constant.  However, though the results are more stable (when the
percentile is changed) than for constant $\epsilon$, results from
simulations show that they still change too much to single out a
definite set of histories.

By looking at the consistency statistics the problem is easy to
understand. Since a projection is made at the earliest possible time
generically once it has been made the DHP jumps up as the most
consistent projection has occurred. The consistency level then falls.
While it is decreasing monotonically any change in $\epsilon$ will
produce a continuous change in the time of the next projection.
However, if $\epsilon$ is too far below its mean level the projection
times will vary discontinuously, and all the projections afterwards
will generically be completely different. Since the mean level of the
DHP depends on the number of histories strongly either $\epsilon$ must
be chosen sufficiently large so as to be above this or it must be
chosen so as to increase with the number of histories. This is
demonstrated by the first few projections as shown in
fig.~(\ref{fig:releg7}) --- a close-up  of fig.~(\ref{fig:releg5}b)
would also show this.

\subsubsection{Other initial conditions}
\pairoffigures{ds}{releg7}{relative consistency, $\delta = 0.02$,
  $\epsilon = 0.05$} 
Simulations with a rank 2 initial state, and all
other parameters remaining the same, produce the same results except
that each of the initial projections is repeated producing two trivial
histories with probability $\delta$.  We can choose $\delta$ according
to eq.~(\ref{preventinitial:rel}) to try to prevent these projections,
that is choose $\delta = O(\epsilon^2)$. Since many of the histories
we expect to generate will have probabilities smaller than this it is
sensible to use a relative non-triviality criterion\footnote{Using a
  relative non-triviality criterion earlier does not qualitatively
  change the results except that the initial trivial history would
  also have been extended --- the results would have been
  qualitatively the same as for rank two initial reduced density
  matrices.}.  The analysis leading to eq.~(\ref{preventinitial:rel})
is only accurate to first order in $t$, therefor
eq.~(\ref{preventinitial:rel}) is only valid when $\delta$ is
sufficiently small. If $\delta$ is too large the consistency level of
a reprojection will start decreasing and when a reprojection
eventually becomes non-trivial it will be consistent. For the example
discussed there were no values of $\delta$ with $\epsilon=0.15$ that
prevent an initial trivial reprojection. Fig.~(\ref{fig:releg7})
demonstrates the start of a simulation with $\epsilon=0.05$ and
$\delta = 0.02 = 8\epsilon^2$. The graph of the consistency statistics
shows the initial projection at $t=0$ and then that there are no
non-trivial extensions until $t\approx 0.08$ by which time the
projection is not consistent. A non-trivial projection is made at
$t\approx 0.32$ and the algorithm then continues as before, with the
trivial projection avoided. Because the projection has not occurred
with probability $\delta$ there is a range of values for $\delta$ that
do not affect the resulting histories --- they are independent of
$\delta$. The first three projection times will obviously vary
continuously for a small range of $\epsilon$. The other projections
that occurred in this simulation all occurred at much more separated
times in a regime where the consistency level was not decreasing
monotonically.  If $\epsilon$ is chosen according to the percentile
distribution in fig.~(\ref{fig:percentile}) --- that is $\epsilon(k)
\approx \epsilon(e) \log k$ where $k$ is the number of histories ---
$\epsilon$ will be small enough initially to prevent any trivial
projections (with an initial density matrix of rank greater than one)
and will allow a full set of histories to be built up at later times.

Simulations with a full rank initial density matrix and
$\epsilon=0.15$ result in a trivial repeated projection for each
initial history. If $\epsilon$ is smaller ($\leq 0.07$) no trivial
reprojections occur. If two initial projections are made uncorrelated
with the Schmidt states then further (Schmidt) projections at $t=0$
will not generically be trivial or consistent. In both
cases, after the initial projections and possible reprojections the
qualitative behaviour is the same as the rank one case. 

\subsection{Results for absolute consistency}
To produce interesting sets of histories from an absolute consistency
criterion three effects need to be balanced against each other.  If
$\epsilon$ is too small the most likely projections will be those that
produce very small probability histories. If $\delta/\epsilon$ is too
small the likelihood of repeated projections (hence trivial histories)
will be high. If $\delta$ is too large the non-triviality criterion
will dominate the
algorithm and only probability $\delta$ (trivial) histories will be
produced. Only an absolute parameterised non-triviality criterion is
considered since a relative criterion will clearly produce almost
nothing except infinitesimal histories. The following results
demonstrate these effects. 

\subsubsection{Rank one initial density matrix}
\begin{wrapfigure}{I}{3.75in}%
  \epsfig{file=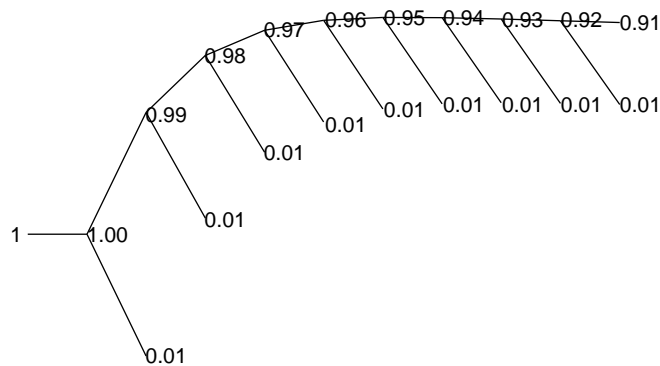}%
  \caption{absolute consistency, $\delta=0.01$, $\epsilon=0.1$}
  \label{fig:abseg1}
\end{wrapfigure}
For example if $\delta < \epsilon$ and the initial reduced density
matrix has rank one the algorithm will generically produce
$\lfloor1/\delta\rfloor$ trivial histories. This is a particularly
simple case of the analysis that suggests that if $\delta/\epsilon <
O(1)$ repeated projections are probable. 
Fig.~(\ref{fig:abseg1}) shows an example of this from a computer
simulation with $\epsilon =.1$ and $\delta=0.01$. Only the first ten
projections are shown. This behaviour remains the same in the limit as
$\epsilon \to 0, \delta \to 0, \delta\leq\epsilon$.

\onefigure{abseg12}{absolute consistency, $\delta=0.01$,
  $\epsilon=0.02$} As the ratio $\delta/\epsilon$ increases and
becomes $O(1)$ the nature of the set of histories changes.
Occasionally when a reprojection becomes non-trivial it will no longer
be consistent and a reprojection will not occur. A significant time
may elapse before the next projection is made which will result in a
non-trivial projection, which will then be followed by more trivial
repeated projections.  This is demonstrated in
fig.~(\ref{fig:abseg12}) where $\epsilon = 0.02$ and $\delta=0.01$.
Though this is an interesting set of histories this range of parameter
values does not give a theory with predictive power since simulations
show that the results vary enormously for small changes in $\epsilon$
and $\delta$.

\pairoffigures{ds}{abseg4}{absolute consistency, $\delta=0.01$,
  $\epsilon=0.001$} As $\delta/\epsilon$ increases past $1$ the number
of histories made with probability $\delta$ decreases to just the
initial projection. Fig.~(\ref{fig:abseg4}) shows that this range of
parameter values produces interesting histories but the projections
are occuring at times when the consistency level is fluctuating
randomly about the mean and so will be unstable to small changes in
$\epsilon$.

\subsubsection{Other initial conditions}
Choosing larger rank initial reduced density matrices or initial
projections does not qualitatively change the analysis. The only
difference is that for $\delta > \epsilon$ and $\epsilon$ sufficiently
small no trivial projections will be made.
 
\section{Conclusions}
The algorithm produces sets of histories with a complicated branching
structure and with many non-trivial projections for a range of
parameter values. Algorithms using the DHC produce results that are
essentially the same for a wide range of $\delta$ (the non-triviality
parameter) including the limit $\delta \to 0$. However, the algorithm
does not make useful predictions when applied to this model since the
results vary erratically with $\epsilon$ and there is no special
choice of $\epsilon$ singled out. Choosing $\epsilon$ as a function of
the number of histories according to fig.~(\ref{fig:percentile}) produces
the least unstable sets of histories and the largest sets of
non-trivial histories, but even in this case the algorithm does not
single out a definite set.

The algorithm is less effective when used with an absolute consistency
criterion: in this case the predictions of the algorithm also vary
erratically with $\delta$ and the resulting sets of histories include
fewer non-trivial histories. 

The results of the simulations agree well with the theoretical
analysis of section~(\ref{sec:rand:analysis}) and demonstrate features
of the algorithm that will also apply to other models --- such as the
analysis of repeated projections. They also demonstrate some of the
difficulties that an algorithm must overcome. These problems can be
related to the discussion of recoherence in
section~(\ref{sec:recoherence}).  The algorithm will only produce
stable results (with respect to $\epsilon$) if the projections occur
when the off-diagonal terms of the decoherence matrix are
monotonically decreasing. This behaviour is only likely in a system
like this for times small compared to the recurrence time of the
system and when the number of histories is small compared to the size
of the environment Hilbert space. The results of the model do show
stability for the first few projections and if much larger spaces were
used this behaviour would be expected for a larger number of
histories. In particular as the size of the environment goes to
infinity it is plausible that the algorithm applied to this model will
produce a large, stable, non-trivial set of histories.

In retrospect it was ambitious to hope that an algorithm applied to
this model would produce large sets of stable histories. A random
model like this, with such a small environment, will generically only
decohere a few histories (see
also~\cite{Albrecht:decoherence,Albrecht:collapsing}.)


\chapter{Maximum information}\label{chap:max:inf}

\section{Introduction}\label{sec:inf:intro} 

Chapters~\ref{chap:spin}~and~\ref{chap:random} have shown some of the
difficulties in formulating a successful set selection algorithm.
Indeed section~\ref{sec:recoherence} shows that in many systems no
algorithm that constructs sets by proceeding forwards in time will
produce the correct physical set. An algorithm must consider the
entire time evolution of a system if it is always to overcome this
problem. This chapter introduces a new algorithm that is global with
respect to time: it considers the class\footnote{\emph{Class} is used
as a synonym for \emph{set} when referring to a \emph{set} of sets of
consistent histories.} of all consistent sets of histories formed from
Schmidt projections and selects from among them the one with the
greatest Shannon information~\cite{Shannon:Weaver}.

Information\footnote{\emph{Entropy} or \emph{information-entropy} are
  used instead by some authors.} is a term often used in the study of
  quantum mechanics and is used in many different senses.
  Hartle~\cite{Hartle:spacetime:information} considers the
  \emph{missing information} of a set of histories in a generalised
  spacetime quantum mechanics --- he defines the missing information S
  of a set of histories $\mathcal{S}$ with initial density matrix
  $\rho$ as
\begin{equation}\label{inf:deficit}
  S(\mathcal{S},\rho) = \max_{\rho'\in\{D(\mathcal{S},\rho') = D(
    \mathcal{S},\rho)\}} E(\rho')\,,
\end{equation}
where $D(\mathcal{S},\rho)$ is the decoherence matrix for the set of
histories $\mathcal{S}$ with initial density matrix $\rho$.
Throughout this chapter $E$ will denote the Shannon information of a
set of probabilities or, in the case of a positive definite Hermitian
matrix, the Shannon information of its eigenvalues\footnote{in
information theory the singularity for zero probabilities is removed
by defining $0 \log 0 = 0$.}. So, for example, $E(\rho') = - \mbox{Tr}
\rho' \log \rho'$ and
\begin{equation}
  E(\mathcal{S},\rho) = \sum_{\alpha \in \mathcal{S}} -
  D_{\alpha\alpha} \log D_{\alpha\alpha}\,,
\end{equation}
where $\{D_{\alpha\alpha}\}$ are the diagonal elements of the
decoherence matrix $D(\mathcal{S},\rho)$.  Note that if a set of
histories $\mathcal{S}$ is medium consistent then $E({\mathcal
S},\rho) = E[D(\mathcal{S},\rho)]$: generically this is not true for
weak consistency criteria.

$S(\mathcal{S},\rho)$ is the information content of a
\emph{maximum-entropy}~\cite{Jaynes:papers} estimation of the initial
density matrix given the set of histories and their probabilities---
it quantifies what can be inferred about the initial density matrix
using the set of histories and their probabilities.  Hartle goes on to
define
\begin{equation}\label{inf:min}
  S(\mathcal{G},\rho) = \min_{\mathcal{S}\in \mathcal{G}}
  S(\mathcal{S},\rho),
\end{equation}
where $\mathcal{G}$ is some class of consistent sets of histories.
Computing $S(\mathcal{G},\rho)$ for different classes enables one to
understand different ways information about a quantum system can be
obtained. For example Hartle suggests comparing whether the same
information is available using homogeneous~\cite{Isham:homog}
histories instead of the more general inhomogeneous histories. When
$\mathcal{G}$ is the class of all consistent sets he calls
$S(\mathcal{G},\rho)$ the \emph{complete information}.

Eq.~(\ref{inf:min}) could be used as the basis for a set selection
algorithm by specifying some class of sets of histories $\mathcal{G}$
and selecting a set of histories that produces the minimum in
eq.~(\ref{inf:min}). This does not work for general classes, since if
the class contains sets of histories which include projections onto
the eigenspaces of $\rho$ (in non-relativistic quantum mechanics)
these projections completely specify $\rho$, so a rather uninteresting
set of histories is selected. However, if the initial state is pure
and a \emph{Schmidt class} (a class of sets of histories formed from
Schmidt projections) is used it will not generically contain a set of
histories that includes a rank one projection onto the initial state,
hence the set of histories selected by eq.~(\ref{inf:min}) might not
be trivial.  For instance the set of histories consisting of
projections $P \otimes I$ and $\overline P \otimes I$, where $P$ is
the projection onto the non-zero system Schmidt eigenspaces, has
missing information $\log \mbox{rank}(P \otimes I)$. It might be
considered unnatural to assume a pure initial state and then make a
maximum entropy calculation over density matrices of other ranks;
however, this idea has a more serious flaw. The aim of set selection
algorithms is to make statements concerning physical events, not
merely to supply initial conditions. This algorithm only searches for
a set of histories that best specifies the initial conditions and
there is no reason to expect it to produce sets that do more than
describe the initial conditions.

Isham and Linden~\cite{Isham:Linden:information} very recently
proposed a different version of missing information, which they call
\emph{information-entropy}, that is simpler and does not use ideas of
maximum entropy.
\begin{equation}\label{IL:information}
  S'(\mathcal{S},\rho) = - \sum_{\alpha \in \mathcal{S}}
  D_{\alpha\alpha} \log
  \frac{D_{\alpha\alpha}}{\mbox{dim}^2(\alpha)}\,,
\end{equation}
where $\mbox{dim}(\alpha) = \mbox{Tr}(C_\alpha)$ when $C_\alpha$ is
considered as an operator in the $n$-fold tensor product
space~\cite{Isham:Linden:temporal,Isham:Linden:Schreckenberg} of
$\mathcal{H}$. For example if the history $\alpha$ is defined by
consecutive projections $\{P_k, k = 1,\ldots,n\}$ then
$\mbox{dim}(\alpha) = \mbox{Tr} (P_1 \otimes \cdots \otimes P_n) =
\mbox{rank}(P_1) \times \cdots \times \mbox{rank}(P_n)$.  Like
Hartle's missing information, $S'$ \emph{decreases} under a refinement
of $\mathcal{S}$ and
\begin{equation}\label{IL:lbound}
  S' (\mathcal{S},\rho) \geq -\mbox{Tr $\rho \log \rho$}\,.
\end{equation}
Isham and Linden show for some examples that
\begin{equation}\label{IL:mininf}
  \min_{\mathcal{S} \in \mathcal{G}} S' (\mathcal{S},\rho) = -\mbox{Tr
  $\rho \log \rho$}
\end{equation}
and conjecture that the bound is attained in general. These are
interesting results and suggest a useful definition of information ---
especially in complicated spacetimes. Isham and Linden also suggest
that information-entropy might help in the development of a set
selection criterion. Although they have not yet made a definite
proposal, they suggest that perhaps the minimisation should be carried
out with respect to a system--environment split. Clearly some
restriction on the class of sets used is necessary since
bound~(\ref{IL:lbound}) contains no mention of the Hamiltonian or time
evolution of the system --- simply minimising information-entropy is
unlikely to produce a good set selection algorithm, since the sets of
histories that describes experimental situations are much more than a
description of the initial conditions.

Gell-Mann and Hartle discuss similar ideas in detail in
ref.~\cite{gmhstrong}. They introduce a measure, which they call
\emph{total information} or \emph{augmented entropy}, $\Sigma$ that
combines algorithmic information (see for example
ref.~\cite{Zurek:algorithmic}), entropy-information and coarse
graining. This is an attempt to provide a quantitative measure of
\emph{quasiclassicality}. They show that minimising $\Sigma$ does not
provide a useful set selection algorithm --- the results are trivial,
histories are selected that consist of nothing but projections onto
the initial state --- but they suggest augmenting the minimisation
with a stronger consistency criterion,
\begin{equation}\label{GMH:strong}
  \langle\alpha| M^\dagger_\alpha M_\beta |\beta\rangle = p_\alpha
  \delta_{\alpha\beta} \mbox{~$\forall \alpha \neq \beta$, $M_\alpha
  \in \mathcal{M}_\alpha$ and $M_\beta \in \mathcal{M}_\beta$,}
\end{equation}
where $\mathcal{M}_\alpha$ and $\mathcal{M}_\beta$ are sets of
operators. This is an interesting idea. So far however, Gell-Mann and
Hartle have not proposed a definite algorithm for choosing the
$\mathcal{M}_\alpha$. Without a concrete scheme for choosing the sets
$\mathcal{M}_\alpha$ the set selection problem of course becomes the
problem of selecting $\mathcal{M}_\alpha$.  Gell-Mann and Hartle
proposal also has the previously mentioned disadvantage of appearing
to favour set of histories that only provide a description of the
initial state and say nothing about the evolution.

The approach we present here starts with a precisely defined class of
quasiclassical sets of histories (formed from Schmidt projections) and
picks the set of histories from this class with the largest
information.

\section{Algorithm}\label{sec:inf:algorithm} 

Let $\mathcal{G}(\mathcal{H},U,|\psi\rangle)$ be the class of all sets
of non-trivial, exactly consistent, branch-dependent\footnote{A
branch-independent version of the algorithm can be formulated
similarly} histories formed from Schmidt projection operators, where
$\mathcal{H} = \mathcal{H}_1 \otimes \mathcal{H}_2$ is a finite
Hilbert space, $U(t)$ a time evolution operator and $|\psi\rangle$ the
initial state.  Note that in this section the set of histories
includes the initial state.  The algorithm selects the set
$\mathcal{S} \in \mathcal{G}$ with the greatest Shannon
information. That is
\begin{equation} \label{maxinf}
   \max_{\mathcal{S} \in \mathcal{G}} E(\mathcal{S}) =
   \max_{\mathcal{S} \in {\mathcal G}} \sum_{\alpha \in \mathcal{S}} -
   p_\alpha \log p_\alpha,
\end{equation}
where $p_\alpha$ is the probability of history $\alpha$. The class
$\mathcal{G}$ could be chosen differently by using any of the
consistency or non-triviality criteria from
chapter~\ref{chap:prediction}. Another variant uses sets of histories
formed by Schmidt projections onto the system eigenspaces of the
individual path-projected-states ($U(t)C_\alpha|\psi\rangle$), not the
total state (see sec.~\ref{sec:branchdep}), so that the choice of
projections is branch-dependent as well as the choice of projection
times. This is likely to be necessary in general to produce realistic
sets.

When the initial state is pure, in a Hilbert space of dimension $d$
($= d_1 d_2$) there can only be $d$ non-trivial, exactly consistent
histories\footnote{There can be $2d$ if weak consistency is used.}. In
realistic examples approximate consistency may have to be considered.
To ensure the algorithm is well defined it is important that the
number of possible history vectors in a set is finite, which will only
be true if we use a parameterised non-triviality criterion or we use a
consistency criterion, such as the DHC, that can only be satisfied by
a finite number of history vectors~\cite{McElwaine:1}. This is a natural
requirement for any set of histories in a finite Hilbert space since
the exactly consistent sets are finite.

Eq.~(\ref{maxinf}) selects an equivalence class of sets of histories
that all have the maximum information. The equivalence relation is
defined by $\mathcal{S}_1 \sim \mathcal{S}_2$ if $E(\mathcal{S}_1) =
E(\mathcal{S}_2)$; that is, sets of histories are equivalent if they
have the same information. Sufficient conditions for
eq.~(\ref{maxinf}) to be well defined are that
$\mathcal{G}/\kern-.35em\sim$ is closed and that $E(\mathcal{S})$ is
bounded. $\mathcal{G}$ itself is not closed, but the only limit sets
of histories it does not include are those containing zero probability
histories, and since zero probability histories contribute zero
information these limit sets are equivalent to sets which are in
$\mathcal{G}$, hence $\mathcal{G}/\kern-.35em\sim$ is closed.
Moreover these limit sets are also physically equivalent to some of
the sets that they are information-equivalent to, since they only
differ by zero probability histories --- excluding the limit sets does
not change anything physical. The information of any set of histories
in $\mathcal{G}$ is bounded, since the number of histories in any set
of histories in $\mathcal{G}$ is bounded and the information of a set
of $n$ probabilities is bounded by $\log n$.  Conditions sufficient to
ensure uniqueness are much more complicated.  Probably the best that
one can hope for in many situations is that a class of physically
equivalent sets is selected.

First we describe some useful properties of this algorithm and then we
apply it to a simple model.

\subsection{Completeness}\label{sec:completeness}

The set of histories selected by the algorithm cannot be extended
(except trivially) because any non-trivial extension increases the
information content. To see this consider the set of histories
${\mathcal S}$ and an extension $\mathcal{S}'$. The probabilities for
the new histories can be written in the form $p_\alpha
q^{(\alpha)}_\beta$ where $\sum_\beta q^{(\alpha)}_\beta=1$ for all
$\alpha$. The information of the new set is
\begin{equation}\label{infadd}
  E(\mathcal{S}') = -\sum_\alpha \sum_\beta p_\alpha
  q^{(\alpha)}_\beta \log p_\alpha q^{(\alpha)}_\beta = E(\mathcal{S})
  + \sum_\alpha p_\alpha E(q^{(\alpha)}_\beta),
\end{equation}
which is strictly greater than $E(\mathcal{S})$ whenever the extension
results in at least one non-zero probability.

\subsection{Additivity}

A set of branch-dependent histories has a probability tree structure,
where each history $\alpha$ refers to a terminal node of the tree and
the unique path from that node to the root node. The nodes themselves
are associated with projection operators and path projected states.
Define $\mathcal{S}_{\alpha k}$ to be the set of all histories
extending from the $k^{\mbox{\scriptsize th}}$ node of history
$\alpha$, normalised so that the total probability is one. This is a
set of histories in its own right which will be consistent if the
entire set of histories is consistent. Consider a simple example where
the first projection produces two histories with probabilities $p$ and
$q$ and the subtrees from these nodes are ${\mathcal S}_p$ and
$\mathcal{S}_q$. The information for the set of histories can then be
written,
\begin{equation}\label{subtreeadd}
  E(\mathcal{S}) = E(\{p,q\}) + p E(\mathcal{S}_p) + q
  E(\mathcal{S}_q).
\end{equation}
This formula is easy to generalise. Each subtree must have maximum
information subject to the constraint that the history vectors span a
space orthogonal to the other history states. That is, a global
maximum must also be a local maximum in each degree of freedom and the
subtrees are the degrees of freedom.

\subsection{Large sets}

One of the problems with the algorithms in
chapter~\ref{chap:prediction} is their tendency to make projections
too early so that they prevent projections at later times. Other
problems also arise with algorithms that produce histories with zero
or small probabilities. The maximum-information algorithm will not
have these problems, since any projection that prevents later
extensions is unlikely to be selected, histories with zero probability
will never be selected (since they contribute no information), and
histories with small probabilities are also unlikely to be selected.
Therefore the algorithm is much more likely to produce large
complicated sets of histories than previous algorithms.

\subsection{Stability}

Another problem with the algorithms in chapter~\ref{chap:prediction}
was demonstrated in chapter~\ref{chap:random}: the predictions are
unstable to perturbations of the Hamiltonian or initial state, and the
predictions vary erratically with the choice of the approximate
consistency parameter. It is difficult to prove any general results
about stability for this algorithm, but it seems much more likely to
produce stable predictions for the following reasons. There is no
reason why any projections should be made that exactly satisfy the
approximate consistency criterion, whereas in an earliest time
algorithm this is the generic result. For example, if a set is
predicted that is exactly consistent the same set could still be
predicted for any reasonably small $\epsilon$. There are also unlikely
to be complications caused by trivial histories, since these
contribute little information and are not likely to be selected by the
algorithm. The sets of histories described by the Schmidt projections
generically will vary continuously with sufficiently small changes in
the initial state and Hamiltonian. Therefore we expect generically
that for a range of approximate consistency parameters the selected
set will also vary continuously. This is a generic result for
continuous optimisation problems: they are stable for small
perturbations.

\section{The spin model}\label{sec:spininf}

A set of histories that maximises information must be complete,
therefore all histories must consist of projections at times
$\{1,\ldots, k-1 ,t ,k: t \in(k-1,k)\}$. First we show that $k$ must
be the same for all histories (that is, the set is branch
independent), then we calculate $k$.

The information content of two subtrees rooted at the same point only
depends on the projection times within each one. Either the two
subtrees have the same information, in which case their projection
times must be the same, or one has more, but since the projection
times used in the subtree with greater information will also be
consistent if used in the subtree with less information these
projection times can be used instead. Therefore in the set with
maximum information all the subtree must have the same projection
times, thus all the histories must have the same projection times ---
the maximal set is branch independent.

Let the projection times be $\{1,\ldots, k-1 ,t ,k: t \in(k-1,k)\}$.
Then from eq.~(\ref{probsb}) and eq.~(\ref{infadd}) the information
content of this set is

\begin{eqnarray} \label{kinfa}
  f[N_k(\theta_k(t))] + f[({\bf u}_{k}.  {\bf u}_{k-1}) N_k^{-1}
    (\theta_k(t))] + \sum_{k>j>0} f({\bf u}_{j-1}. {\bf u}_j)
\end{eqnarray}
where
\begin{equation}
  f(x) = - \frac{1+x}{2} \log \frac{1+x}{2} - \frac{1-x}{2} \log
  \frac{1-x}{2}.
\end{equation}
Maximising eq.~(\ref{kinfa}) with respect to $t$ yields
\begin{equation}
  E(\mathcal{S}_k) = E_k = 2f(|{\bf u}_{k}.  {\bf u}_{k-1}|^{1/2}) +
  \sum_{k>j>0} f({\bf u}_{j-1}. {\bf u}_j),
\end{equation}
where $\mathcal{S}_k$ is the branch independent set consisting of
projections at times $\{1,\ldots,k-1,t_k,k\}$.  This is usually
maximised by $k=n$ but depending on the relationships between the
${\bf u}_j$ any value of $k$ may be possible. For example, consider
${\bf u}_{j-1}.{\bf u}_j = 1-\epsilon$ for all $j \neq k$ and ${\bf
u}_{k-1}. {\bf u}_k = \epsilon$ and $\epsilon$ is small.
\begin{equation}
  E_m = \left\{
    \begin{array}{lr}
      O(\epsilon\log \epsilon), & \mbox{for $m < k$}, \\ 2 \log 2 +
      O(\epsilon\log \epsilon), & \mbox{for $m = k$}, \\ \log 2 +
      O(\epsilon\log \epsilon) & \mbox{for $m > k$},
    \end{array}\right.
\end{equation}
which for small $\epsilon$ is maximised by $E_k$.

The precise relationship between the $\{{\bf u}_j\}$ that ensure $E_n
> E_k$ for all $k < n$ is complicated in detail, but simple
qualitatively. Roughly speaking, $E_n < E_k$ only if $|{\bf u}_{j-1}.
{\bf u}_j| \gg |{\bf u}_{k-1}. {\bf u}_k|$ for all $j > k$, that is
all the measurement directions must be approximately parallel after
the $k^{\mbox{\scriptsize th}}$.  Monte Carlo integration over $\{{\bf
  u}_i\}$ (with the $SO(3)$ invariant measure) shows that for $n=3$
set $\mathcal{S}_n$ is selected $85.7\%$ of the time, for $n=4$ it is
selected $84.3\%$ of the time, and for all $n>4$ it is selected
$84.2\%$ of the time.  When the vectors are approximately parallel,
that is $|{\bf u}_{j-1}.  {\bf u}_j| = 1 - O(\epsilon)$, set $S_n$ is
selected with probability $1-O(\epsilon)$.  If however all the
measurement spins are approximately parallel ($|\mathbf{u}_{j-1}.
\mathbf{u}_j| > 1 - \epsilon$, and $-n\epsilon\log\epsilon<4\log2$)
then for some orientations of the initial system spin (${\bf v} = {\bf
  u}_0$) $E_1 > E_k$ for all $E_k$ so set $\mathcal{S}_1$ is selected.
That is, the maximal set consists only of a projection during the
first interaction and at the end of the first interaction.

Though the results of the algorithm may seem counterintuitive the
following discussion shows why this is not a problem.

First consider the case when the system is genuinely closed. All the
projections before the last interaction are natural\footnote{The
adjective \emph{natural} is used to describe sets or projections that
agree with our intuitive understanding of a quantum mechanical
system.}. It is only the projections during the last interaction,
which occur when the set of histories is nearly complete, that are
unnatural. Our intuition about the system and the result we believe to
be correct relies on the experiment being imbedded in a larger system
in which the sets of histories considered are always far from
complete.

Second consider the case where the system is approximately closed.
Then the sets $\mathcal{S}_k$ should describe the first projections of
a maximum-information solution in a larger Hilbert space. For reasons
explained below, no non-trivial projections onto the system space will
result in consistent extensions of the sets $\mathcal{S}_k$, even if
the system interacts with new degrees of freedom in the environment.
This shows that though it is a maximum-information set for a
subsystem, it is unlikely to be part of the maximum-information set
for the entire system. The set most likely to be part of the
maximum-information set is the natural set, the set that consists of
projections only at the end of each interaction.

The set of normalised histories (in the Schr\"odinger picture at time
$k$, that is the path-projected states) is
\begin{equation}
    \mathcal{S}_k = \{ |\alpha_0{\bf v}_k \rangle \otimes
    |\alpha_1(\uparrow), \ldots, \alpha_{k-1}(\uparrow),
    \alpha_k(\rightarrow), \uparrow_{k+1}, \ldots, \uparrow_n\rangle
    \forall \alpha \in Z_2^{k+1}\},
\end{equation}
where $\alpha$ is a string of $2^{k+1}$ plusses and minuses,
$+(\uparrow) = \uparrow$, $-(\uparrow) = |\downarrow\rangle$ and
$\pm(\rightarrow)$ are orthogonal vectors depending on
$\mathbf{u}_{k-1}$ and $\mathbf{u}_k$. This set of histories cannot be
non-trivially extended with Schmidt projections (see
sec.~\ref{sec:spin:analysis}).  The reason for this is clear.
Consider two of the histories $|{\bf \pm v}_k \rangle \otimes
|e\rangle$ where $|e\rangle$ is the environment state. These histories
are only orthogonal because of the orthogonality of the system part of
the states. There can be no future non-trivial extensions unless there
is an exact degeneracy, because consistency terms between these two
histories will contain terms like $|\langle{\bf v}| P({\bf w}) |{\bf
v}\rangle| = \sqrt{1/2(1+{\bf v}.{\bf w})}$, which is only zero when
$\mathbf{w}=-\mathbf{v}$. In contrast if projections are only made at
the end of interactions all the histories are orthogonal in the
environment Hilbert space of the finished interactions. Unless these
interactions are ``undone'' these histories will always remain
orthogonal and cannot interfere. This argument suggests that the true
maximum-information set for the total Hilbert space starts of with
projections at the end of every interaction but at \emph{no} interior
times.

This suggests that an algorithm designed to produce a
maximum-information set for a subsystem could be constructed by
requiring that all the histories in a set were orthogonal in the
environment space, that is the reduced density matrices in the
environment Hilbert space for each history are orthogonal. This is
related to the strong consistency criterion eq.~(\ref{GMH:strong})
when the sets $\mathcal{M}_\alpha$ are all chosen to be the set $\{P
\otimes I: \forall P^2=P\,,P^\dagger=P\}$.

\section{Other algorithms}\label{sec:otheralg}

Let $\mathcal{G}(\mathcal{H},U,|\psi\rangle)$ be the class of all sets
of non-trivial\footnote{It is important that each set of histories is
  finite. If trivial histories are to be allowed the limit DHC could
  be used.}, exactly consistent, branch-dependent histories formed
from Schmidt projection operators in the spin model.  Consider an
algorithm that selects the set in $\mathcal{G}$ that minimises Isham
and Linden's information-entropy~(\ref{IL:information}). Due to the
special symmetries of the spin model the selected set will be branch
independent --- the argument at the start of
section~(\ref{sec:spininf}) is valid.

Consider the set of projections at $m$ times, so that the dimension of
each history is $(d/2)^m$, where $d$ is the dimension of the total
Hilbert space.  Information-entropy for this set is
\begin{equation}
  S' = - \sum_{\alpha \in \mathcal{S}}
  p_{\alpha} \log \frac{p_\alpha}{(d/2)^{2m}} = 2m \log (d/2) -
  \sum_{\alpha \in \mathcal{S}}  p_{\alpha} \log p_\alpha\,.
\end{equation}
This is minimised for $m=0$ --- the selected set consists of no
projections at all --- which is not a natural set.  The problem with
directly using minimum-information-entropy as a set selection
criterion is clear; the information-entropy will consist of two terms,
one roughly proportional to the number of projection times and another
that is the Shannon information of the probabilities, both of which
are minimised by the set with the fewest histories.

Isham and Linden show that information-entropy decreases when the set
is refined by replacing a projective decomposition by a finer one.
However, if the set is extended by adding projections at a new time
then the information-entropy increases, and since the proposal is to
minimise information-entropy this algorithm will always select the set
of histories in the class with the fewest histories. There are two
obvious approaches to this difficulty, which are largely equivalent.
One could extend (with the identity projection) all the sets of
histories in the class so that projections occur at the same times in
all sets. Adding a projection would then be a refinement of the
identity and the information-entropy would decrease. However, it is
not all obvious that this is well defined in general, though there are
no problems in this example.  When it is well defined it yields the
same results as the more natural modification of
using the \emph{normalised dimension}, $\widehat {\mbox{dim}}$, in the
formula~(\ref{IL:information}) information-entropy.  For a homogenous
history
\begin{equation}
  \widehat {\mbox{dim}}\ \alpha = \frac{\mbox{dim } P_{\alpha_1} \times
    \cdots \times \mbox{dim } P_{\alpha_m}}{\mbox{dim } I \times
    \cdots \times \mbox{dim } I} = (\mbox{dim } \alpha) / d^m\,,
\end{equation}
where $d = \mbox{dim } I$ the dimension of the Hilbert space. 
With this definition information-entropy decreases for all refinements
and extensions.

The information-entropy with normalised dimension is
\begin{equation}\label{revisedILIE}
  S'' = -2m \log 2  - \sum_\alpha p_\alpha \log p_\alpha\,.
\end{equation}
Using the notation of the previous section this can be written
\begin{equation}
  S'' = -\sum_{m > k > 0} [2 \log 2 - f(\alpha_k)]\,,
\end{equation}
where the $\alpha_k$ depend on the projection times and vary between
$-1$ and $1$.  Since $f(x)\leq \log 2$ each term in the sum is always
negative so the minimum occurs for $m = n+1$, and the selected set
consists of projections at the end of every interaction and a
projection either at the end or the beginning of the last interaction
--- the algorithm has selected a natural set. The revised (using
normalised dimension) minimum-information-entropy algorithm selects a
set with as many projections as possible, and among these sets it
selects the set whose probabilities have the lowest Shannon
information.  Though using $S''$ has worked well here this approach
would probably not be successful in general as the algorithm will
tend towards making a large number of repeated projections and these
will be greatly favoured over more complicated sets. The set of
normalised dimensions $\{\hat d_\alpha\}$ has an interpretation as a
set of probabilities. Perhaps a more natural selection scheme would be
to minimise the difference in Shannon information between the $\{\hat
d_\alpha\}$ and the $\{p_\alpha\}$. An approach like this may work in
particular situations but all the suggestions in this
section~(\ref{sec:otheralg}) appear ad hoc and unsuitable as a part of
a fundamental theory.

\section{Conclusions}

This chapter defines a precise algorithm for making probabilistic
predictions for closed quantum systems. The algorithm considers the
class of all non-trivial, exactly consistent, branch-dependent sets of
histories defined by Schmidt projections with respect to a fixed split
of the Hilbert space and selects from among them the set with the
largest Shannon information. The algorithm avoids many of the problems
of the algorithms we considered earlier in
chapter~\ref{chap:prediction}. Because it considers the entire time
evolution of a system -- roughly speaking it is global in time,
whereas the algorithms in chapter~\ref{chap:prediction} are local ---
it does not make unphysical projections in systems where recoherence
occurs and it produces complete sets of histories that describe the
correlations between the system and the environment. Trivial and very
small probability histories, which cause problems for some of the
algorithms considered earlier by preventing later physical
projections, are unlikely to be selected since they contribute little
information. The algorithm is also likely to be stable under
perturbations in the initial conditions, the Hamiltonian and the
parameters, since it involves maximising a continuous function.

Section~(\ref{sec:spininf}) has shown that the algorithm selects a
natural set for a simple spin model.  It would be interesting to test
out the algorithm on more realistic examples; however, it seems
difficult to apply the algorithm directly, because of the large size
and complicated nature of $\mathcal{G}$.  Analytic calculations are
only possible when the system is very simple and in more realistic
examples computer simulations will be necessary.  However, it should
be possible at least to get some insight into the algorithm's
predictions by maximising subject to constraints, that is by
considering a more computationally tractable subset of $\mathcal{G}$.
For example, we could choose a time interval $T$ that is greater than
the time of individual interactions (within the particular system) and
larger than any timescale over which recoherence occurs.  This would
be used as a moving time-window over which to perform the
maximisation. The earliest projection within each time-window would be
selected and the next time-window would commence from that time. Such
algorithms should select the same set as a global algorithm if $T$ is
large enough, and are also independently interesting.

Because the algorithm predicts the probabilities for events \emph{and}
the set of possible events the algorithm is falsifiable. For example
the algorithm is wrong if it selects any sets that do not agree with
our experiences. The algorithm can also be applied to situations where
we have no experience of what the natural sets of histories are: for
example, a (finite) closed system of electrons and photons --- and
perhaps ultimately could be applied to theories of quantum cosmology.
The algorithm, like other tentative proposals in the
literature~\cite{Isham:Linden:information,gmhstrong}, has not yet been
tested in a wide range of realistic physical examples: further
investigations would clearly be worthwhile.


\appendix
\chapter{Probability distributions}\label{sec:probdists}

The notation $1_{\mathbf{x} \in A}$ is used to denote an indicator
function in this dissertation, that is
\begin{equation}
  \renewcommand{\arraystretch}{.7} 1_{\mathbf{x} \in A} = \left\{
  \begin{array}[c]{lr}
    1 & \mbox {if $\mathbf{x} \in A$} \\ 0 & \mbox {if $\mathbf{x}
    \not\in A$}
  \end{array}\right.\,.
\end{equation}
\section{Sum of $k$ components of a random vector in~$S^{d-1}$}
Let $A$ be the random variable
\begin{equation}
  A = \sum_{k\geq i \geq 1} X_i^2\,,
\end{equation}
where $X_i$ are the $d$ components of a random vector uniformly
distributed in $S^{d-1}$. That is, $A$ is the sum of the first $k$
components of a random unit vector in $R^d$.  $\mathbf{X}$ can be
written $\mathbf{X} = \mathbf{Y}/|\mathbf{Y}|$, where $Y_i$ are
identical, independent, mean zero, normal random variables, since the
probability density function is proportional to
$e^{-|\mathbf{Y}|^2}$. Then
\begin{equation}\label{onevecksuma}
  P( A < \lambda) =
  P\left( \frac{\sum_{k \geq i \geq 1} Y_i^2}{%
      \sum_{d \geq i \geq 1} Y_i^2} < \lambda\right) = P\left(
  \frac{\chi^2_k}{\chi^2_k + \chi^2_{d-k}} < \lambda\right),
\end{equation}
where $\chi^2_k$ and $\chi^2_{d-k}$ are independent, Chi-squared
random variable with $k$ and $d-k$ degrees of freedom
respectively. This ratio of Chi-squared variable is a beta variable
with distribution function
\begin{equation}\label{betadist}\label{ksumreal}
  B[\lambda;k/2,(d-k)/2] =
  \frac{\Gamma(d/2)}{\Gamma(k/2)\Gamma[(d-k)/2]} \int_0^\lambda dx\,
  x^{k/2-1} (1-x)^{(d-k)/2-1},
\end{equation}
which has mean $k/d$ and variance $\frac{k(d-k)}{d^2(d/2+1)}$.  A
complex unit vector $\mathbf{Z} \in CS^{d-1}$ can be regarded as a
real vector in $S^{2d-1}$ so
\begin{equation}\label{ksumcomplex}
  P\left( \sum_{k\geq i \geq 1} |Z_i|^2 < \lambda \right) =
   B[\lambda;2k/2,(2d-2k)/2] = B(\lambda;k,d-k).
\end{equation}

\section{Maximum of $k$ components of a random vector in~$S^{d-1}$}
\label{sec:maxk}
Let $A$ be the random variable
\begin{equation}
  A = \max_{k\geq i \geq 1} X_i^2\,,
\end{equation}
where $X_i$ are the $d$ components of a random vector uniformly
distributed in $S^{d-1}$ and $f(\lambda) = P(A<\lambda)$. The
probability density function for $\mathbf{X}$ is
$N_d\,\delta(1-|\mathbf{X}|^2)$, where $N_d =
\Gamma(d/2)/(2\pi^{d/2})$.  Therefore
\begin{equation}\label{maxoneveca}
  f(\lambda) = N_d \int d\mathbf{x}\, \delta(1-|\mathbf{x}|^2)
  \prod_{k \geq i \geq1} 1_{x_i^2 < \lambda} \,.
\end{equation}
After changing variables $\mathbf{x} \to \mathbf{x}/\sqrt{r}$ and
$\mathbf{\lambda} \to \mathbf{\lambda}/r$ eq.~(\ref{maxoneveca}) is
\begin{equation}\label{maxonevecb}
   f(\mathbf{\lambda}/r) = r^{-d/2} N_d \int d\mathbf{x}\,
  \delta(1-|\mathbf{x}|^2/r) \prod_{k \geq i \geq1} 1_{x_i^2 <
  \lambda} \,.
\end{equation}
Multiply both sides by $r^{n/2-1}$ and integrate with respect to $r$
from $0$ to $\infty$.
\begin{equation}\label{maxonevecc}
   \int_0^\infty dr\, r^{n/2-1} e^{-r} f(\mathbf{\lambda}/r) = N_d
  \int d\mathbf{x}\, e^{-|\mathbf{x}|^2} \prod_{k \geq i \geq1}
  1_{x_i^2 < \lambda} = \Gamma(d/2) \prod_{k \geq i \geq1}
  \mbox{erf}(\sqrt{\lambda})\,.
\end{equation}
Rescale $\mathbf{\lambda} \to p\mathbf{\lambda}$ and $r \to rp$.
\begin{equation}\label{maxonevecd}
   p^{d/2} \int_0^\infty dr\, r^{d/2-1} e^{-rp} f(\mathbf{\lambda}/r)
 = \Gamma(d/2) \prod_{k \geq i \geq1} \mbox{erf}(\sqrt{p\lambda})\,.
\end{equation}
The l.h.s.\ is a Laplace transform of the function $ r^{d/2-1}
f(\mathbf{\lambda}/r)$ with respect to the variable $p$, so using the
Laplace inversion theorem
\begin{equation}\label{maxonevece}
     r^{n/2-1} f(\mathbf{\lambda}/r) = \frac{\Gamma(d/2)}{2\pi i}
 \int_{\gamma-i\infty}^{\gamma+i\infty} dp\, e^{rp} p^{-d/2} \prod_{k
 \geq i \geq1} \mbox{erf}(\sqrt{\lambda p})\,.
\end{equation}
Finally let $r=1$
\begin{equation}\label{maxonevecf}
  f(\mathbf{\lambda}) = \frac{\Gamma(d/2)}{2\pi i}
  \int_{\gamma-i\infty}^{\gamma+i\infty} dp\, e^{p} p^{-d/2} \prod_{k
  \geq i \geq1} \mbox{erf}(\sqrt{\lambda p})\,.
\end{equation}
This integral is very complicated. However, when $d\lambda = O(1)$ and
$k^2/d = o(1)$ we can perform an asymptotic expansion in $d$ using the
method of steepest descents. The saddle point is at $p = \sqrt{d/2}$
and the steepest descent contour is $p(x) = -W(-e^{x^2-1})$, where $W$
is Lambert's W function. The first two terms are
\begin{equation}
  \mbox{erf}^k(\sqrt{d\lambda/2})\left[ 1 +
    \frac{k\sqrt{\lambda}(d\lambda-3)e^{-d\lambda/2}}{%
      2\sqrt{2\pi d}\, \mbox{erf}(\sqrt{d\lambda/2})} -
    \frac{k(k-1) \lambda e^{-d\lambda}}{%
      2 \pi \, \mbox{erf}^2(\sqrt{d\lambda/2})} + O(1/d^2)\right],
\end{equation}
which equals $\mbox{erf}^k(\sqrt{d\lambda/2})[1 + O(k^2/d)]$.  This
shows that if we are looking at fewer than $\sqrt d$ components of a
unit vector we can accurately approximate the distribution by
considering $k$ independent normal variable with variance $1/d$ for
$\lambda < 1/k$.

The complex case is more straightforward. Let $A'$ be the random
variable
\begin{equation}
  A' = \max_{k\geq i \geq 1} |Z_i|^2\,,
\end{equation}
where $Z_i$ are the $d$ components of a random vector uniformly
distributed in $CS^{d-1}$ and $f'(\lambda) = P(A'<\lambda)$.  The same
method can be used in the complex case to get
 \begin{equation}\label{maxonevecfcomplex}
     f'(\mathbf{\lambda}) = \frac{\Gamma(d)}{2\pi i}
 \int_{\gamma-i\infty}^{\gamma+i\infty} dp\, e^{p} p^{-d} (1-e^{-\lambda p})^k\,,
\end{equation}
and then
\begin{eqnarray}\label{maxonevecgcomplex}
  f_k'(\lambda) &=& \frac{\Gamma(d)}{2\pi i} \sum_m (-1)^m {k \choose
  m} \int_{\gamma-i\infty}^{\gamma+i\infty} dp\, 1_{m\lambda<1}
  e^{p(1-m\lambda)} p^{-d} \\ &=& \sum_m (-1)^m {k \choose m}
  (1-m\lambda)^{d-1} 1_{m\lambda<1}
\end{eqnarray}
If $d\lambda^2k^2 = o(1)$ then eq.~(\ref{maxonevecgcomplex}) becomes
\begin{equation}
  \sum (-1)^m {k \choose m} e^{-d m\lambda + O(m^2/d)} = (1 - e^{-d
  \lambda + o(1)})^k ]\,.
\end{equation}

Note that the form of the answer is essentially the same in the
real and complex cases.

\section{Two orthonormal vectors in $S^{d-1}$}
Consider two orthonormal random vectors $\mathbf{X}$ and $\mathbf{Y}$
in $S^{d-1}$. The probability density function is proportional to
$\delta(1-\mathbf{x}^2) \delta(1-\mathbf{y}^2) \delta(\mathbf{x}.
\mathbf{y})$ so
\begin{equation}\label{twoorthveca}
  P(\mathbf{X} \in A, \mathbf{Y} \in B) = \int d\mathbf{x}\,
  d\mathbf{y}\, 1_{\mathbf{x} \in A} 1_{\mathbf{y} \in B}
  \delta(1-\mathbf{x}^2) \delta(1-\mathbf{y}^2) \delta(\mathbf{x}.
  \mathbf{y}).
\end{equation}
Define
\begin{equation}
  a_m = \sum_{m \geq i \geq 1} x_iy_i\,, \quad b_m = 1- \sum_{m \geq i
  \geq 1} x_i^2\,, \quad c_m = 1- \sum_{m \geq i \geq 1} y_i^2\,.
\end{equation}
Suppose that the regions $A$ and $B$ only restrict the first $k$
components of $\mathbf{x}$ and $\mathbf{y}$. Change variables to
\begin{eqnarray}
  r = x_d^2 + x_{d-1}^2\,, & s = y_d^2 + y_{d-1}^2\,, \\ t = x_d y_d +
  x_{d-1} y_{d-1}\,, & q = x_d y_d - x_{d-1} y_{d-1}\,,
\end{eqnarray}
then
\begin{equation}
    dx_d\,dx_{d-1}\,dy_d\,dy_{d-1}\, = \sqrt{sr-q^2} \sqrt{sr-t^2}
    dr\,ds\,dt\,dq\,.
\end{equation}
The integral over $x_n$, $x_{n-2}$, $y_n$ and $y_{n-2}$ in
eq.~(\ref{twoorthveca}) becomes
\begin{eqnarray}\label{twoorthvecb}
  &&\int \frac{dr\,ds\,dt\,dq\, \delta(b_{n-2}-r)
  \delta(c_{n-2}-s)\delta(a_{n-2}+t)}{ \sqrt{sr-q^2} \sqrt{sr-t^2}} \\
  &&\propto \int_0^{\sqrt{b_{n-2}c_{n-2}}} \frac{dq\,1_{a_{n-2}^2 \leq
  b_{n-2}c_{n-2}}}{\sqrt{b_{n-2}c_{n-2}-q^2} \sqrt{b_{n-2}c_{n-2} -
  a_{n-2}^2}} \\ && \propto \frac{1_{a_{n-2}^2 \leq b_{n-2}c_{n-2}}}{
  \sqrt{b_{n-2}c_{n-2} - a_{n-2}^2}}\,.
\end{eqnarray}
The remaining free variables can be integrated out pairwise in a
similar fashion to leave the density
\begin{equation}
  [b_kc_k-a_k^2]^{\frac{d-k-3}{2}} 1_{a_k^2 \leq b_kc_k} =
  [(1-\mathbf{x}^2)(1-\mathbf{y}^2) -
  (\mathbf{x}.\mathbf{y})^2]^{\frac{d-k-3}{2}}
  1_{(\mathbf{x}.\mathbf{y})^2 \leq
  (1-\mathbf{x}^2)(1-\mathbf{y}^2)}\,,
\end{equation}
where $\mathbf{x}$ and $\mathbf{y}$ have been redefined to be the
first $k$ components of the random variables $\mathbf{X}$ and
$\mathbf{Y}$.
\section{Maximum of DHC}\label{subsec:maxDHC}
Let $A$ be the random variable
\begin{equation}\label{ltprobaa}
  A = \frac{({\bf v}_i^TP{\bf u})^2}{|P{\bf u}|^2} + \frac{({\bf
  v}_i^T\overline P{\bf u})^2}{|\overline P{\bf u}|^2}\,,
\end{equation}
where $\{{\bf v}_i,{\bf u},i=1\ldots k\} $ are a uniformly distributed
set of orthonormal vector in $S^{d-1}$ and $P$ is an uncorrelated
projection operator of rank $m$. $A$ can be simplified to
\begin{equation}\label{ltproba}
  A = \frac{|{\bf v}_i^TP{\bf u}|^2}{|P{\bf u}|^2\, |\overline P{\bf
  u}|^2}.
\end{equation} 
Let ${\bf u} = \sqrt r {\bf w} + \sqrt{1-r} {\bf z}$ such that $P{\bf
w} = {\bf w}$ and $P{\bf z} = 0$.  ${\bf w}$ and ${\bf z}$ are
distributed like orthonormal elements of $S^{d-1}$ with density
$f({\bf w},{\bf z})$ and $r$ has distribution $B[m/2,(d-m)/2]$ density
$h(r)$. Let $({\bf v}_i)_j = \delta_{ij}$. Then $P(A<\lambda)$ is
proportional to
\begin{equation}\label{ltprobb}
  \int dr\, d{\bf w}\, d{\bf z}\, f_d({\bf w},{\bf z}) g(r) \prod_{k
  \geq i \geq 1} 1_{|w_i|^2 + |z_i|^2 < \lambda} \delta(\sqrt r w_i -
  \sqrt{1-r} z_i).
\end{equation}
The integral over $w_i$ and $z_i$ for $i > k$ can be done using $ \int
\prod_{d \geq i > k} dw_i\, dz_i f_d({\bf w},{\bf z}) $ is
proportional to $ f_{dk}({\bf w},{\bf z}) = [(1-w^2)(1-z^2) - ({\bf
w}.{\bf z})^2]^{(d-k-3)/2} $, where ${\bf w}$ and ${\bf z}$ are now
vectors in $R^k$. Eq.~(\ref{ltprobb}) becomes
\begin{equation}\label{ltprobc}
   \int dr\, d{\bf w}\, d{\bf z}\, f_{dk}({\bf w},{\bf z}) g(r)
   \prod_i 1_{|w_i|^2 + |z_i|^2 < \lambda} \delta(\sqrt r w_i -
   \sqrt{1-r} z_i).
\end{equation}
Change variables ${\bf w} \to \sqrt{1-r} {\bf w}$ and ${\bf z} \to
\sqrt{r} {\bf z}$
\begin{equation}\label{ltprobd}
   \int dr\, d{\bf w}\, d{\bf z}\, f_{dk}( \sqrt{1-r}{\bf w},\sqrt{r}
   {\bf z}) g(r) \prod_i 1_{(1-r) |w_i|^2 + r|z_i|^2 < \lambda}
   \delta(w_i - z_i).
\end{equation}
The integral over ${\bf w}$ is now straightforward
\begin{equation}\label{ltprobe}
   \int dr\, d{\bf z}\, f_{dk}( \sqrt{1-r}{\bf z},\sqrt{r}{\bf z})
   g(r) \prod_i 1_{|z_i|^2 < \lambda},
\end{equation}
which equals
\begin{equation}
   \int dr\, g(r) \int d{\bf z} (1-z^2)^{(d-k-3)/2} \prod_i 1_{|z_i|^2
   < \lambda}\,.
\end{equation}
So the probability is proportional to
\begin{equation}
   \int d{\bf z} (1-z^2)^{(d-k-3)/2} \prod_i 1_{|z_i|^2 < \lambda},
\end{equation}
Which is the probability that $k$ components of a vector in $S^{d-2}$
are all less than $\epsilon$. This has been calculated in
section~\ref{sec:maxk}. The complex case is more difficult as the
probability density function for two complex orthogonal vectors cannot
easily be integrated. However, the distribution is expected to be
similar.

\section{Time evolution of Schmidt operators}\label{sec:schmidtevol}

Consider a continuously differentiable Hermitian operator $A(t)$. All
quantities are functions of $t$ but the dependence will not be
explicitly written.  $A$ can be written $\sum_n p_n {\bf u}_n {\bf
u}_n^\dagger$ where $p_n$ are real and $\{{\bf u}_n\}$ an orthonormal
set. Assume the $\{p_n\}$ are all distinct. ${\bf u}_n^\dagger {\bf
u}_m = \delta_{mn}$ implies $\dot{{\bf u}}_n^\dagger {\bf u}_m + {\bf
u}_n^\dagger \dot{{\bf u}}_m = 0$, where $\dot{}$ denotes
differentiation with respect to $t$. Then
\begin{equation}\label{Adot}
  \dot{A} = \sum_n( \dot{p}_n {\bf u}_n {\bf u}_n^\dagger + p_n
  \dot{{\bf u}}_n {\bf u}_n^\dagger + p_n {\bf u}_n \dot{{\bf
  u}}_n^\dagger),
\end{equation}
hence ${\bf u}_n^\dagger \dot{A} {\bf u}_n = \dot{p}_n + p_n {\bf
  u}_n^\dagger \dot{{\bf u}}_n + p_n \dot{{\bf u}}_n^\dagger {\bf
  u}_n$,
\begin{equation}\label{pevol}
  \mbox{so} \quad \dot{p}_n = {\bf u}_n^\dagger \dot{A} {\bf u}_n.
\end{equation}
When the $\{p_n\}$ are distinct the operators
\begin{equation}
  O_n = \sum_{m\neq n} \frac{{\bf u}_m {\bf u}_m^\dagger}{p_n - p_m}
\end{equation}
are well defined. Multiplying ${\bf u}_n$ by eq.\ (\ref{Adot}) and
then $O_n$ gives
\begin{eqnarray}
  \sum_{m\neq n} \frac{{\bf u}_m {\bf u}_m^\dagger \dot{A} {\bf u}_n}{
    p_n - p_m} &=& \sum_{m\neq n} \frac{{\bf u}_m {\bf u}_m^\dagger}{
    p_n - p_m} \dot{p}_n {\bf u}_n + p_n \sum_{m\neq n} \frac{{\bf
    u}_m {\bf u}_m^\dagger \dot{{\bf u}}_n}{ p_n - p_m} + \sum_{m\neq
    n} \frac{ p_m {\bf u}_m \dot{{\bf u}}_m^\dagger {\bf u}_n}{p_n -
    p_m} \\ &=& \sum_{m\neq n} \frac{(p_n - p_m) {\bf u}_m {\bf
    u}_m^\dagger \dot{{\bf u}}_n}{ p_n - p_m} \hspace{\arraycolsep} =
    \hspace{\arraycolsep} \dot{{\bf u}}_n - {\bf u}_n {\bf
    u}_n^\dagger\dot{{\bf u}}_n\label{uevol}
\end{eqnarray} 
The eigenvectors are only unique up to a complex phase so without
loss of generality one can take ${\bf u}_n^\dagger\dot{{\bf u}}_n =
0$\footnote{let ${\bf v} = e^{i\theta} {\bf u}$. Then ${\bf v}^\dagger
\dot{\bf v} = {\bf u}^\dagger \dot{\bf u} + i \dot \theta$. Since
${\bf u}^\dagger \dot{\bf u} + \dot{\bf u}^\dagger {\bf u} = 0$
$\theta$ can be chosen real such that ${\bf v}^\dagger \dot{\bf v} =
0$.}.  The Hermitian operator
\begin{equation} \label{Bdef}
  B = i\sum_n \sum_{m \neq n} \frac{Q_m \dot{A} Q_n}{p_n - p_m},
\end{equation}
where $Q_n = {\bf u}_n {\bf u}_n^\dagger$, then generates the
evolution of all the eigenvectors through the equation $ i\dot{\bf
u}_n = B {\bf u}_n$. The matrix elements of $B$ are
\begin{equation}
  {\bf u}_m^\dagger B {\bf u}_n = i \frac{{\bf u}_m^\dagger \dot A
    {\bf u}_n}{p_n-p_m}.
\end{equation}

The Schmidt states are the eigenvectors of the reduced density matrix
$\rho_r$. Since they are positive semi-definite hermitian matrices
eq.\ (\ref{uevol}) gives their time evolution. The corresponding
Schr\"odinger picture projection operators $Q_n = {\bf u}_n {\bf
u}_n^\dagger$ satisfy
\begin{equation}
  \dot{Q_n} \hspace{\arraycolsep} = \hspace{\arraycolsep} \dot{\bf
    u}_n {\bf u}_n^\dagger +{\bf u}_n \dot{\bf u}_n^\dagger
    \hspace{\arraycolsep} = \hspace{\arraycolsep} (-iB {\bf u}_n) {\bf
    u}_n^\dagger + {\bf u}_n (-iB {\bf u}_n)^\dagger
    \hspace{\arraycolsep} = \hspace{\arraycolsep} -i [B,Q_n],
\end{equation}
where $B$ is defined by eq.\ (\ref{Bdef}) with $\dot A = \dot\rho_r$.
This equation also holds for repeated eigenvalues in which case the
$Q_n$'s have rank equal to the multiplicity of the corresponding
eigenvalues.

In consistent histories it is frequently convenient to discuss
Heisenberg picture projection operators. Suppose the initial state is
$|\psi\rangle$ the time evolution operator is $U(t)$ and Schmidt
projectors of the form $P_n = Q_n\otimes I_2$ are considered. Then the
Heisenberg picture projection operators are $P_{Hn} = U^\dagger
Q_n\otimes I_2 U$ and satisfy the equation
\begin{eqnarray}\label{schmidtPHevol} 
  \dot{P}_{Hn} &=& \dot{U}^\dagger Q_n \otimes I_2 U + U^\dagger
  Q_n\otimes I_2 \dot{U} + U^\dagger \dot{Q_n}\otimes I_2 U \\ &=&
  \dot{U}^\dagger U P_{Hn} + P_{Hn} \dot{U} U^\dagger -i U^\dagger
  [B,Q_n] \otimes I_2 U \\ &=& i [H_H - U^\dagger B \otimes I_2 U ,
  P_{Hn}] \label{schrPevol}
\end{eqnarray}
where $H_H = -i\dot{U}^\dagger U $ is the (possibly time dependent)
Hamiltonian in the Heisenberg picture with units such that $\hbar =
1$, and $\dot{\rho}_{Hr} = U^\dagger \dot{\rho}_r \otimes I_2 U$.

\section{Degenerate eigenspaces}
The generator of the Schmidt evolution (eq.~\ref{Bdef}) is undefined
when two eigenvalues coalesce. This section shows that if an operator
is continuous and two eigenvectors are equal at a point then the
eigenspaces can be continuously defined through this point --- the
singularity in the evolution equation can be removed.

Consider a continuously parametrised Hermitian matrix $A(t)$ with two
eigenvalues $p(t) \pm \lambda(t)$ such that $\lambda(0)=0$ and with no
other degenerate eigenvalues. Let projection operators onto the
eigenspaces be denoted by $P_\pm(t)$.
\begin{eqnarray}\label{degone}
  A &=& (p+\lambda) P_1 + (p-\lambda) P_2 + C\\ &=& pX + \lambda Y +
  C.
\end{eqnarray}
$C$ contains the rest of the operator $A$, $X = P_1 + P_2$ and $Y =
P_1 - P_2$, so $X^2 = Y^2 = X$ and $XY = Y$. Let $d_1 =
\mbox{Rank}(P_1)$ and $d_2 = \mbox{Rank}(P_2)$. Note $X(t)$ is
continuous for all $t$ because $A(t)$ is continuous. To show that the
$P_1$ and $P_2$ are well defined it is necessary to find an expression
for $Y$ in terms of $X$ and $A$ only. Multiplying eq.~\ref{degone} by
$X$ from the left and right and rearranging it becomes
\begin{equation}\label{degtwo}
  Y = \frac{XAX - pX}{\lambda}\,,
\end{equation}
since $CXC = 0$.

Define $D = XAX - X\mbox{Tr}(XA) (d_1+d_2)^{-1}$, the traceless part
of $A$ in the degenerate eigenspace. Now
\begin{eqnarray*}
  XAX &=& D + \frac{X\mbox{Tr}(XA)}{(d_1+d_2)}\,,\\ \mbox{Tr}(XA) &=&
  p (d_1+d_2) + \lambda (d_1-d_2) \,, \\ \mbox{Tr}(XAXA) &=&
  (\lambda^2 + p^2) (d_1+d_2) + 2 p\lambda (d_1-d_2)\,,\\
  \mbox{Tr}(D^2) &=& \mbox{Tr}\left[XAXAX + \frac{X\mbox{Tr}^2(XA)}{
  (d_1+d_2)^2} - \frac{2XAX \mbox{Tr}(XA)}{d_1+d_2}\right] \\ &=&
  \mbox{Tr}(XAXAX) - \frac{\mbox{Tr}^2(XA)}{d_1+d_2} \\ &=& (\lambda^2
  + p^2) (d_1+d_2) + 2 p\lambda (d_1-d_2) - \frac{[p (d_1+d_2) +
  \lambda (d_1-d_2)]^2}{d_1+d_2} \\ &=& \lambda^2 \frac{(d_1+d_2)^2 -
  (d_1-d_2)^2}{d_1+d_2} \quad = \quad \frac{4\lambda^2
  d_1d_2}{d_1+d_2}.
\end{eqnarray*}

Substitute these results into eq.~(\ref{degtwo})
\begin{eqnarray}
  Y &=& [D + Xp + X\lambda\frac{d_1-d_2}{d_1+d_2} - pX] \lambda^{-1}
  \\ &=& X\frac{d_1-d_2}{d_1+d_2} + \frac{D}{\lambda} \\ &=&
  X\frac{d_1-d_2}{d_1+d_2} + \frac{2D \sqrt{d_1d_2)}}{
  \sqrt{\mbox{Tr}(D^2) (d_1+d_2)}}.
\end{eqnarray}
This provides a definition for $Y$ (and hence $P_1$ and $P_2$) which
is continuous for all $t$ including $t=0$, provided $D/\|D\|_{HS}$ is
continuous, which is a necessary and sufficient condition.

\section{The Gaussian Unitary Ensemble}\label{app:gue}
In the GUE the matrix elements are chosen according to the
distribution
\begin{eqnarray*}
  p(A) & = & \frac{2^{n/2}}{[(2\pi)^{1/2}\sigma]^{n^2}}
  \exp\left\{-\frac{\mbox{Tr}(A^2)}{4\sigma^2}\right\}\\ & = &
  \frac{2^{n/2}}{[(2\pi)^{1/2}\sigma]^{n^2}} \prod_{n\geq j,k \geq
  1}\exp\left\{-\frac{A_{jk}A_{kj}}{4\sigma^2}\right\}\\ & = &
  \frac{2^{n/2}}{[(2\pi)^{1/2}\sigma]^{n^2}} \prod_{n\geq j
  \geq=1}\exp\left\{-\frac{X_{jj}^2}{4\sigma^2}\right\} \prod_{n\geq k
  \geq j \geq 1} \exp\left\{-\frac{X_{jk}^2}{2\sigma^2}\right\}
  \prod_{n\geq k \geq j \geq
  1}\exp\left\{-\frac{Y_{jk}^2}{2\sigma^2}\right\}
\end{eqnarray*}
where $ A_{jk} = X_{jk} + i Y_{jk}$, $X_{jk} = X_{kj}$ and $Y_{jk} =
-Y_{kj}$. Therfore all the elements are independently, normally
distributed, the diagonal with variance $2\sigma$ and the real and
imaginary off-diagonal with variance $\sigma$. Some expectations for a
normal variable with variance $\sigma$ are
\begin{eqnarray*}
  \mbox{E}[|X|^{n}] & = &
  \frac{2^{n/2}\sigma^n\Gamma(\frac{p+1}{2})}{\sqrt{\pi}}
\end{eqnarray*}
and in particular $\mbox{E}[|X|] = \sqrt{(2/\pi)}\sigma$,
$\mbox{E}[X^2] = \sigma^2$ and $\mbox{E}[X^4] = 3\sigma^2$.  Since
$X_{ij}$ is independent of $X_{kl}$ unless $i=k$ and $j=l$, or $i=l$
and $j=k$
\begin{eqnarray*}
  \mbox{E}[X_{ij}X_{kl}] & = & \sigma^2(\delta_{il}\delta_{jk} +
  \delta_{ik}\delta_{jl}),\\ \mbox{E}[Y_{ij}Y_{kl}] & = &
  \sigma^2(\delta_{ik}\delta_{jl} - \delta_{il}\delta_{jk}).
\end{eqnarray*}
Therefore, for the elements of $A$
\begin{eqnarray*}
  \mbox{E}[A_{ij}] & = & 0, \\ \mbox{E}[A_{ij}A_{kl}] & = &
  2\sigma^2\delta_{il}\delta_{jk}, \\
  \mbox{E}[A_{ij}A_{kl}A_{mn}A_{op}] & = & 4\sigma^4
  (\delta_{il}\delta_{jk}\delta_{mp}\delta_{no} +
  \delta_{in}\delta_{jm}\delta_{kp}\delta_{lo} +
  \delta_{ip}\delta_{jo}\delta_{kn}\delta_{lm}).
\end{eqnarray*}
Applying these results to vectors and projection operators,
\begin{eqnarray}\label{expectgue}
  \mbox{E}[{\bf n}^\dagger A{\bf m}] & = & 0, \\\nonumber
  \mbox{E}[|{\bf n}^\dagger A{\bf m}|^2] & = & 2\sigma^2 |{\bf
  n}|^2|{\bf m}|^2,\\\nonumber \mbox{E}[{\bf n}^\dagger APA{\bf m}] &
  = & 2d\sigma^2{\bf n}^\dagger{\bf m},\\\nonumber \mbox{E}[|{\bf
  n}^\dagger APA{\bf m}|^2] & = & 4\sigma^4 [ d^2|{\bf n}^\dagger{\bf
  m}|^2 + ({\bf n}^\dagger P{\bf n}) ({\bf m}^\dagger P{\bf m}) + d
  |{\bf n}|^2|{\bf m}|^2],
\end{eqnarray}
where $d$ is the rank of $P$. For the real part or the imaginary part
just take half of the above since $|z|^2 = [\mbox{Re}(z)]^2 +
[\mbox{Im}(z)]^2$.


\chapter{Sphere-packing bounds}\label{upperboundappendix}

\section{Upper bounds using zonal spherical harmonic polynomials} 

Various authors~\cite{Kabatyanski:Levenshtein,Delsarte} have
constructed upper bounds for $M$ by using the properties of zonal
spherical harmonic polynomials, which for many spaces are the Jacobi
polynomials $P_n^{(\alpha,\beta)}(x)$. The bounds
\begin{equation}
  \label{Delsarte:bound}
  M({\bf S}^{d-1},\, |{\bf u}^T{\bf v}| \leq \epsilon) = N[(d-3)/2,
  -1/2, 2\epsilon^2-1],
\end{equation}
for $d \geq 3$, and
\begin{equation} 
  M({\bf CS}^{d-1},\, |{\bf u}^\dagger{\bf v}| \leq \epsilon) = N(d-2,
  0, 2\epsilon^2-1),
\end{equation}
for $d \geq 2$, have been proved by Kabatyanski et
al.~\cite{Kabatyanski:Levenshtein} and (\ref{Delsarte:bound}) also by
Delsarte et al.~\cite{Delsarte}.  Here $N(\alpha, \beta, s)$ is
defined as the solution to the following optimisation problem.

Consider $s$ as a given number $-1 \leq s < 1$. Let ${\cal R}
(\alpha,\beta,s)$ be the set of polynomials of degree at most $k$ with
the following properties:
\begin{eqnarray*}
  f(t) & = & \sum_{i=0}^k f_i P_i^{(\alpha,\beta)}(t), \\ f_i & \geq &
  0, \quad i = 0,1,\ldots,k, \quad \mbox{and} \quad f_0 > 0,\\ f(t) &
  \leq & 0 \quad \mbox{for} \quad -1 \leq t \leq s.
\end{eqnarray*}
Then
\begin{displaymath}
  N(\alpha, \beta, s) = \inf_{f(t) \in {\cal R}(\alpha, \beta, s)}
  f(1)/f_0.
\end{displaymath}
This can be converted to a linear program by defining
\begin{displaymath}
  \tilde{P}_i^{(\alpha,\beta)}(t) = P_i^{(\alpha,\beta)}(t) /
  P_i^{(\alpha,\beta)}(1).
\end{displaymath}
Then $ N(\alpha, \beta, s) = 1 + \sum_{i=1}^k f_i, $ where $
\sum_{i=1}^k f_i $ is minimised subject to $ f_i \geq 0 $ and $
\sum_{i=1}^k f_i \tilde{P}_i^{(\alpha,\beta)}(t) \leq -1, $ for $-1
\leq t \leq s$.
This formulation is discussed in Conway and Sloane~\cite{Conway}, but
no exact solutions are known. However, any $f(t)$ satisfying the
constraints does provide a bound, though it may not be optimal. I show
in appendix~\ref{alpha:half} that
\begin{displaymath}
  \tilde{P}^{(\alpha,-1/2)}_n(x) > \tilde{P}^{(\alpha,-1/2)}_1(x),
\end{displaymath}
if $-1 < x < -(2\alpha+3)(2\alpha+5)^{-1}$, $n > 1$ and $\alpha \geq
1$. So if $s$ is less than $-(2\alpha+3)/(2\alpha+5)$ then
$\tilde{P}^{(\alpha,-1/2)}_1(t)$ is more negative than any other of
the $\tilde{P}^{(\alpha,-1/2)}_i(t)$, and since
$\tilde{P}^{(\alpha,-1/2)}_1(t)$ is increasing the solution is
\begin{displaymath} 
  f_i = 0, \quad i = 2,3,\ldots,\quad f_1 =
  -1/\tilde{P}^{(\alpha,-1/2)}_1(s) \quad \forall k.
\end{displaymath}
So the optimal bound using zonal spherical harmonics is
\begin{eqnarray}\nonumber
  M({\bf CS}^{d-1}, \mbox{Re}({\bf u^\dagger v}) \leq \epsilon) & = &
  M({\bf S}^{2d-1},\, |{\bf u}^T{\bf v}| \leq \epsilon) \\\nonumber &
  \leq & N(d-3/2, -1/2, 2\epsilon^2-1), \\\nonumber & = & 1 -
  1/\tilde{P}^{(d-3/2,-1/2)}_1(2\epsilon^2-1) \\ \label{R:upper:bound}
  & = & \frac{2d(1-\epsilon^2)}{1-2d\epsilon^2},
\end{eqnarray}
if $\epsilon^2 \leq 1/(2d+2)$ and $d \geq 3$.  I prove a similar
inequality in appendix~\ref{alpha:zero} for $\alpha = 0$. So for the
medium DHC
\begin{eqnarray}\nonumber
  M({\bf CS}^{d-1}, |{\bf u^\dagger v}| \leq \epsilon) & \leq & N(d-2,
  0, 2\epsilon^2-1),\\ & = &
\label{M:upper:bound}
\frac{d(1-\epsilon^2)}{1-d\epsilon^2},
\end{eqnarray}
if $\epsilon^2 \leq 1/(d+1)$ and $d \geq 2$.

\section{Shannon's lower bound}\label{shannonap}
In a pioneering paper~\cite{Shannon} Shannon proved
\begin{theorem}\mbox{}
  \begin{equation}
    M({\bf S}^{d-1},|{\bf u}^T{\bf v}| \leq \cos\theta) \geq
    \sin^{1-d}\theta.
  \end{equation}
\end{theorem}
Let
\begin{equation} 
  S_d(r) = d r^{d-1}\pi^{d/2}/\Gamma[(d+2)/2]
\end{equation}
be the surface area of a sphere in Euclidean $d$-space of radius $r$,
and let $A_d(r,\theta)$ be the area of a $d$-dimensional spherical cap
cut from a sphere of radius $r$ with half angle $\theta$. It is not
hard to show that
\begin{equation} 
A_d (r,\theta) = \frac{(d-1)r^{d-1}\pi^{d-1/2}}{\Gamma[(d+2)/2]}
\int^{\theta}_{0}\sin^{d-2}\phi\, d \phi.
\end{equation}

Consider the largest possible set of rays through the origin
intersecting a sphere at points points ${\bf u} \in {\bf
S}^{d-1}$. About each point ${\bf u}$, consider the spherical cap of
all points on the sphere within $\theta$ degrees. Now, the set of all
such caps about each point ${\bf u}$ must cover the entire surface of
the sphere, otherwise we could add a new ray passing through the
uncovered areas. Since the area of each cap is $A_d(r,\theta)$, we
have
\begin{equation}
  2\, A_d(r,\theta) \,M({\bf S}^{d-1},|{\bf u}^T{\bf v}| \leq
  \cos\theta)\,\geq\, S_d(r) \mbox{.}
\end{equation}
But a spherical cap, $A_d(r,\theta)$, is contained within a hemisphere
of radius $r\sin\theta$, $A_d(r,\theta) \leq 1/2\,S_d(r\sin\theta)
$\footnote{This is easy to prove by changing variables in the integral
to $\sin\phi = \sin\theta\,\sin\psi$} , so
\begin{equation}
  M({\bf S}^{d-1},|{\bf u}^T{\bf v}| \leq \cos\theta) \geq
  S_d(r)/S_d(r\sin\theta) = \sin^{1-d}\theta
  \,\,{\rule[0ex]{1.5ex}{1.5ex}}
\end{equation}
or
\begin{equation}
  M({\bf CS}^{d-1},\mbox{Re}({\bf u^\dagger v)} \leq \cos\theta) \geq
  \sin^{1-2d}\theta.
\end{equation}

The straightforward extension of the proof to the complex case does
not appear to exist in the literature. It is slightly simpler as it is
easy to calculate the integral $A_d(r,\theta)$ exactly.
\begin{theorem}\mbox{}
  \begin{equation}
    M({\bf CS}^{d-1},|{\bf u^\dagger v}| \leq \cos\theta) \geq
    \sin^{2-2d}\theta
  \end{equation}
\end{theorem}
The area of a unit sphere in ${\bf CS}^{d-1}$ is $S_{2d}(1)$. Let
$A_d(1,\theta)$ now be the area of a cap defined by
\begin{equation}
  \{{\bf u} \in {\bf CS}^{d-1} : |u_1|^2 \geq \cos\theta\}.
\end{equation}
We can choose coordinates for a vector ${\bf u}$ in ${\bf CS}^{d-1}$
by defining
\begin{eqnarray*}
  \mbox{Re}{(u_1)} & = & \cos\phi_1, \\ \mbox{Im}{(u_1)} & = &
  \sin\phi_1\,\cos\phi_2, \\ \vdots & \vdots & \vdots, \\
  \mbox{Re}{(u_d)} & = &
  \sin\phi_1\,\sin\phi_2\,\sin\phi_3\ldots\sin\phi_{2d-2}\cos\psi,\\
  \mbox{Im}{(u_d)} & = &
  \sin\phi_1\,\sin\phi_2\,\sin\phi_3\ldots\sin\phi_{2d-2}\sin\psi,
\end{eqnarray*}
where $\phi_n \in [0,\pi)$ and $\psi \in [0,2\pi)$.Then, by
integrating over $\phi_2,\phi_3,\ldots,\phi_{2d-2}$ and $\psi$, we get
\begin{eqnarray}\nonumber
  A_d(1,\theta) &=& S_{2d-2}(1)\,
  \!\!\!\!\!\!\!\!\!\!\!\!\!\!\!\!\!\!\!\!\!\!\!
  \mathop{\int\!\!\int}_{\cos^2\phi_1 +\sin^2\phi_1\cos^2\phi_2 \geq
  \cos\theta} \!\!\!\!\!\!\!\!\!\!\!\!\!\!\!\!\!\!\!\!\!\!\!
  \sin^{d-2}\phi_1\sin^{d-3}\phi_2\, d \phi_1\, d \phi_2\\ &=&
  \frac{\pi S_{2d-2}(1) \sin^{2d-2}\theta}{d-2}.
\end{eqnarray} 
Hence, using Shannon's argument again,
\begin{eqnarray}\nonumber
  M({\bf CS}^{d-1},|{\bf u^\dagger v}| \leq \cos\theta) &\geq& \frac
  {(d-2)\,S_{2d}(1)} {\pi S_{2d-2}(1)\sin^{2d-2}\theta}\\ &\geq&
  \sin^{2-2d}\theta \,\,{\rule[0ex]{1.5ex}{1.5ex}}
\end{eqnarray}

Expressed in terms of $\epsilon = \cos\theta$ the bounds are
\begin{eqnarray*}
  M({\bf CS}^{d-1}, \mbox{Re}({\bf u^\dagger v}) \leq \epsilon) & \geq
  & (1-\epsilon^2)^{1/2-d}, \\ \mbox{and} \qquad M({\bf CS}^{d-1},
  |{\bf u^\dagger v}| \leq \epsilon) & \geq & (1-\epsilon^2)^{1-d}.
\end{eqnarray*}

\section{Jacobi polynomials}
I have used trivial properties of the Jacobi polynomials without
citation. All of these results can be found in chapter IV of
Szeg\"o~\cite{Szego}, which provides an excellent introduction to, and
reference source for, the Jacobi polynomials.

\subsection{$S^{d-1}$, $\beta = -1/2$}\label{alpha:half}
In ${\bf S}^{d-1}$ the zonal spherical polynomials are
$P^{(\alpha,-1/2)}_n(x)$ with $\alpha = (d-3)/2$.
\begin{theorem}\mbox{}
  \begin{equation} 
    \label{inequalitybhalf}
    \tilde{P}^{(\alpha,-1/2)}_n(x) > \tilde{P}^{(\alpha,-1/2)}_1(x)
  \end{equation}
  for $-1<x<-(2\alpha+3)(2\alpha+5)^{-1}$, $n > 1$ and $\alpha \geq
  1$, where
  \begin{displaymath}
    \tilde{P}^{(\alpha,-1/2)}_n(x) =
    P^{(\alpha,-1/2)}_n(x)/P^{(\alpha,-1/2)}_n(1)\,.
  \end{displaymath} 
\end{theorem}
I begin by considering two special cases, $n = 2$ and $n = 3$. The
first four polynomials are:
\begin{eqnarray*}
  \tilde{P}^{(\alpha,-1/2)}_0 (x) & = & 1 \\
  \tilde{P}^{(\alpha,-1/2)}_1 (x) & = & \frac{2\alpha + 1 + (2\alpha +
  3)x}{4(\alpha+1)}\\ \tilde{P}^{(\alpha,-1/2)}_2 (x) & = &
  \frac{4\alpha^2 - 13 + 2(2\alpha+1)(2\alpha+5)x
  +(2\alpha+5)(2\alpha+7)x^2}{ 16(\alpha+1)(\alpha+2)}\\
  \tilde{P}^{(\alpha,-1/2)}_3 (x) & = &
  \frac{(2\alpha+1)(4\alpha^2-8\alpha-57) +
    3(2\alpha+7)(4\alpha^2-21)x}{}\\&& 
  \frac{\mbox{} +3(2\alpha+1)(2\alpha+7)(2\alpha+9)x^2}{}\\&&
  \frac{\mbox{} +(2\alpha+7)(2\alpha+9)(2\alpha+11)x^3}{64(\alpha+1)
  (\alpha+2)(\alpha+3)}.
\end{eqnarray*}
So
\begin{eqnarray}\label{p2-p1-1/2} 
  \tilde{P}^{(\alpha,-1/2)}_2 (x) - \tilde{P}^{(\alpha,-1/2)}_1 (x)
  &=& \frac{-(2\alpha+7)(1-x)[2\alpha+3 + (2\alpha+5)x]}{
  16\,(\alpha+1)(\alpha+2)}
\end{eqnarray}
\begin{eqnarray}\nonumber
  \tilde{P}^{(\alpha,-1/2)}_3 (x) - \tilde{P}^{(\alpha,-1/2)}_1 (x)
  &=& \frac{-(2\alpha+9)(1-x)[(2\alpha+1)(6\alpha+17)}{}\\&&
  \frac{\mbox{} + 2\,(2\alpha+7)(4\alpha+7)x +
    (2\alpha+7)(2\alpha+11)x^2]}{64\,(\alpha+1)(\alpha+2)(\alpha+3)}.\mbox{~~~~}
\label{p3-p1-1/2}
\end{eqnarray}
Equation (\ref{p2-p1-1/2}) is positive for \mbox{$x
  <-(2\alpha+3)(2\alpha+5)^{-1} $} (hence the range chosen for
  (\ref{inequalitybhalf}).)  Equation (\ref{p3-p1-1/2}) is positive
  where the quadratic factor
\begin {equation}
  (2\alpha+1)(6\alpha+17) + 2(2\alpha+7)(4\alpha+7)x
  +(2\alpha+7)(2\alpha+11)x^2 \label{quadratic}
\end{equation}
is negative. Since (\ref{quadratic}) is positive for large $|x|$ if it
is negative at any two points it will be negative in between. At $x =
-1$ it is $-4\,(2\alpha+1)$, and at $x = -(2\alpha+3)(2\alpha+5)^{-1}$
it is $ -16\,(\alpha+2)(2\alpha+11)(5+2\alpha)^{-2}$, which is
negative for $\alpha > -2$.  So the inequality (\ref{inequalitybhalf})
holds for $n=2$ and $n=3$.

For $n > 3$ the inequality is easily proved, by bounding the solutions
of the Jacobi differential equation,
\begin{equation}
  (1-x^2) y''(x)+[\beta - \alpha - (\alpha+\beta+2)x] y'(x) +
  n(n+\alpha+\beta+1) y(x) = 0,
\label{jacobiode1}
\end{equation} 
where $y(x)=P^{(\alpha,\beta)}_n(x)$.  Define $w(s) = (1-s^2)^\alpha
y(2s^2-1)$, $s \in [0,1]$. Substituting $\beta = -1/2$ into equation
(\ref{jacobiode1}) it becomes
\begin{eqnarray}
  \left[\frac{w'(s)}{(1-s^2)^{\alpha-1}}\right]' +
  \frac{2(\alpha+n)(1+2n)w(s)}{(1-s^2)^{\alpha}} = 0,
\label{jacobiode2}
\end{eqnarray}
which is of the form
\begin{displaymath}
  [k(s)w'(s)]'+\phi(s)w(s) = 0
\end{displaymath}
with $k(s)$ and $\phi(s)$ positive, and $k(s)\phi(s)$ increasing, if
$\alpha$ and $n$ are positive. These are the necessary conditions for
the Sonine-P\"olya theorem~(\ref{SP:theorem}), which states that the
local maxima of $|w(s)|$ will be decreasing. From its definition
$|w(s)|$ has a local maximum at $s = 0$, since $w(0)w''(0) < 0 $, and
a local minimum at $s = 1$, since $w(0) = 0$. $w(s)$ is continuous so
it is bounded by its local maxima, hence $|w(s)| \leq |w(0)|$, for $s
\in [0,1]$. In the original variables this is
\begin{eqnarray}
  \left(\frac{1-x}{2}\right)^\alpha
    \left|P^{(\alpha,-1/2)}_n(x)\right| & \leq &
    \left|P^{(\alpha,-1/2)}_n(-1)\right| \label{ode-1/2bound}.
\end{eqnarray}
Substituting in the values of $P^{(\alpha,-1/2)}_n(-1)$ and
$P^{(\alpha,-1/2)}_n(1)$ this becomes\footnote{The Pochhammer symbol
$(a)_n = \Gamma(a+n)/\Gamma(a)= a(a+1)\ldots(a+n-1)$.}
\begin{eqnarray}
  |\tilde{P}^{(\alpha,-1/2)}_n(x)| & \leq &
  \frac{(1/2)_n}{(\alpha+1)_n} \left(\frac{2}{1-x}\right)^\alpha ,
\end{eqnarray}
for $-1 \leq x \leq 1$. The right hand side is decreasing with $n$ if
$\alpha > -1/2$.  So for $n \geq 4$
\begin{eqnarray}
  |\tilde{P}^{(\alpha,-1/2)}_n(x)| & < &
   \frac{105/16}{(\alpha+1)_4}\left(\frac{2}{1-x}\right)^\alpha.
\end{eqnarray}
This is increasing with x so achieves its maximum at $x =
-(2\alpha+3)/(2\alpha+5)$. Thus
\begin{eqnarray}
  |\tilde{P}^{(\alpha,-1/2)}_n(x)| & \leq &
  \frac{105/16}{(\alpha+1)_4}
  \left(\frac{2\alpha+5}{2\alpha+4}\right)^\alpha\,.
\end{eqnarray}
For $\alpha \geq 1$ this is strictly bounded by
\begin{eqnarray}
  \frac{1}{(\alpha+1)(2\alpha+5)} & = &
  \left|\tilde{P}^{(\alpha,-1/2)}_1\left(-\frac{2\alpha+3}{2\alpha+5}\right)
  \right|,
\end{eqnarray}
and since it is decreasing and $x \leq -(2\alpha+3)(2\alpha+5)^{-1}$
\begin{eqnarray}
  |\tilde{P}^{(\alpha,-1/2)}_n(x)| & < &
  |\tilde{P}^{(\alpha,-1/2)}_1(x)|.
\end{eqnarray}
But $\tilde{P}^{(\alpha,-1/2)}_1(x)$ is negative on the range of $x$
so
\begin{eqnarray}
  \tilde{P}^{(\alpha,-1/2)}_n(x) & > & \tilde{P}^{(\alpha,-1/2)}_1(x)
  \,\,{\rule[0ex]{1.5ex}{1.5ex}}
\end{eqnarray}

\subsection{$CS^{d-1}$, $\beta = 0$}
\label{alpha:zero}
In ${\bf CS}^{d-1}$ the zonal spherical polynomials are
$P^{(\alpha,0)}_n(x)$, where $\alpha = d - 2$, and a similar theorem
exists.
\begin{theorem}\mbox{}
  \begin{equation}
    \label{inequalityb0}
    \tilde{P}^{(\alpha,0)}_n(x) > \tilde{P}^{(\alpha,0)}_1(x),
  \end{equation}
  for $-1 < x < -(\alpha+1)(\alpha+3)^{-1}$, $n > 1$ and $\alpha \geq
  2$, where $\tilde{P}^{(\alpha,0)}_n(x) =
  P^{(\alpha,0)}_n(x)/P^{(\alpha,0)}_n(1)$.
\end{theorem}
I begin by considering two special cases, $n=2$ and $n=3$.  The first
four polynomials are
\begin{eqnarray*}
  \tilde{P}^{(\alpha,0)}_0(x) & = & 1 \\ \tilde{P}^{(\alpha,0)}_1(x) &
  = & \frac{\alpha + (\alpha+2)x}{2(\alpha+1)} \\
  \tilde{P}^{(\alpha,0)}_2(x) & = &
  \frac{\alpha^2-\alpha-4+2\alpha(\alpha+3)x +
  (\alpha+3)(\alpha+4)x^2}{ 4(\alpha+1)(\alpha+2)} \\
  \tilde{P}^{(\alpha,0)}_3(x) & = & \frac{\alpha(\alpha^2-3\alpha-16)
  + 3(\alpha-3)(\alpha+2)(\alpha+4)x + 3\alpha(\alpha+4)(\alpha+5)x^2
  }{}\\&&
  \frac{\mbox{}+ (\alpha+4)(\alpha+5)(\alpha+6)x^3}{
  8(\alpha+1)(\alpha+2)(\alpha+3)}
\end{eqnarray*}
So
\begin{eqnarray}
  \label{p2-p1}
  \tilde{P}^{(\alpha,0)}_2(x)-\tilde{P}^{(\alpha,0)}_1(x) & = &
  -\frac{(\alpha+4)(1-x)(\alpha+1+(3+\alpha)x)}{
  4(\alpha+1)(\alpha+2)} \\ \label{p3-p1}
  \tilde{P}^{(\alpha,0)}_3(x)-\tilde{P}^{(\alpha,0)}_1(x) & =
  &\nonumber
  -\frac{(\alpha+5)(1-x)[\alpha(3\alpha+8)+2(\alpha+4)(2\alpha+3)x
  }{}\\&&\frac{\mbox{}+(\alpha+4)(\alpha+6)x^2]}{8(\alpha+1)(\alpha+2)(\alpha+3)}
\end{eqnarray}
Equation (\ref{p2-p1}) is positive for $x <
-(\alpha+1)(\alpha+3)^{-1}$ (hence the range chosen for
(\ref{inequalityb0}).)  Equation (\ref{p3-p1}) is positive when the
quadratic factor
\begin{equation}
  \label{quadraticfactor2}
  \alpha(3\alpha+8)+2(\alpha+4)(2\alpha+3)x+(\alpha+4)(\alpha+6)x^2,
\end{equation}
is negative.  At $ x = - 1 $ equation (\ref{quadraticfactor2}) equals
$-4\alpha$, and at $ x = -(\alpha+1)/(\alpha+3)$ it equals
$-4(\alpha+2)(\alpha+6)(\alpha+3)^{-2}$ both of which are negative for
$\alpha > 0$. Therefore inequality (\ref{inequalityb0}) holds for $n =
2$ and $n = 3$.

The method in the previous appendix cannot be used unless $\beta =
\pm1/2$, since the differential equation has a singular point at $x =
-1$.  There is a simple method\footnote{Szeg\"o proves that for
polynomials $p(s)$ orthogonal with weight function $w(s)$, that if
$w(s)$ is non-decreasing then $[w(s)]^{1/2}|p(s)|$ is non-decreasing
also. The weight measure over which $P^{(\alpha,0)}$ are orthogonal is
$(1-x)^\alpha \mbox{dx}$. After changing variable to $x=2s^2-1$ the
new measure is proportional to $s^{(2\alpha+1)}\mbox{ds}$, which is
non-decreasing.} for the special case of $\beta = 0$, based on a
result from Szeg\"o (7.21.2)~\cite{Szego}
\begin{equation}\label{temp2}
  \left[(1-x)/2\right]^{\alpha/2+1/4}\left|P^{(\alpha,0)}_n(x)\right|
    \leq 1,
\end{equation}
when $-1\leq x \leq 1$ and $\alpha \geq -1/2$.  Substituting
$P^{(\alpha,0)}_n(1) = (\alpha + 1)_n/n!{}$ into~(\ref{temp2}) it
becomes
\begin{eqnarray}
  \left|\tilde{P}^{(\alpha,0)}_n(x)\right| & \leq &
    \frac{n!}{(\alpha+1)_n}
    \left(\frac{2}{1-x}\right)^{(\alpha/2+1/4)}.
\end{eqnarray}
For $\alpha > 0 $ the right hand side is decreasing with $n$, so for
$n \geq 4$
\begin{eqnarray}
  \left|\tilde{P}^{(\alpha,0)}_n(x)\right| & \leq &
    \frac{4!}{(\alpha+1)_4} \left(\frac{2}{1-x}
    \right)^{(\alpha/2+1/4)}.
\end{eqnarray} 
This is decreasing with $x$ so achieves its maximum at $x \leq
-(\alpha+1)(\alpha+3)^{-1}$. Thus
\begin{eqnarray}
  \left|\tilde{P}^{(\alpha,0)}_n(x)\right| & \leq &
    \frac{4!}{(\alpha+1)_4} \left(\frac{\alpha+2}{\alpha+3} \right)
    ^{(\alpha/2+1/4)}.
\end{eqnarray}
For $\alpha \geq 2$ this is strictly bounded by
\begin{eqnarray}
    \frac{1}{(\alpha+1)(\alpha+3)} &=&\left|
    \tilde{P}^{(\alpha,0)}_1\left(-\frac{\alpha+1}{\alpha+3}
    \right)\right|.
\end{eqnarray}
Since $|\tilde{P}^\alpha_1(x)|$ is monotonic increasing
\begin{eqnarray}
  \left|\tilde{P}^{(\alpha,0)}_n(x)\right| & < &
    \left|\tilde{P}^{(\alpha,0)}_1(x)\right|.
\end{eqnarray}
But $\tilde{P}^\alpha_1(x)$ is negative on the range so
\begin{displaymath}
  \tilde{P}^{(\alpha,0)}_n (x) > \tilde{P}^{(\alpha,0)}_1(x)
  \,\,{\rule[0ex]{1.5ex}{1.5ex}}
\end{displaymath}

\section{Sonine-P\"olya theorem}
\label{sonine}
This standard theorem is referred to in~\cite[7.31.2]{Szego}.
\begin{theorem}\label{SP:theorem}
  Let $y(x)$ be a solution of the differential equation
  \begin{displaymath} 
    [k(x)y'(x)]'+\phi(x)y(x) = 0 \mbox{.}
  \end{displaymath}
  If $k(x)$ and $\phi(x)$ are positive, and $k(x)\phi(x)$ is
  increasing (decreasing) and its derivative exists, then the local
  maxima of $|y(x)|$ are decreasing (increasing).
\end{theorem}
Let
\begin{displaymath}
  f(x) = [y(x)]^2 + [k(x)y'(x)]^2[k(x)\phi(x)]^{-1}
\end{displaymath}
then $f(x) = [y(x)]^2$ if $y'(x)=0$, and
\begin{displaymath}
  f' = 2y' \left\{ y + \frac{[ky']'}{k\phi}
  -\frac{[k\phi]'y'}{2\phi^2} \right\} = -\frac{y'^2 [k\phi ]
  '}{\phi^2}.
\end{displaymath}
So $\mbox{sgn} f'(x) =
-\mbox{sgn}[k(x)\phi(x)]'\,\,{\rule[0ex]{1.5ex}{1.5ex}}$

\chapter{An example of large probability violation}\label{dheg}
Consider the $2n$ vectors
\begin{eqnarray}
  ({\bf u}_i)_j = \frac{a\delta_{ij}-1}{\sqrt{a^2-2a+n}} \in
  \mathcal{H}_1,\\ ({\bf v}_i)_j =
  \frac{b\delta_{ij}+1}{\sqrt{b^2+2b+n}}\in \mathcal{H}_2,
\end{eqnarray}
where
\begin{eqnarray}
  a = \frac{1+\epsilon+\sqrt{(1+\epsilon)(1+\epsilon-n\epsilon)}}
  {\epsilon},\\ b =
  \frac{1-\epsilon+\sqrt{(1-\epsilon)(1-\epsilon+n\epsilon)}}
  {\epsilon}
\end{eqnarray}
and $\epsilon \leq 1/(n-1)$.  Then
\begin{equation}
  {{\bf u}_i}^\dagger{\bf u}_j = \delta_{ij}(1+\epsilon)-\epsilon
  \quad\mbox{and}\quad {{\bf v}_i}^\dagger{\bf v}_j =
  \delta_{ij}(1-\epsilon)+\epsilon.
\end{equation}
Define
\begin{equation}
  {\bf w}_i = ({\bf u}_i \oplus {\bf u}_i)/\sqrt{2} \in \mathcal{H}_1
  \oplus \mathcal{H}_2,
\end{equation}
so
\begin{equation}
  {\bf w}_i^\dagger{\bf w}_j = \delta_{ij}.
\end{equation}
Let the initial state be
\begin{equation}
  \psi = \frac{1}{\sqrt{n}}\sum_{i=1}^n{\bf w}_i \in \mathcal{H}_1
  \oplus \mathcal{H}_2.
\end{equation}
Then use $\{{\bf w}_i{\bf w}_i^\dagger\}$ as a set of projectors to
get the history states
\begin{equation}
  \left\{{\bf w}_1/\sqrt{n},\ldots,{\bf w}_n/\sqrt{n}\right\}
\end{equation}
Then make a projection onto $\mathcal{H}_1$ and $\mathcal{H}_2$ to get
the history states
\begin{equation}
  \left\{{\bf u}_1/\sqrt{2n},\ldots,{\bf u}_n/\sqrt{2n},{\bf
      v}_1/\sqrt{2n},\ldots,{\bf v}_n/\sqrt{2n}\right\}.
\end{equation}
The decoherence matrix can be written
\begin{equation}
  \frac{1}{2n} \left(
    \begin{array}{cccccccc}
             1 & -\epsilon & \ldots & -\epsilon & 0 & \ldots & \ldots
             & 0 \\ -\epsilon & \ddots & \ddots & \vdots & \vdots &
             \ddots & \ddots & \vdots \\ \vdots & \ddots & \ddots &
             -\epsilon & \vdots & \ddots & \ddots & \vdots \\
             -\epsilon & \ldots & -\epsilon & 1 & 0 & \ldots & \ldots
             & 0 \\ 0 & \ldots & \ldots & 0 & 1 & \epsilon & \ldots &
             \epsilon \\ \vdots & \ddots & \ddots & \vdots & \epsilon
             & \ddots & \ddots & \vdots \\ \vdots & \ddots & \ddots &
             \vdots & \vdots & \ddots & \ddots & \epsilon \\ 0 &
             \ldots & \ldots & 0 & \epsilon & \ldots & \epsilon & 1
    \end{array}
  \right).
\end{equation}
The MPV for this set is $|-n(n-1)\epsilon/(2n)| = (n-1)\epsilon/2
\approx d\epsilon/4$. It is achieved by coarse-graining all the ${\bf
u}_i$'s (or ${\bf v}_i$'s) together.

\chapter{Quantum Zeno effect}\label{zenoapp}
The Quantum Zeno effect is often discussed in the interpretation of
quantum mechanics, but has had no quantitative analysis in the
consistent histories formalism.

Consider a two dimensional Hilbert space.  Define the vectors
\begin{equation}
  {\bf u}^n_+ = \left(\begin{array}{c}\cos(n\epsilon)\\
\sin(n\epsilon)\end{array}\right), {\bf u}^n_- =
\left(\begin{array}{c}-\sin(n\epsilon)\\\cos(n\epsilon)\end{array}\right),
  \label{vectordef}
\end{equation}
and the projectors
\begin{equation}
  P^n_+ = {\bf u}^n_+ {{\bf u}^{n}_+}^\dagger, P^n_- = {\bf u}^n_-
  {{\bf u}^{n}_-}^\dagger.
  \label{projectordef}
\end{equation}
For any $n$, $P^n_+$ and $P^n_-$ are a complete set of projectors.
Consider the set of histories formed by using strings of these
projectors on the initial state ${\bf u}^0_+$.
\begin{equation}
  C_\alpha = P^n_{\alpha_n} \ldots P^1_{\alpha_1}.
\end{equation}
The histories $\alpha$ are string of $n$ pluses or minuses.

Define $|\alpha|$ to be the number of transitions from plus to minus
or vice versa in the string $\{\alpha_1,\ldots,\alpha_n,+\}$. Then
\begin{equation}
  C_\alpha {\bf u}^0_+ = {\bf
    u}^n_{\alpha_n}(-1)^{\lfloor\frac{|\alpha| +1}{2}\rfloor}
    \cos^{n-|\alpha|}\epsilon\sin^{|\alpha|}\epsilon,
\end{equation}
and there will be
$n\choose|\alpha|$ identical histories states.  The non-zero
decoherence matrix elements are those with $|\alpha| = |\beta|
\bmod{2}$ and are
\begin{equation} \label{egdm}
  D_{\alpha\beta} = (-1)^{\lfloor\frac{|\alpha| +1}{2}\rfloor}
  (-1)^{\lfloor\frac{|\beta| +1}{2} \rfloor}
  \cos^{2n-|\alpha|-|\beta|}\epsilon\sin^{|\alpha|+|\beta|}\epsilon,
\end{equation}

Because of the simple form of (\ref{egdm}) all of the following
calculations can be done exactly, but for simplicity I shall let
$\epsilon = \theta/n$ and work to leading order in $1/n$.  The largest
probability violation for this decoherence matrix will be achieved by
coarse-graining together all the histories with a positive sign into
one history and all those with a negative sign into another.  Let $X$
denote the histories $|\alpha| = 0,3 \bmod 4$ and $Y$ the histories
$|\alpha| = 1,2 \bmod 4$. Then the probability violations for these
sets are,
\begin{eqnarray*}
  \sum_{\alpha\neq\beta\in X}D_{\alpha\beta} & = & 1/2 \cosh^2\theta +
  1/2 \cos\theta \cosh\theta \\&&\mbox{}- 1/2 \sin\theta \sinh\theta -
  1+ O(1/n)
\end{eqnarray*}
for $X$ and
\begin{eqnarray*}
  \sum_{\alpha\neq\beta\in Y}D_{\alpha\beta} &=& 1/2\cosh^2\theta -
  1/2\cos\theta\cosh\theta \\&&\mbox{}+ 1/2\sin\theta\sinh\theta+
  O(1/n)
\end{eqnarray*}
for $Y$.  The off-diagonal elements in the decoherence matrix
(\ref{egdm}) are all less than $\theta^2/n^2$ yet the MPV is order
$\exp{(2\theta)}$, so by choosing $n \gg \theta \gg 1$ the
off-diagonal elements can be made arbitrarily small whilst the MPV is
arbitrarily large. This proves the following theorem.
\begin{theorem}\label{proveuseless} 
  For all Hilbert spaces of dimension $\geq 2$, $\epsilon > 0$ and $x
  > 0 $ there exist finite sets of histories such that
  \begin{equation}
    |D_{\alpha\beta}| \leq \epsilon, \quad \forall \alpha \neq \beta,
  \end{equation}
  and with $\mbox{MPV} > x$.
\end{theorem}
Now suppose the limit $n \to \infty$ is taken. Then all the elements
of the decoherence matrix (\ref{egdm}) are zero except for
$D_{\alpha\alpha} = 1$, $\alpha = \{+\cdots+\}$. A naive argument
would be to say that since all the off-diagonal elements are zero the
set is consistent, but this is false. The set is pathologically
inconsistent.

This shows that care must be taken with infinite sets of histories. It
is incorrect to take the limit of a set of histories and then apply
consistency criteria. Instead the order must be reversed and the limit
of the criteria taken. This does not always seem to have been
recognised in the literature. For instance
Halliwell~\cite{Halliwell:fluctuations} states : ``In particular, it
[$|D_{\alpha\beta}| \leq (D_{\alpha\alpha}D_{\beta\beta})^{1/2}$]
implies that consistency is automatically satisfied if the system has
one history with $D_{\alpha\alpha} = 1$, and $D_{\beta\beta} = 0$ for
all other histories.'' He says this after a similar limit has been
taken, and I have shown above that this is not necessarily true.

The DHC trivially rejects this family of histories as grossly
inconsistent since
\begin{equation}
  \frac{D_{\alpha\beta}}{(D_{\alpha\alpha} D_{\beta\beta})^{1/2}} = 1,
  \quad \mbox{whenever $|\alpha| = |\beta| \bmod2$}.
\end{equation}


\chapter{Computer programs}\label{app:programs}

A suite of computer programs was written in MATLAB and C that can run
different variants of the algorithm in order to analyse its
predictions. The programs can run the spin model or the random
Hamiltonian model. For the spin model the interaction can overlap and
the form of the $\theta(t)$ functions can be varied. The programs can
simulate an algorithm which makes a projection at the earliest
possible time or one that looks for the consistency of a double
projection or a local information maximum.

There are five main programs: \texttt{createdata.m}~(\ref{createdata.m})
which creates the input file containing all the parameter values, the
Hamiltonian and the initial state, \texttt{model.m}~(\ref{model.m})
which reads in the input file, and repeatedly searches forwards for
the next projection time and then creates the output file,
\texttt{calcdecoh.m}~(\ref{calcdecoh.m}) which is called by
\texttt{model.m} and calculates at a particular time the decoherence
matrices for all non-trivial extensions of all histories,
\texttt{analyse.m}~(\ref{analyse.m}) which performs checks on the output
file and calculates statistics about the simulation and then creates
the analyse file, and \texttt{displaydata.m}~(\ref{displaydata.m}) which
reads in the input, output and analyse files and plots graphs of
requested variables.

The program \texttt{model.m}~(\ref{model.m}) operates in the
Heisenberg picture.  At a time when there are no possible consistent
extensions the function searches forward picking time steps using the
Newton-Raphson method. Once a time has been found where a consistent
extension is possible the function searches back to find the earliest
such time. In the standard algorithm a projection is then made. In
more complicated variants (when $\texttt{maxinf} = 1$) the function
then searches forward until other criteria are satisfied. The function
finishes when \texttt{maxhists} histories have been produced,
\texttt{maxsteps} program steps have occurred, or \texttt{tmax} has
been reached.

\texttt{calcdecoh.m}~(\ref{calcdecoh.m}) is called by
\texttt{model.m}~(\ref{model.m}) at every time step. It calculates the
decoherence matrix for every projection combination on every history.
If requested it also calculates the limit DHP for repeated projections
and the rate of change of the probabilities by using exact expressions
for the time derivative.

\begin{singlespace}
  \section{Main simulation}
  \includeprogram{createdata}
  \includeprogram{model}
  \includeprogram{calcdecoh}
  \includeprogram{tevol}
  \includeprogram{tfuncts}
  \includeprogram{analyse}
  \includeprogram{displaydata}
  \includeprogram{displaytree}
  \includeprogram{checkconsistency}
  \includeprogram{dhc}
  \section{Variables}
  \includeprogram{invariables}
  \includeprogram{outvariables}
  \includeprogram{globalvariables}
  \section{Auxiliary functions}
  \includeprogram[c]{condsum}
  \includeprogram{randunitary}
  \includeprogram{randvector}
  \includeprogram{schmidt}
  \section{Statistics}
  \includeprogram{calclongtime}
  \includeprogram{mcpercentileplot}
\end{singlespace}

\bibliography{/home/quark/jnm11/tex/papers,%
/home/quark/jnm11/tex/books,/home/quark/jnm11/tex/conversations}
\bibliographystyle{/home/quark/jnm11/tex/prsty}
\end{document}